\documentclass{elsart}
\usepackage{graphicx,amssymb,lineno}

\makeatletter
\def\elsartstyle{%
    \def\normalsize{\@setfontsize\normalsize\@xiipt{14.5}}
    \def\small{\@setfontsize\small\@xipt{13.6}}
    \let\footnotesize=\small
    \def\large{\@setfontsize\large\@xivpt{18}}
    \def\Large{\@setfontsize\Large\@xviipt{22}}
    \skip\@mpfootins = 18\p@ \@plus 2\p@
    \normalsize}
\@ifundefined{square}{}{\let\Box\square}
\makeatother

\begin{document}

\begin{frontmatter}
\title{Dipole and quadrupole solitons in optically-induced two-dimensional defocusing photonic lattices}

\author[umass]{H.\ Susanto},
\author[umass]{K.\ Law},
\author[umass]{P.G.\ Kevrekidis\corauthref{cor}},
\corauth[cor]{Corresponding author.}
\ead{kevrekid@math.umass.edu}
\author[teda]{L.\ Tang},
\author[teda]{C. Lou},
\author[teda,sfsu]{X. Wang}
\author[teda,sfsu]{Z.\ Chen}
\address[umass]{Department of Mathematics and Statistics, University of Massachusetts, Amherst, MA 01003, USA}
\address[teda]{Key Laboratory for Weak-Light Nonlinear Photonics Materials, Ministry of Education,and TEDA Applied Physical School, Nankai University, Tianjin 300457, China}
\address[sfsu]{Department of Physics and Astronomy, San Francisco State University,\\ San Francisco, CA 94132, USA}

\begin{abstract}
Dipole and quadrupole solitons in a two-dimensional optically induced
defocusing photonic lattice are theoretically predicted and experimentally
observed. It is shown that in-phase nearest-neighbor and
out-of-phase next-nearest-neighbor dipoles exist and can be stable in the
intermediate intensity regime. There are also different types of dipoles that are
always unstable.
In-phase nearest-neighbor quadrupoles are also numerically obtained,
and may also be linearly stable. Out-of-phase,
nearest-neighbor quadrupoles are found to be typically unstable.
These numerical results are found to be aligned with the
main predictions obtained analytically in the discrete nonlinear
Schr{\"o}dinger model.
 Finally, experimental results are presented for both dipole and quadrupole
structures, indicating that self-trapping of such structures in the
defocusing lattice can be realized for the length of the nonlinear
crystal (10 mm).
\end{abstract}

\end{frontmatter}

\section{Introduction}
\label{intro}

Self-trapping of light in photonic lattices optically induced in
nonlinear photorefractive crystals,
such as
strontium barium niobate (SBN) has attracted a considerable amount
of attention, ever since its theoretical inception \cite{efrem}
and experimental realization \cite{moti1,neshevol03,martinprl04}.
This is perhaps primarily because it constitutes a setting where
it is very natural to consider the competing effects of nonlinearity
with diffraction and to study the effects of periodic ``potentials''
on solitary waves; here the role of the ``potential'' is played by the
ordinary polarization of light forming a waveguide array in which the
extra-ordinarily polarized probe beam evolves.

This setting has provided a fertile ground for the experimental
realization and detailed examination of many interesting concepts
of nonlinear wave physics, including, for instance, the formation
of discrete dipole \cite{yang04}, necklace \cite{neck} solitons and
even stripe patterns \cite{multi}, rotary solitons \cite{rings},
discrete vortices \cite{vortex} or the realization of photonic
quasicrystals \cite{moti22} and Anderson localization \cite{moti23}.
It is clear from these findings that this setting can serve
not only as a host for the examination of localized structures
that may be usable as carriers and conduits for data transmission
and processing in all-optical communication schemes; it is also
relevant as an experimentally tunable playground where
numerous fundamental issues of solitons and nonlinear waves
can be explored.

It is interesting to mention in passing that in parallel (and
nearly concurrently) to this
direction of photorefractive crystal lattices, a number of other
venues were developed in optical and atomic physics, where the
interplay of nonlinearity with periodicity is important for the
observed dynamics. Such contexts involve on the optical end,
the numerous developments on the experimental and theoretical
investigation of optical waveguide arrays; see e.g.
\cite{review_opt,general_review} for relevant
reviews. On the atomic physics end, the confinement of
dilute alkali vapors in optical lattice potentials \cite{konotop}
has similarly offered the opportunity to examine many fundamental
phenomena involving spatial periodicity,
including the manifestation of modulational instabilities, Bloch
oscillations, Landau-Zener tunneling and gap solitons among others;
see \cite{markus2} for a recent review.

In the present study, motivated by SBN crystals and photorefractive
lattices, we focus on two-dimensional periodic, nonlinear media.
In such settings, the vast majority of studies has centered around
focusing nonlinearities; the reason for this is two-fold:
on the theoretical end, the topic of focusing solitary waves is
interesting due to the collapse (and
concomitant arrest by the lattice) phenomena in the case of
cubic nonlinearity \cite{bb1}. On the other hand, in the photorefractive
case, collapse is absent due to the saturable nature of the nonlinearity;
however, in the latter case, it is technically easier to work with
voltages that are in the regime of focusing rather than in that of
the defocusing nonlinearity (in the latter case, sufficiently large
voltage, which is tantamount to large nonlinearity, may cause damage
to the crystal). Hence, the only coherent structure that appears
to be explored in the defocusing regime experimentally appears
to be that of fundamental gap solitons excited in the vicinity of the edge
of the first Brillouin zone \cite{moti1}. More complex gap structures
appear not to have been studied systematically,
perhaps partially due to the above reasons. In an earlier work \cite{ourpre},
motivated by the above lack of results, we considered more complicated
multipole (i.e., dipole and quadrupole\footnote{It should be pointed out here that although in the standard electromagnetic
convention, one thinks of dipoles as bearing two opposite charges, and
quadrupoles as bearing two positive and two negative charges (in our
case, phases of the excited sites), here we will use a different
notational convention. More specifically, following the etymology
of the words, we will refer to any two-site excitation as a dipole,
and to any four-site excitation as a quadrupole herein.}) and
vortex structures in the defocusing case for a standard
discrete dynamical lattice (namely, the discrete nonlinear Schr{\"o}dinger
equation with cubic nonlinearity \cite{pgk_dnls}). This provided us with an
analytically tractable and numerically accessible roadmap for the study of
some of the relevant solutions. In the present work,
we employ a continuum model involving a periodic potential and
a saturable nonlinearity as associated
with the SBN crystals. In particular,
we demonstrate both experimentally and numerically
multipole (dipoles and quadrupoles) solitons in 2D square lattices
induced with a self-defocusing nonlinearity.
We numerically
analyze both the existence and the stability of these
structures and follow their dynamics, in the cases where we find them
to be unstable. We also qualitatively
compare our findings with the roadmap provided by the {\it discrete} model
\cite{ourpre}.

Our presentation is structured as follows. In section 2,
we present our theoretical setup (related to our experiments).
Dipole
solutions with the two excited sites in adjacent wells of the periodic
potential (nearest-neighbor dipoles) are studied in section 3.
Subsequently, we do the same for next-nearest-neighbor dipoles
(excited in two diagonal sites of the square lattice)
in section 4. Section 5 addresses the case of quadrupoles, and
section 6 presents our experimental results for the structures examined.
Finally,
in section
7, we summarize our findings, posing some interesting questions for future
study.

\section{Setup}

\begin{figure}[tbh]
\begin{center}
\includegraphics[width=0.5\textwidth]{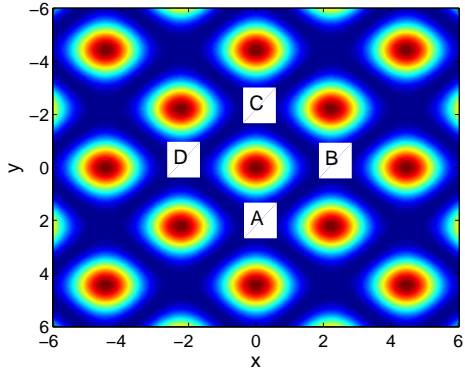}
\end{center}
\caption{(Color online) A spatial (x-y) contour plot of the ordinary polarization standing
wave [lattice beam in Eq.\ (\ref{eq1})]. The localized pulses will be sitting
at the minima of the lattice field, as opposed to the focusing nonlinearity
lattice field, where they reside at the maxima. Points $A,\, B,\, C,$ and $D$
are used for naming the dipole configurations. $A$ is a nearest-neighbor
minimum of $B$ and $D$ (in our ``diagonally-oriented'' lattice), while it is a
next-nearest-neighbor of $C$. Because of that, dipoles whose two lobes are
at $A$ and $B$ (or at $C$ and $D$) will be called nearest-neighbor dipole
solitons. The ones that are sitting at $A$ and $C$ (or at $B$ and $D$) will
be called next-nearest-neighbor dipoles.}
\label{lattice}
\end{figure}

For our theoretical considerations,
we use the non-dimensionalized version of the
photorefractive model with the saturable nonlinearity,
as developed in detail in \cite{yang04_3,yang04_4},
in the following form:
\begin{equation}
iU_z=-\left(U_{xx}+U_{yy}\right)-\frac{E_0}{1+I_{ol}+|U|^2}U.
\label{eq1}
\end{equation}
In the above expression $U$ is the slowly varying amplitude of the probe beam normalized by the dark irradiance of the crystal $I_d$, and
\begin{equation}
I_{ol}=I_0\cos^2\left(\frac{x+y}{\sqrt2}\right)\cos^2\left(\frac{x-y}{\sqrt2}\right),
\end{equation}
is a square optical lattice intensity function in units of $I_d$. Here $I_0$ is the lattice peak intensity, $z$ is the propagation distance (in units of $2k_1D^2/\pi^2)$, $(x, y)$ are transverse distances (in units of $D/\pi$), $E_0$ is the applied DC field voltage (in units of $\pi^2(k^2_0n^4_eD^2r_{33})^{-1}$), $D$ is the lattice spacing, $k_0 = 2\pi/\lambda_0$ is the wavenumber of the laser in the vacuum, $\lambda_0$ is the wavelength, $n_e$ is the refractive index along the extraordinary axis, $k_1 =k_0n_e$, and $r_{33}$ is the electro-optic
coefficient for the extraordinary polarization. In line with our experiment,
we choose the lattice intensity $I_0 = 5$ (in units of $I_d$).
A plot of the optical lattice is shown in Fig.\ \ref{lattice}, also for
illustrative purposes regarding the location where our localized pulses
will be ``inserted''. In addition, we choose other physical parameters
consistently with the experiment as
\[
D = 25\ \mu\textrm{m},\quad\lambda_0 = 0.5\ \mu\textrm{m},\quad n_e = 2.3,\quad r_{33} = 280\ \textrm{pm/V}.
\]
Thus, in this paper, one $x$ or $y$ unit corresponds to 7.96 $\mu$m, one $z$ unit corresponds to 3.66 mm, and one $E_0$ unit corresponds to 12.76 V/mm in physical units.
In the experiments, the applied voltage is $-550$V/$5$mm, which gives $E_0=8.62$ in our numerical simulations.

It should also be noted here that in our experimental results the
diffraction length can be approximately evaluated (for the beams widths
and wavelengths that we typically used) as being $2.5$mm. As our
crystal extends over $10$mm in the z-direction, it is clear that the
patterns that we observe are over a few (roughly 4) diffraction lengths
and hence if they are self-supported within such length scales, this
will indicate that they are indeed self-trapped beams. On the other hand,
as we will see in what follows (in our numerical simulations), for
all the configurations that we will find to be unstable, the instability
development will arise typically for dimensionless propagation distances
of $10 < z < 100$. Since these distances are considerably longer (in
dimensional units) than the propagation distance in our crystal, all
of the patterns presented in sections 3-5
below (even the most unstable ones) should,
in principle, be experimentally observable in our setting. This is
corroborated by our experimental results in section 6.

The numerical simulations are done with a uniform spatial mesh with $\Delta x=1/3$ and domain size $30 \times 30$, i.e.\ $91\times91$ grid points
for most configurations (see Fig.\ \ref{lattice} for a schematic of
the spatial configurations). For some of the configurations,
a larger domain was required, in which case a domain of size $60
\times 60$, i.e.\ $181\times181$ grid points, was used.
Regarding the typical dynamics of a soliton when it is unstable, we simulate the z-dependent behavior using a Runge-Kutta fourth-order method using
a step $\Delta z=0.00025$.

Stationary solutions of Eq. (\ref{eq1}) are sought in the form of $U(x,y,z)=u(x,y)e^{i \mu z}$, where $\mu$ is the propagation constant and $u$ is a real valued function satisfying
\begin{equation}
\mu u-\left(u_{xx}+u_{yy}\right)-\frac{E_0}{1+I_{ol}+u^2}u=0.
\label{sta_eq}
\end{equation}
The localized states $u(x,y)$ of (\ref{sta_eq}) were obtained using the
Newton-GMRES fixed point solver nsoli from \cite{kell03} and
continuation was used
as a function of $\mu$, to follow the relevant branches of solutions.

The propagation constant $\mu$ we consider in this report is in the
first spectral gap. Using Hill's method for the 2D problem
\cite{bernard}, for parameter values mentioned above, we find the band gap to be $4.2\lesssim\mu\lesssim5.46$\footnote{We note that the linear spectrum for
$\mu$ quoted here and shown in the following images was the most accurate we
computed using Hill's method based on consistency over $\Delta x$ in the limit
$\Delta x \rightarrow 0$.  The linear spectrum for the discretized problem
with our chosen discretization is slightly different however, and in particular,
the first band edge occurs at a slightly larger value of $\mu$.  It is
interesting to note that all saddle-node bifurcations occur at the accurate band edge,
while those solutions which degenerate to linear Bloch modes actually degenerate
beyond this value at the band edge particular to the discretization (not shown).}.

The power or the norm of the solitary waves is defined as follows:
\begin{equation}
P = \left[\int_{-\infty}^{\infty}\int_{-\infty}^{\infty}|U|^2\,dx\,dy\right]^{1/2}.
\end{equation}

We analyze the linear stability of soliton solutions $u(x,y)$ of
(\ref{sta_eq}) by applying an infinitesimal perturbation to them. Writing
$U(x,y,z)=e^{i \mu z} \left(u(x,y)+e^{\lambda z}\tilde{u}(x,y)\right)$, the perturbation $\tilde{u}(x,y)$ will then satisfy the following linearized equation
\begin{equation}
i\lambda\tilde{u}=\mu \tilde{u}-\left(\tilde{u}_{xx}+\tilde{u}_{yy}\right)-\frac{E_0}{\left(1+I_{ol}+u^2\right)^2}\left[\left(1+I_{ol}\right)\tilde{u}-u^2\tilde{u}^*\right],
\label{lin_eq}
\end{equation}
where the superscript $*$ denotes complex conjugation. We solve the above linear eigenvalue problem using MATLAB's standard eigenvalue solver package.

At this point, it is also relevant to summarize the results of the
{\it discrete} NLS defocusing model of \cite{ourpre} that we will use for
comparison with the findings below. These results are incorporated
in Table 1, where the stability of all the possible combinations
of in-phase and out-of-phase, nearest-neighbor and next-nearest-neighbor
configurations is quantified in terms of their relevant eigenvalues
of linearization. The configurations are dubbed unstable when they
possess (for all parameter values) real eigenvalue pairs, whereas
they are considered marginally stable, when they do not always have
such pairs. However, the latter configurations typically, in this setting,
possess
imaginary eigenvalues with negative Krein signature (see e.g. \cite{kks}
and references therein), which practically means that if these collide
with other eigenvalues, as $\mu$ is varied, complex eigenvalue quartets
will emerge out of Hamiltonian-Hopf bifurcations \cite{vdm}, destabilizing
the relevant solution. Hence, such solutions are not {\it always} linearly
unstable, but {\it may become} unstable for some parameter ranges.

\begin{table*}[t]
\begin{center}
\begin{tabular}{|l|c|c|}
\hline
& NN Stability & NNN Stability \\ \hline
IP Dipole & Stable ($N_i^{-}=1$) & Unstable  ($N_r=1$) \\ \hline
OOP Dipole & Unstable ($N_r=1$) & Stable ($N_i^{-}=1$) \\ \hline
IP Quadrupole & Stable ($N_i^{-}=3$) & Unstable ($N_r=3$)  \\ \hline
OOP Quadrupole & Unstable ($N_r=3$) & Stable ($N_i^{-}=3$) \\ \hline
\end{tabular}
\end{center}
\label{t1}
\caption{Summary of the stability results of the discrete model with
cubic nonlinearity, studied
in \cite{ourpre}, for all the in-phase (IP) and out-of-phase (OOP),
nearest-neighbor (NN), as well as next-nearest-neighbor (NNN) configurations.
$N_r$ denotes real eigenvalue pairs and $N_i^{-}$ denotes imaginary
eigenvalue pairs with negative Krein signature.}
\end{table*}

\section{Nearest-neighbor Dipole Solitons}

In this section, we report dipole solitons where the two lobes of the
wave are located in two nearest-neighbor (NN)
lattice sites in the 2D square lattice
potential shown in Fig.\ \ref{lattice}. The lobes can have the
same phase or $\pi$ phase difference and are, accordingly, hereafter termed
in-phase (IP) dipoles and out-of-phase (OOP) dipoles, respectively.
Notice that due to the (diagonal) nature of our
lattice, the nearest-neighbor configurations that we consider are
``built'' along the diagonal direction.

\subsection{In-phase Nearest-neighbor Dipole Solitons}

\begin{figure}[tbp!]
\begin{center}
\includegraphics[width=0.4\textwidth]{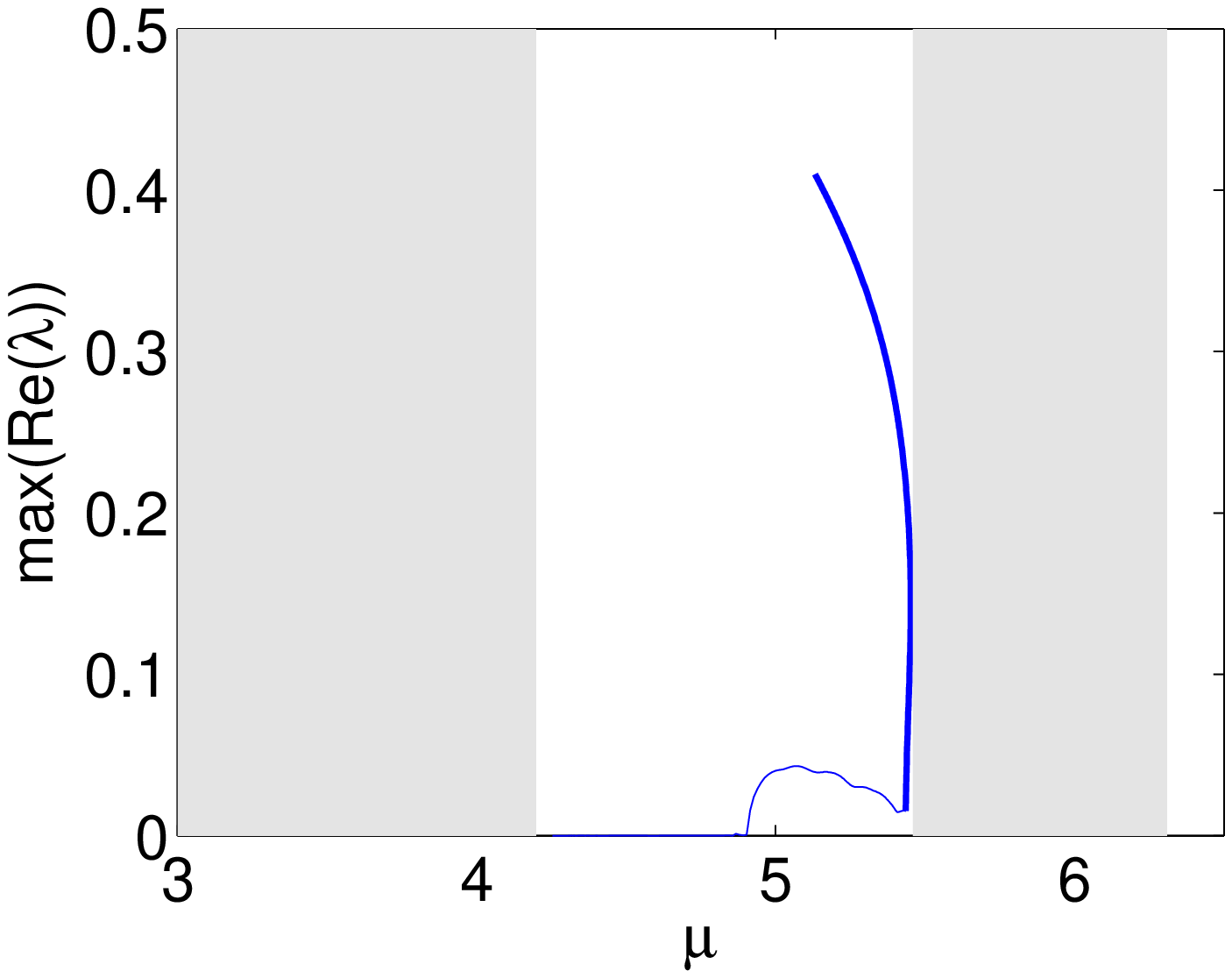}
\includegraphics[width=0.4\textwidth]{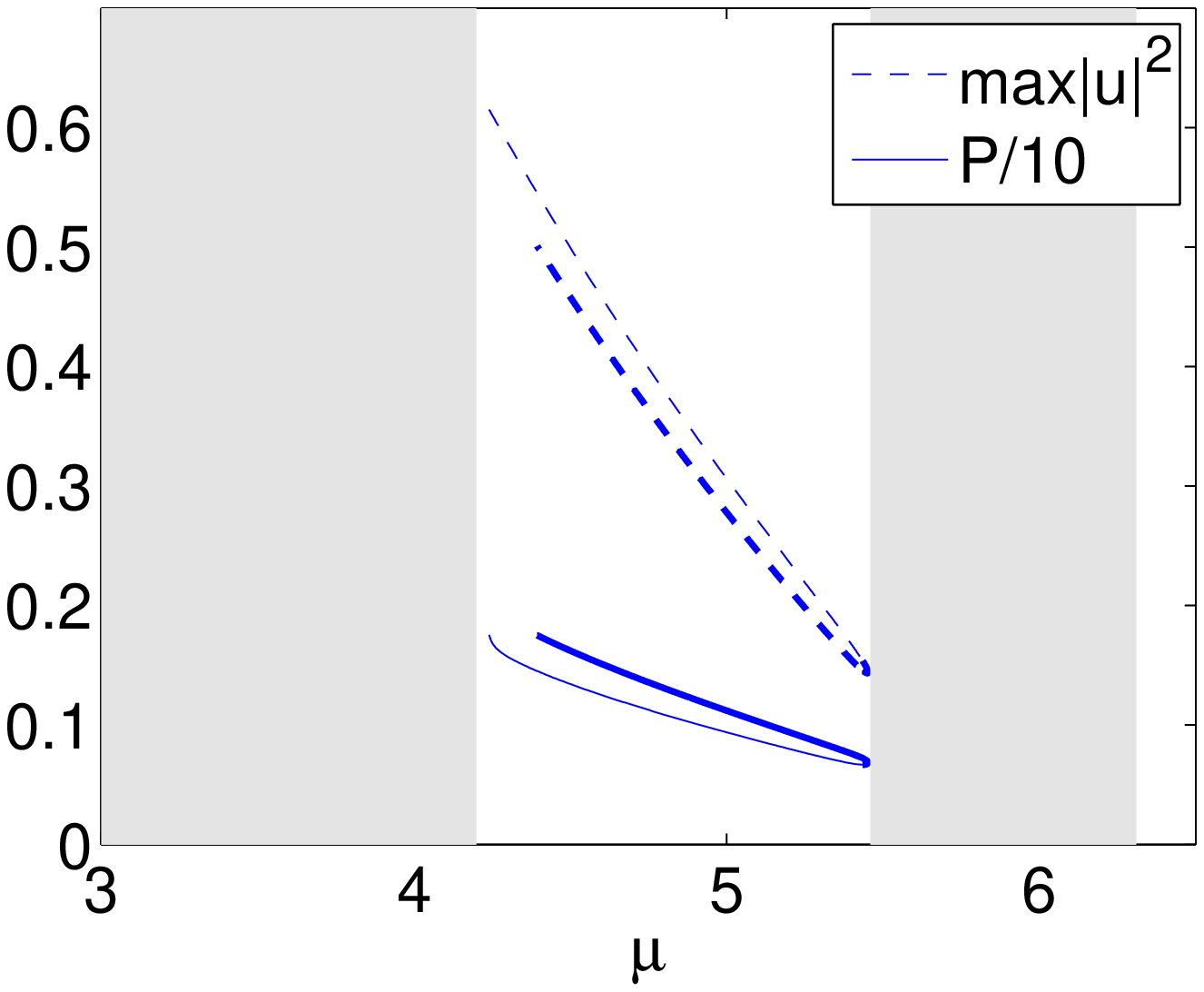}\\
\includegraphics[width=0.4\textwidth]{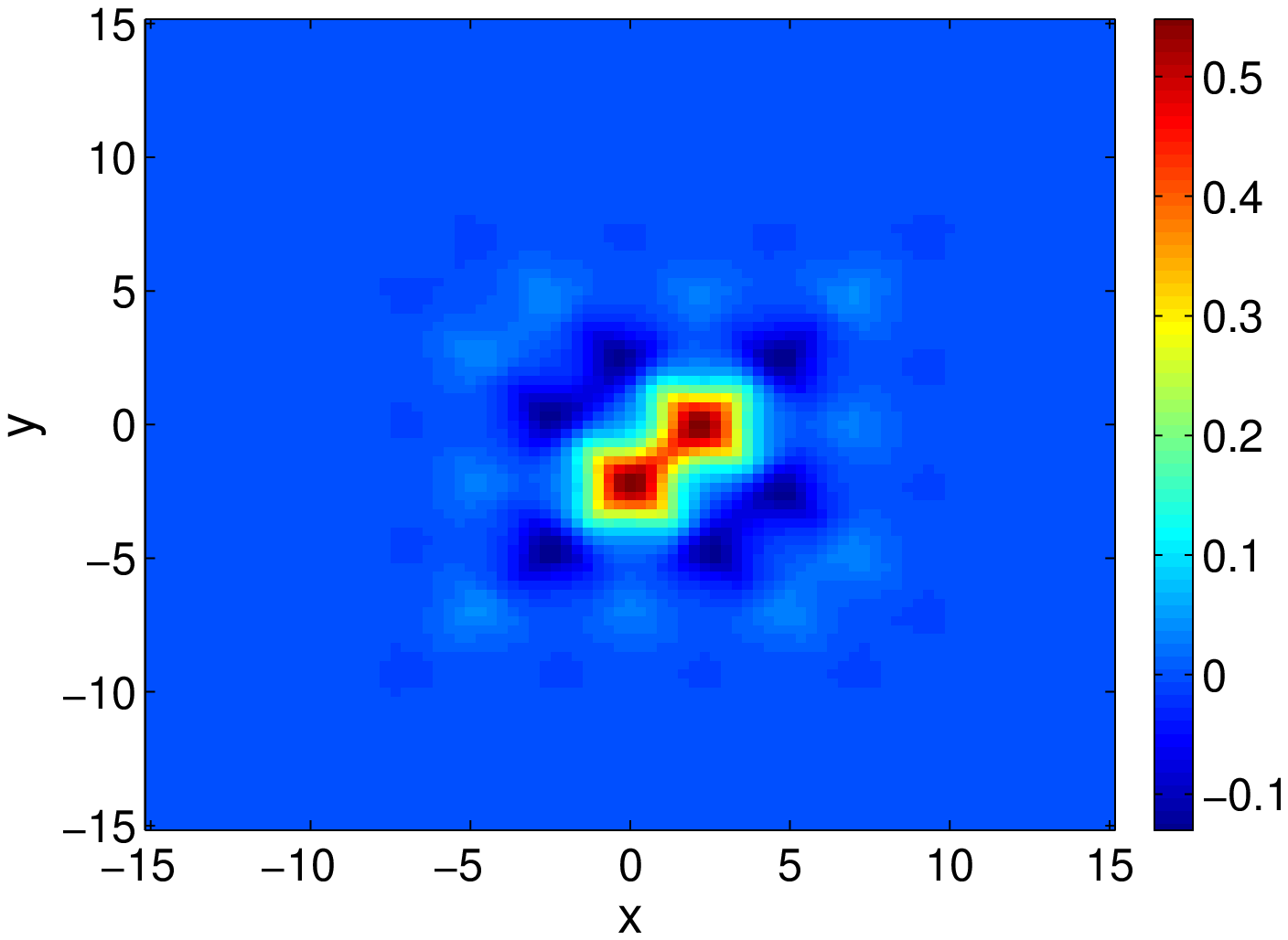}
\includegraphics[width=0.4\textwidth]{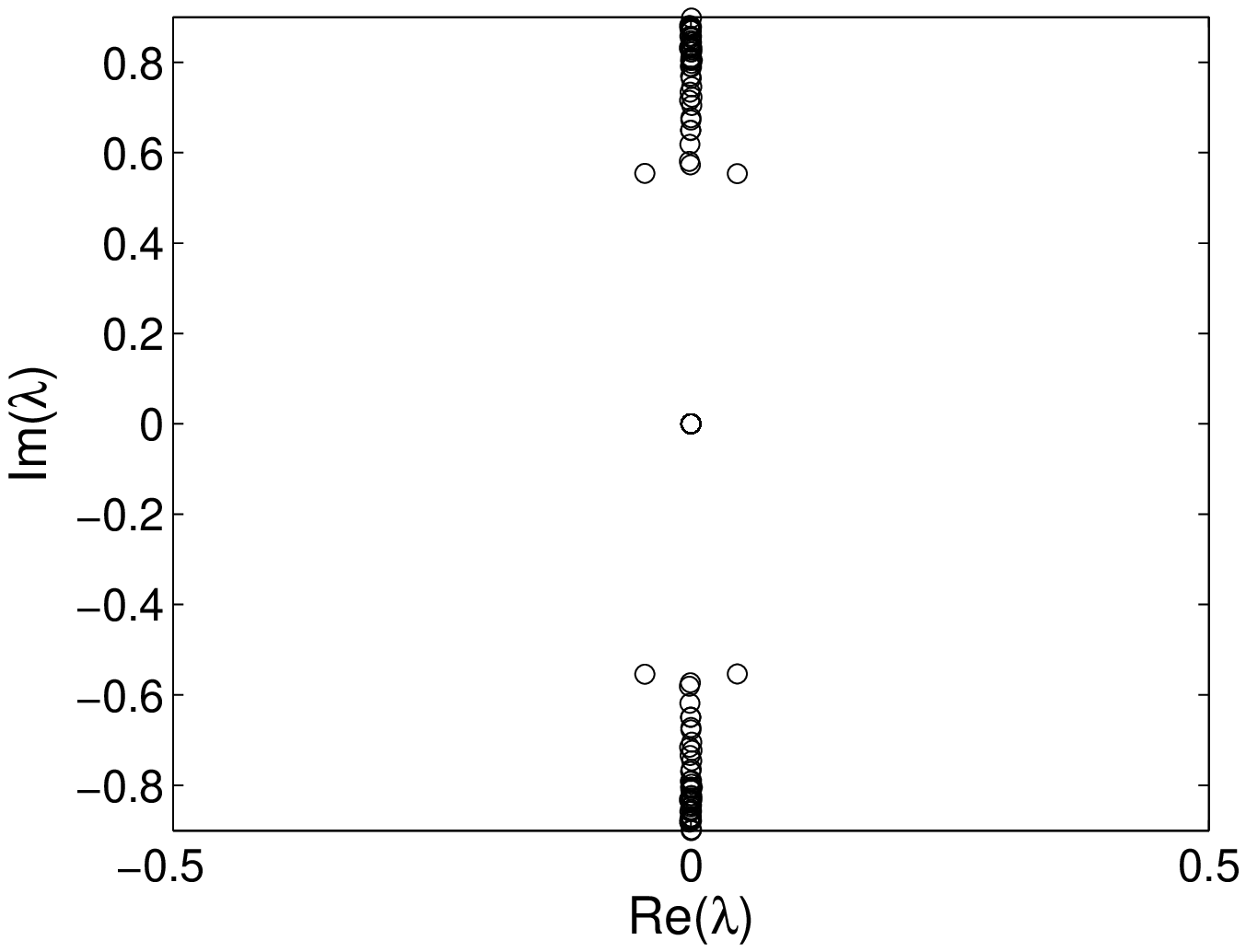}
\includegraphics[width=0.4\textwidth]{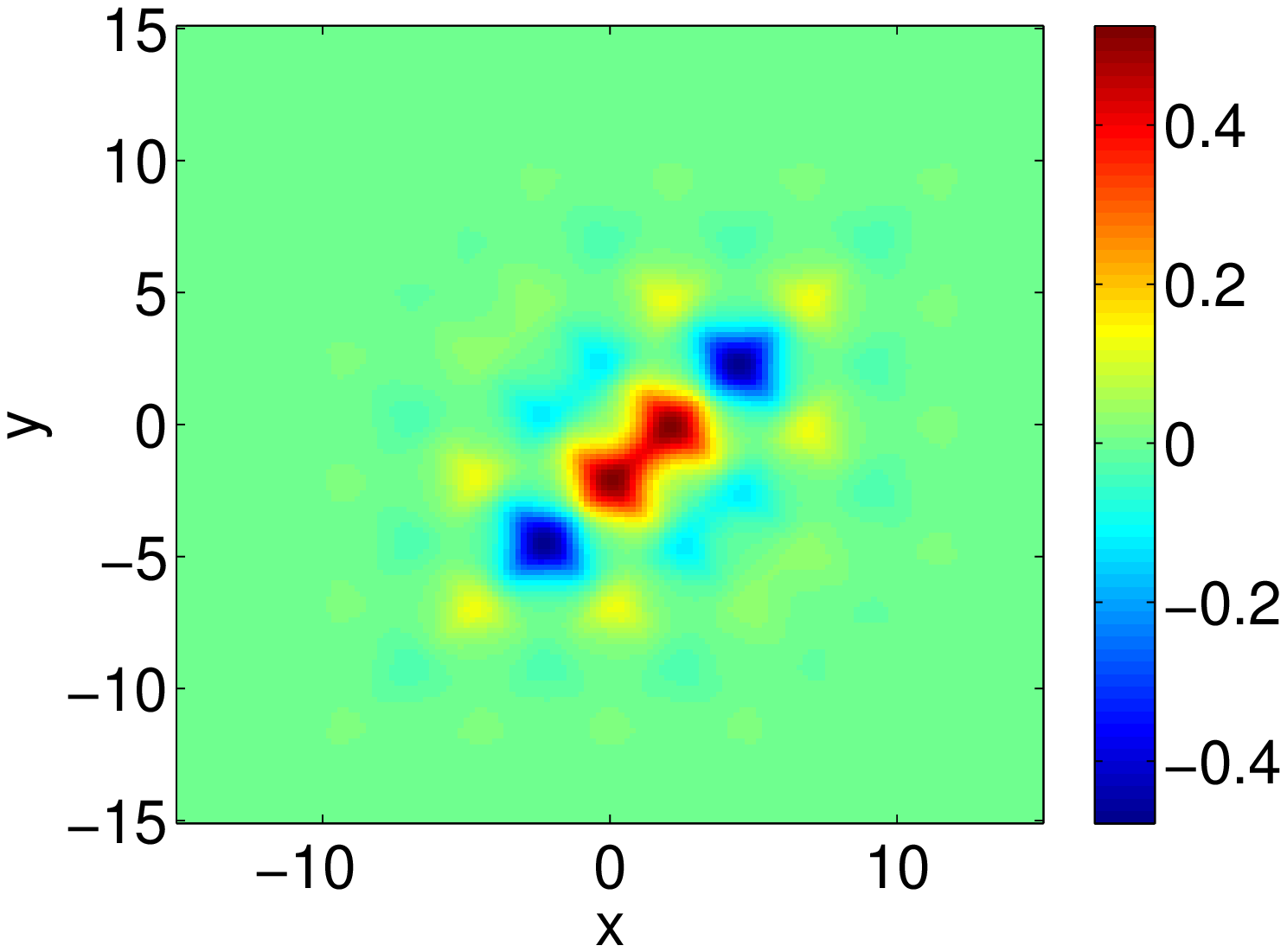}
\includegraphics[width=0.4\textwidth]{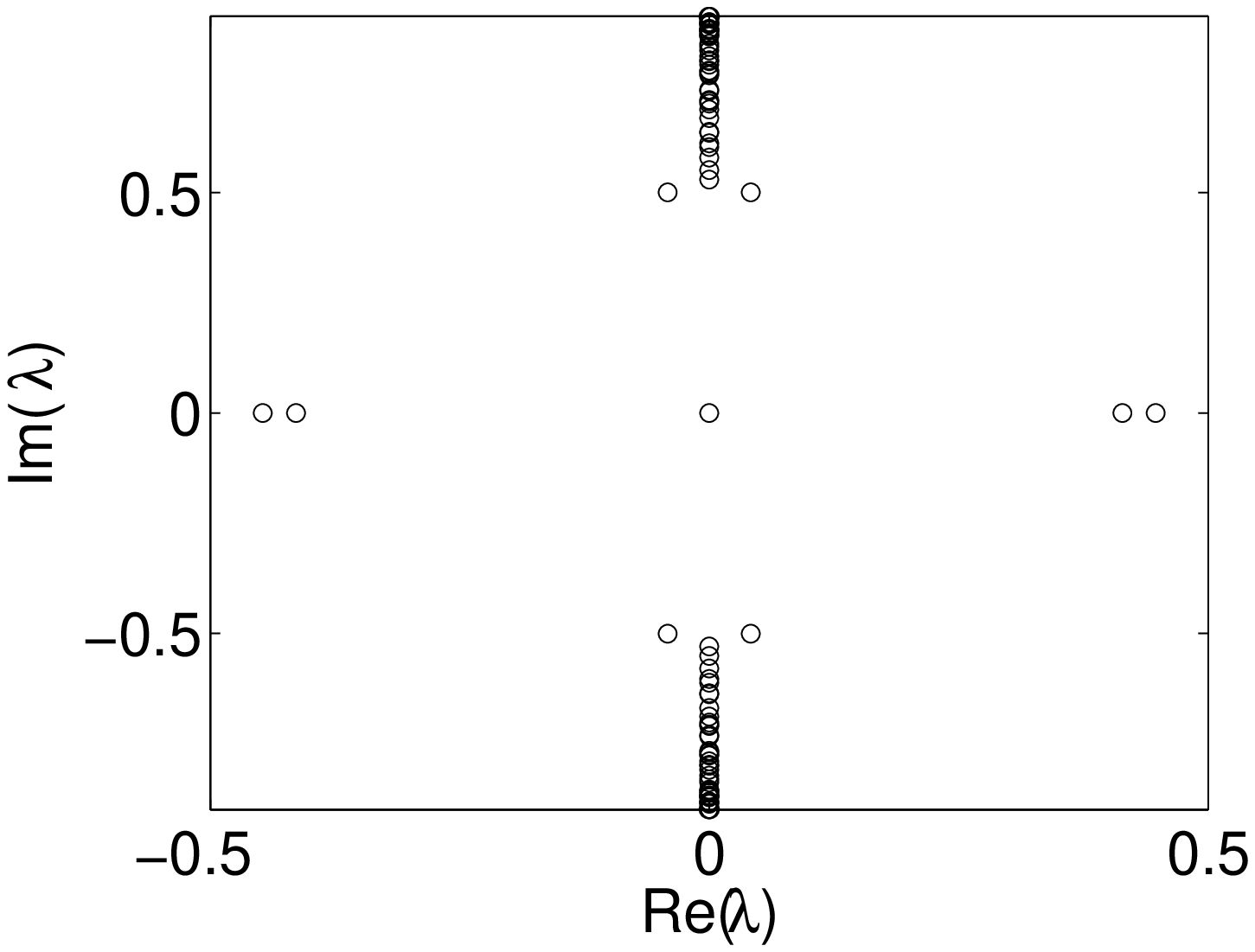}\\
\end{center}
\caption{(Color online) The top left panel shows the stability of the dipoles as a
function of the propagation constant $\mu$. It is stable when the spectra is
purely imaginary (i.e., when
$\max({\rm Re}(\lambda))=0$). The top right panel depicts the peak intensity and the
power of the dipoles. The thin line corresponds to the solution
represented in the middle row, while the
bold line corresponds to the waveform
illustrated in the bottom row. These two branches of solutions collide
and mutually annihilate in a saddle-node bifurcation.
The shaded areas in both of these panels represent the bands of the periodic
potential.
The middle (resp. bottom) left and right panels show the profile
$u$ of the branch indicated by thin (resp. bold) line in the top row
at $\mu=5$ and the corresponding complex spectral
plane $({\rm Re}(\lambda),{\rm Im}(\lambda))$ of the eigenvalues
$\lambda={\rm Re}(\lambda) + i {\rm Im}(\lambda)$.}
\label{diaIP}
\end{figure}

\begin{figure}[bth]
\begin{center}
\includegraphics[width=0.4\textwidth]{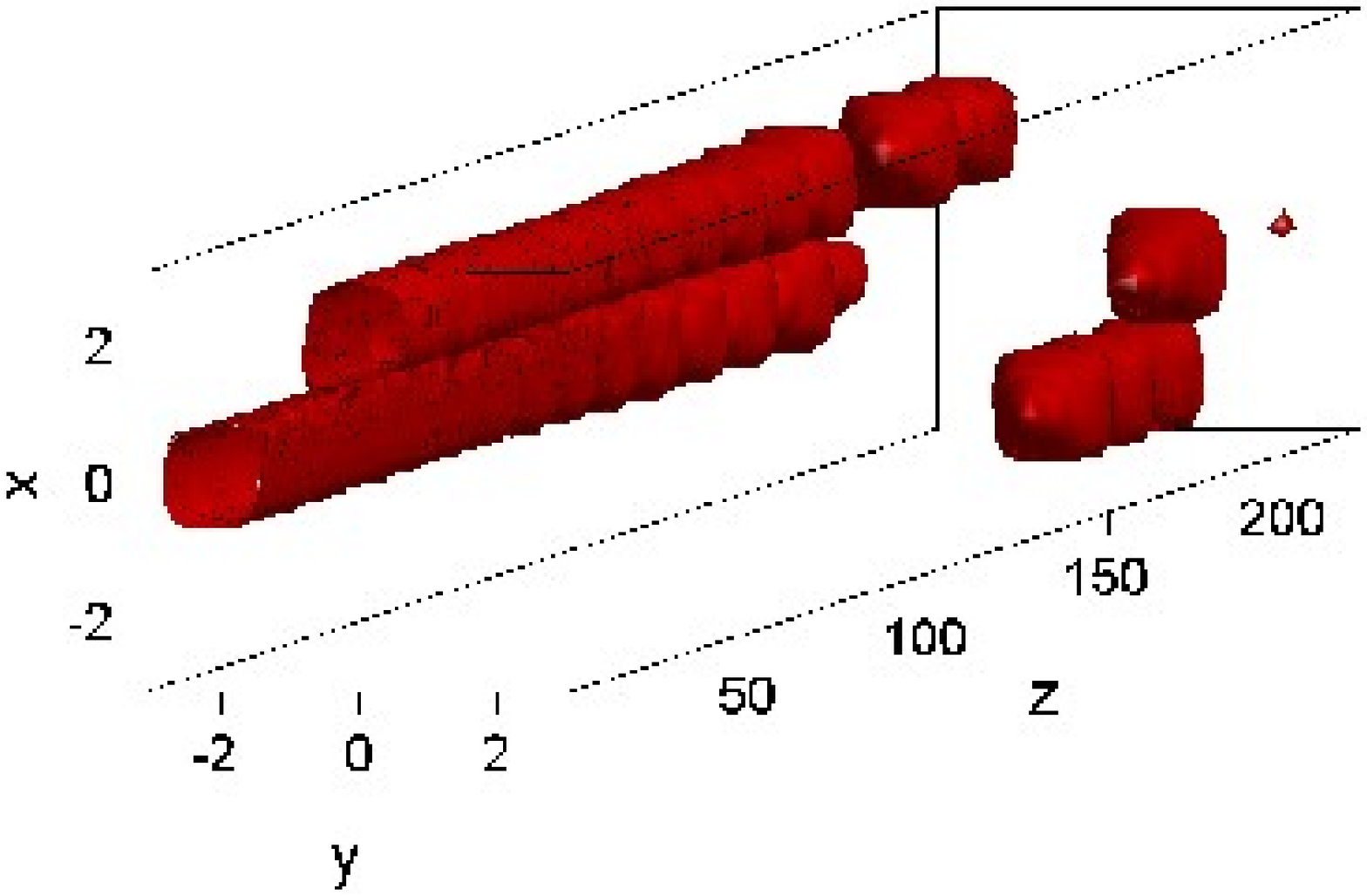}
\includegraphics[width=0.4\textwidth]{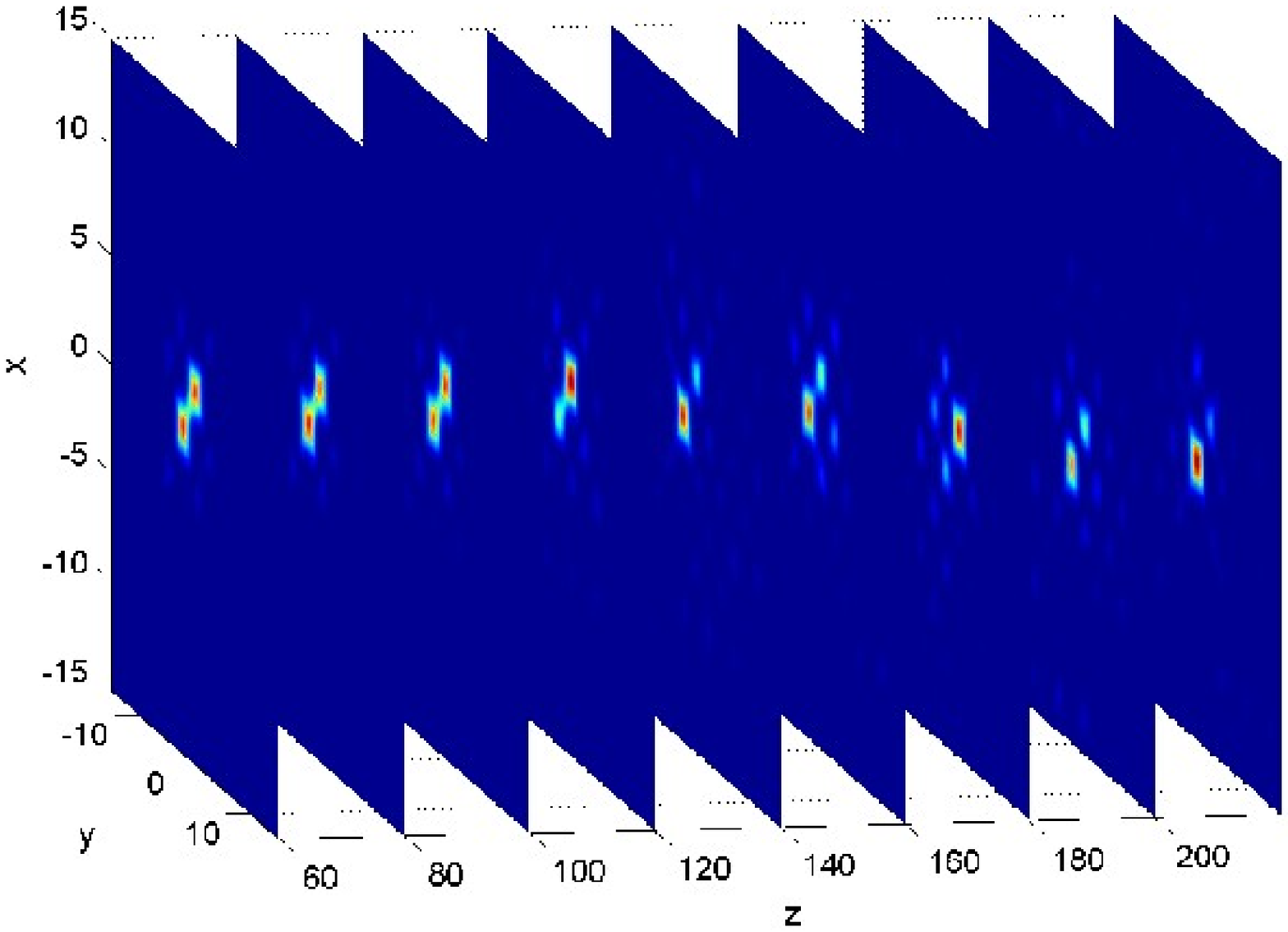}\\
\includegraphics[width=0.4\textwidth]{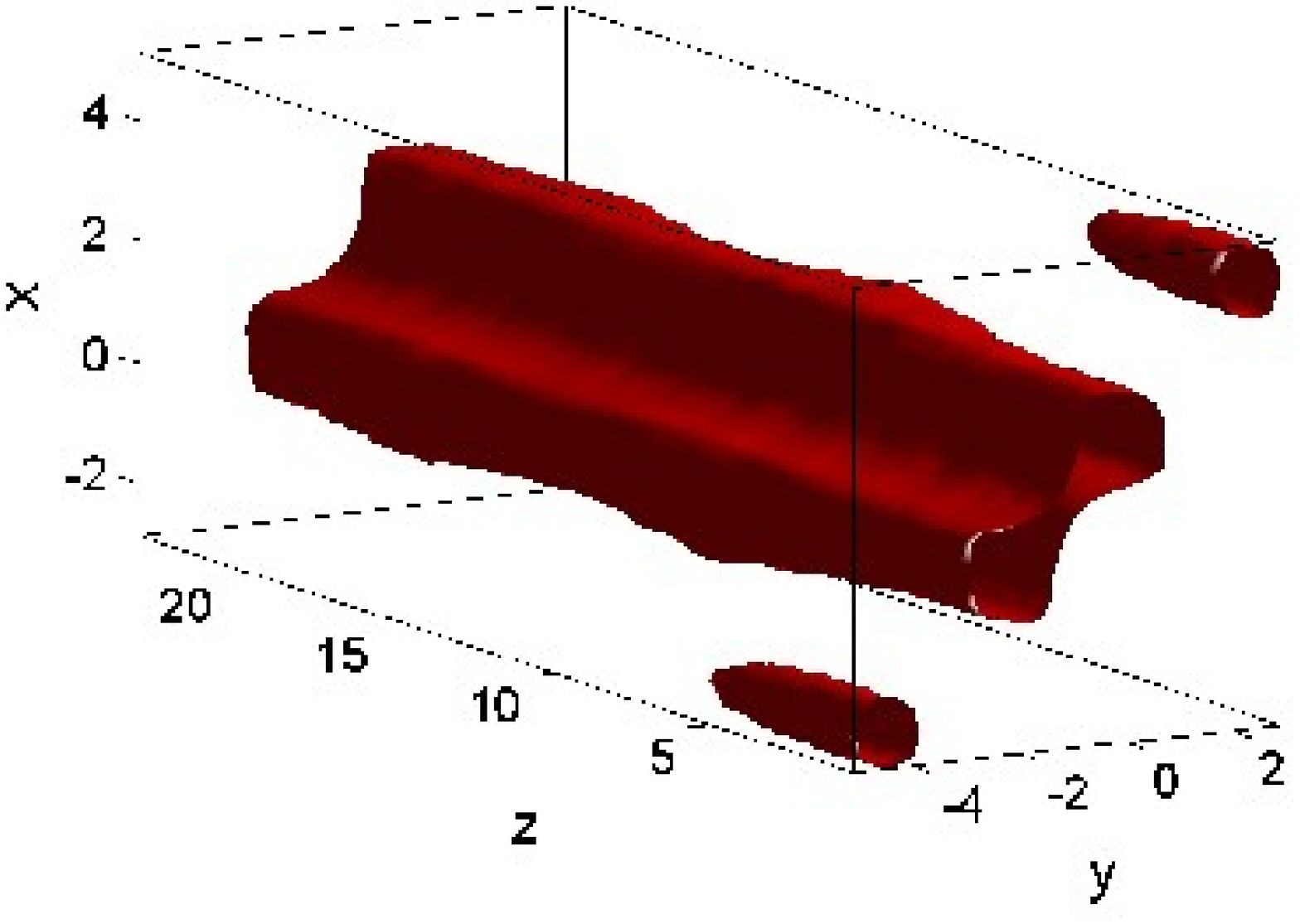}
\includegraphics[width=0.4\textwidth]{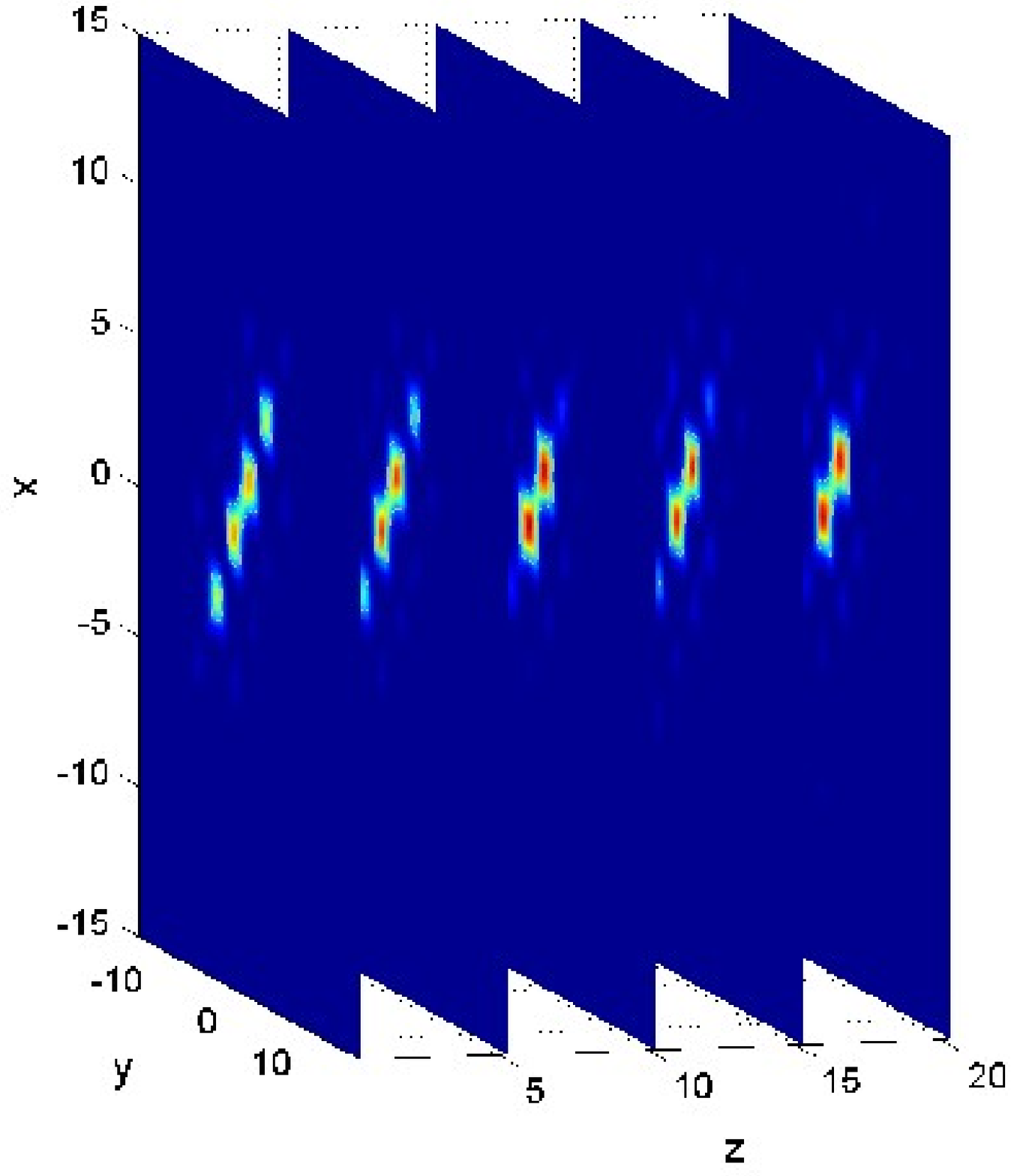}
\end{center}
\caption{(Color online) The evolution of the dipoles shown in Fig.\ \ref{diaIP} perturbed
by a random noise of maximum intensity $0.25\%$ of the soliton peak
intensity. Presented in the figure
are the isosurfaces (left panels) of height 0.2 for the first configuration
(top panels) and of height 0.1 for the second one (bottom panels) and their
slices at some instances (right panels). }
\label{dyn_diaIP}
\end{figure}

We have found IP dipoles in a large interval of propagation constants $\mu$
for the given voltage $E_0$. We found that the solitons exist for
$\mu$ smaller than 5.46, or for peak intensities larger than 0.144.
We note that the intensity of the dipoles cannot be arbitrary low, a
result similar to the
observed results of the focusing case \cite{yang04,yang04_4}.
The relevant findings are summarized in Fig.\ \ref{diaIP}.

The top left panel of Fig.\ \ref{diaIP} shows the stability of the dipoles
as a function of the propagation constant $\mu$, by illustrating the
maximal growth rate (maximum real part over all perturbation eigenvalues)
of perturbations. When $\max({\rm Re}(\lambda))=0$, this implies stability
of the configuration, while the configuration is unstable if
$\max({\rm Re}(\lambda)) \neq 0$ in this Hamiltonian system.
 We found that the stability region of this type of dipoles is given by
$4.2\leq\mu\leq4.91$, the left hand limit corresponding to the Bloch band. The top right panel depicts the peak intensity and the
power of the dipoles.


The middle left and right panels show the profile $u$ of a dipole at
$\mu=5$ and the corresponding spectra at the complex plane, respectively.
We see that the soliton is unstable due to an oscillatory instability.
This is the typical instability in this case, in line with the
discrete cubic model results. Clearly, there is an imaginary eigenvalue
with negative Krein signature \cite{kks} which upon collisions with the
continuous spectrum results in Hamiltonian-Hopf bifurcations and concomitant
oscillatory instabilities.


As we increase $\mu$ further, the dipoles disappear in a saddle-node bifurcation. The bifurcation diagram is depicted in the top panels of Fig.\ \ref{diaIP}. At the bifurcation point, $dP/d\mu\to\infty$, as $\mu\rightarrow 5.46$, i.e.,\
at the boundary of the first Bloch band.  At this point, the
IP NN configuration collides with a configuration shown at the bottom
panel of Fig. \ref{diaIP} (where the two nearest-neighbors -along the axis
of the dipole- of the two populated wells become out-of-phase with them)
and disappears in a saddle-node bifurcation.
The corresponding profile and spectral plane for the saddle branch
at $\mu=5$ is shown in the bottom right panel of the same figure,
illustrating the strong instability of the latter.


We have also simulated the dynamics of the solitary waves
when they are unstable.
In Fig.\ \ref{dyn_diaIP} we present the evolution of the unstable dipoles
shown in Fig.\ \ref{diaIP}. The dipoles are perturbed by a random noise with
maximum intensity $0.25\%$ of the soliton peak intensity. Shown in
Fig.\ \ref{dyn_diaIP} are the isosurfaces and the slices of the soliton along
the propagation direction.

The dynamics of the soliton shown in the middle panel of Fig.\ \ref{diaIP} is presented in the top panels of Fig.\ \ref{dyn_diaIP}.
One can see that even with that strong perturbation,
at $z=100$ the soliton still resembles its initial configuration. Physically,
this corresponds to a propagation distance of approximately 366 mm. This means
that the instability is very unlikely to be observed in the
photorefractive crystal lattice used in
our experiments. For longer propagation distances, the oscillatory instability
sets in and finally rearranges the dipole into a fundamental soliton type
configuration principally centered around a single site.

For the dipole shown in the bottom panel of Fig.\ \ref{diaIP}, we present its dynamics as bottom panels of Fig.\ \ref{dyn_diaIP}.
We found that the instability is strong as predicted above such that even
after a relatively short propagation distance, the configuration turns into
the more stable solution of the IP NN dipole branch. We did not present the further dynamics of the dipole, as it is similar to the upper panels.

\subsection{Out-of-phase Nearest-neighbor Dipole Solitons}

\begin{figure}[tbh]
\begin{center}
\includegraphics[width=0.4\textwidth]{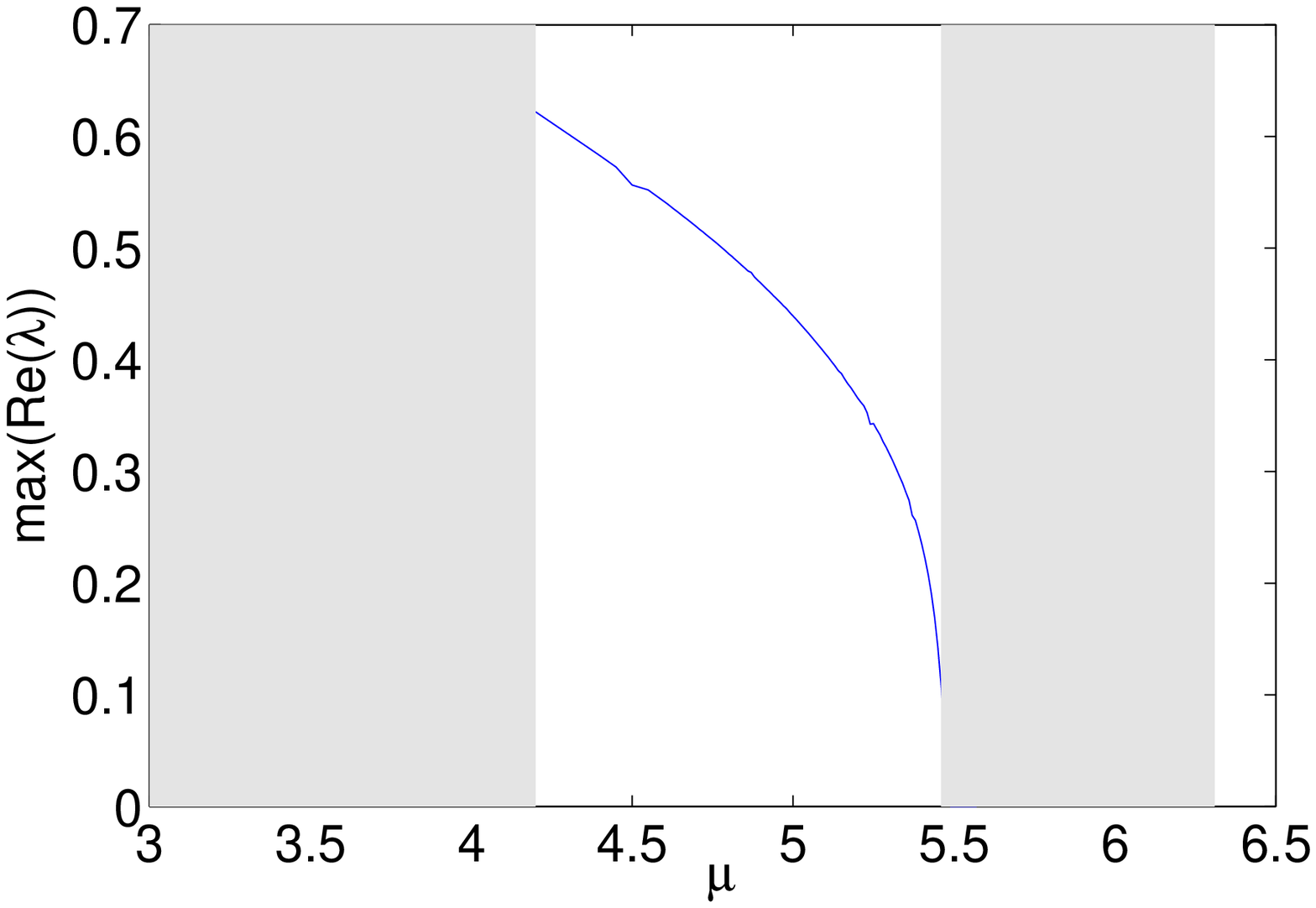}
\includegraphics[width=0.4\textwidth]{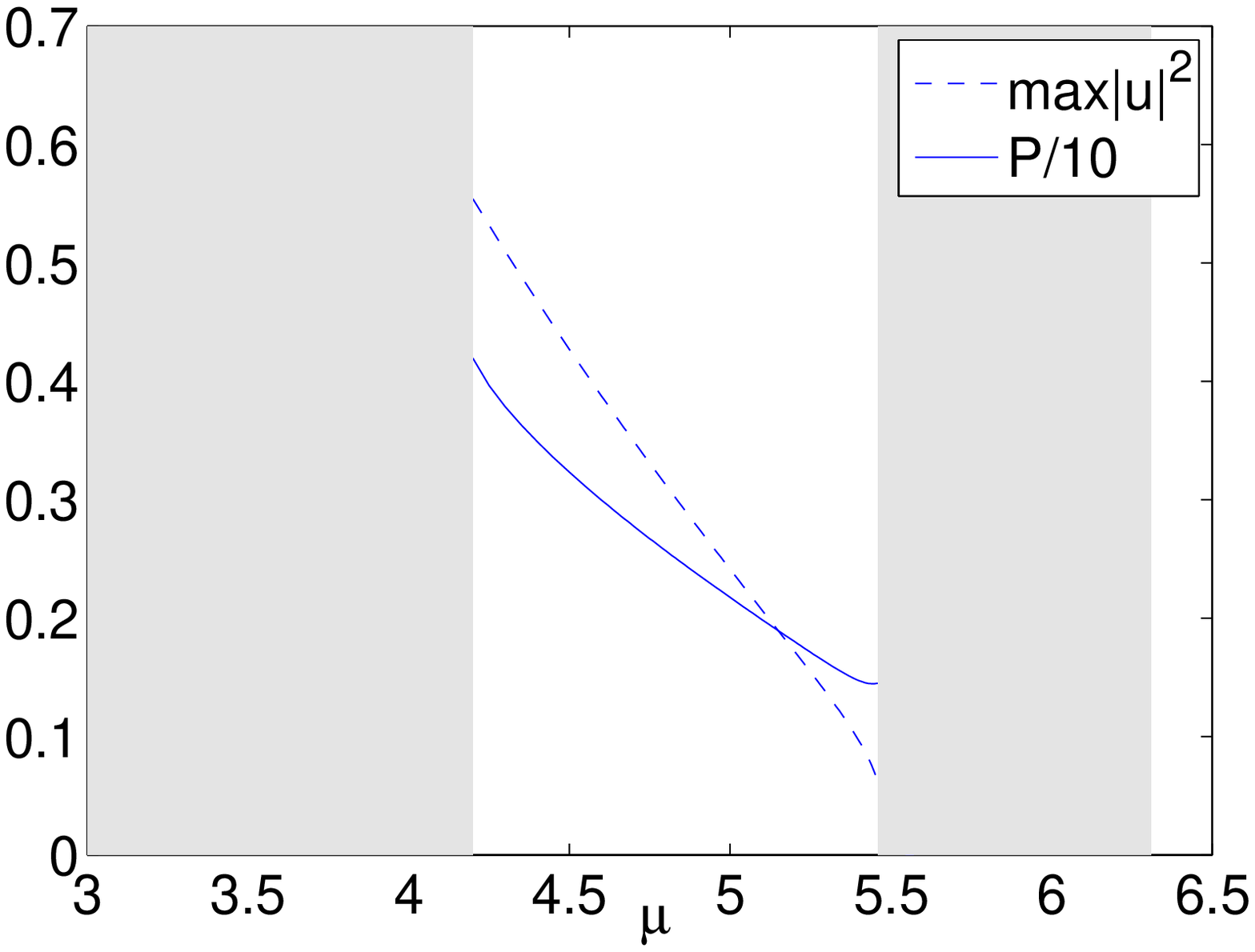}\\
\includegraphics[width=0.4\textwidth]{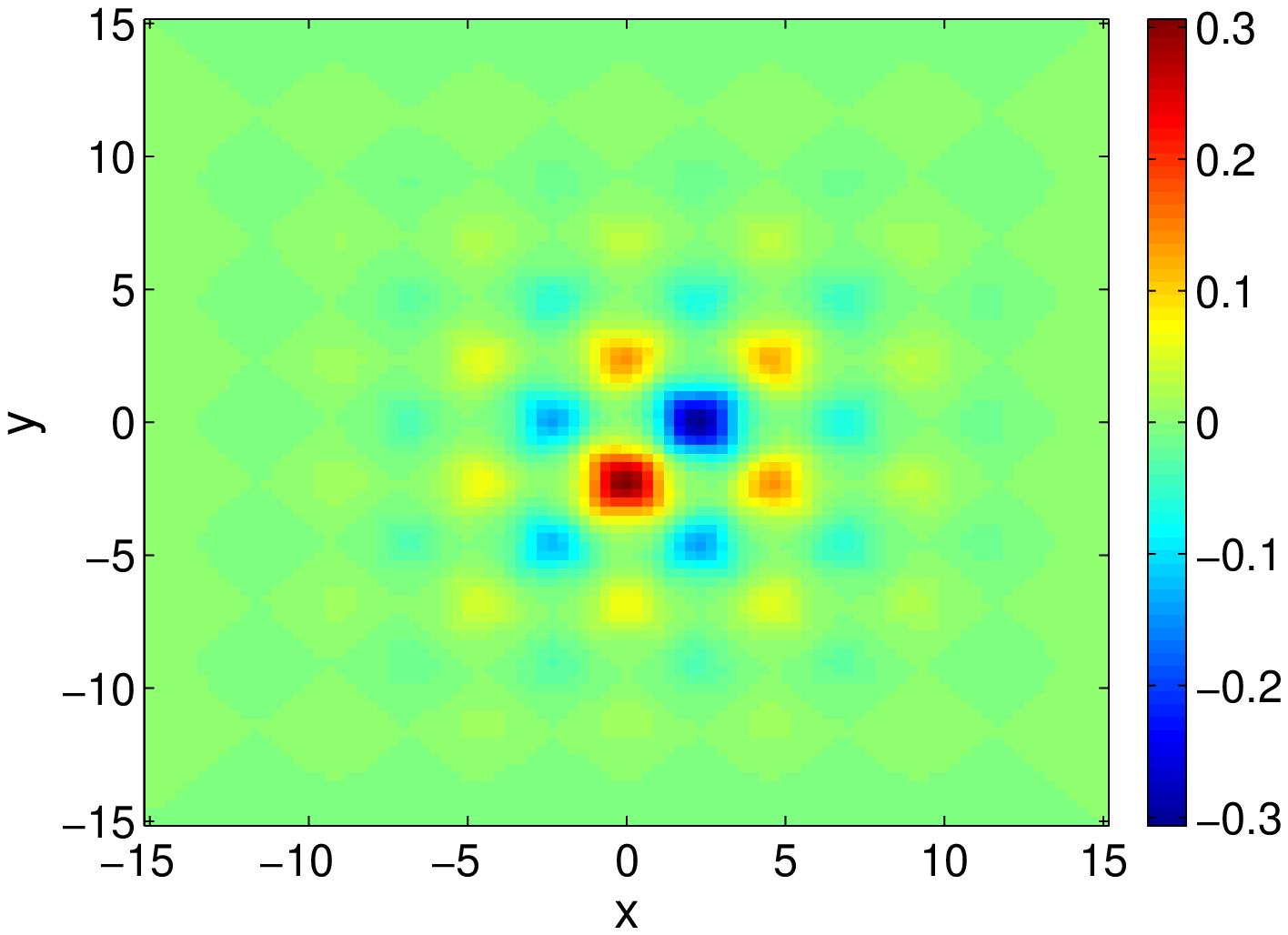}
\includegraphics[width=0.4\textwidth]{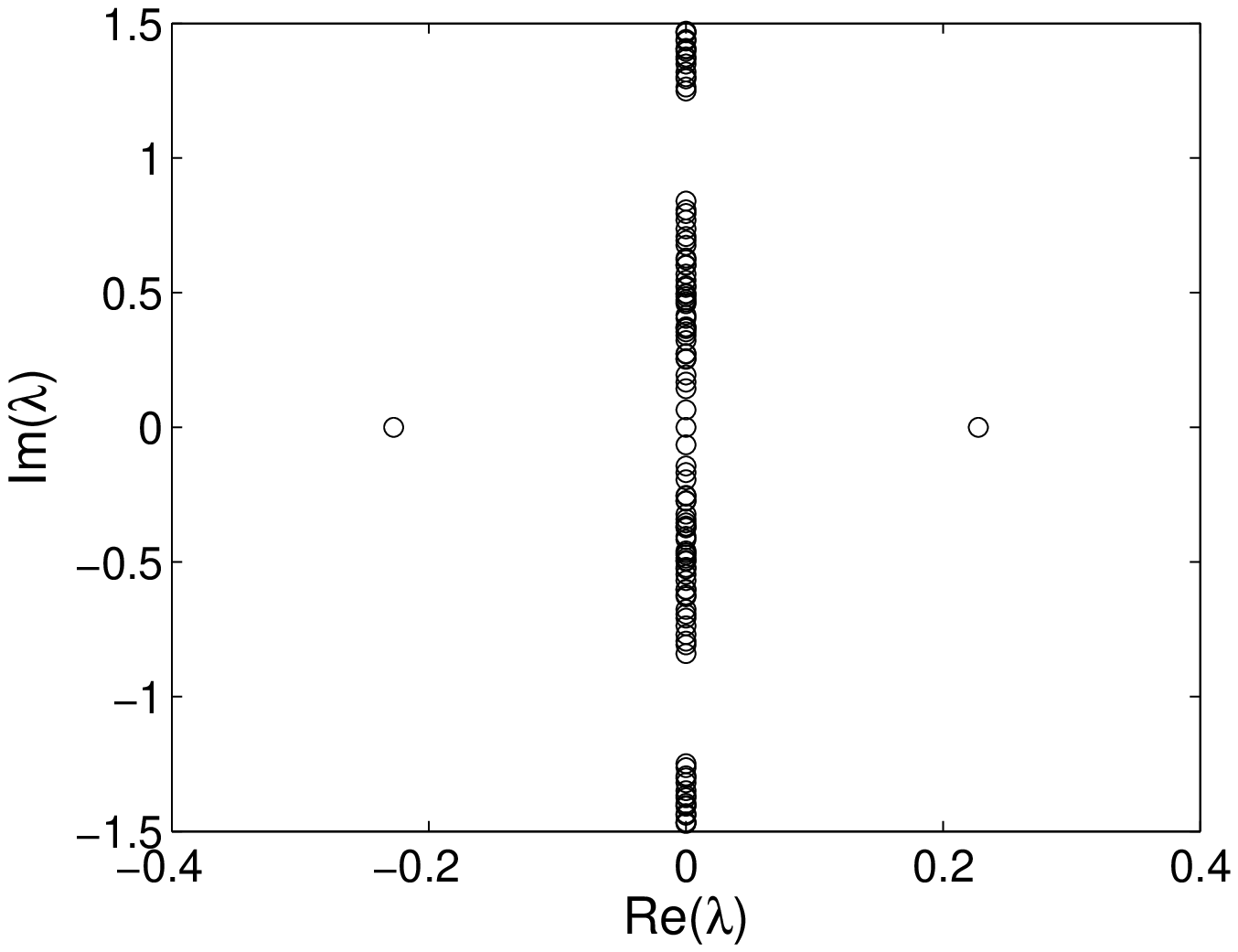}\\
\end{center}
\caption{(Color online) The top panels correspond to the same panels of Fig.\ \ref{diaIP}
but for OOP NN dipole solitons. The bottom panel shows the
profile $u$ and the corresponding spectra in the complex plane of the
dipoles at $\mu=5.4$.}
\label{diaOOP}
\end{figure}

We have also found OOP dipoles arranged in nearest-neighboring lattice wells.
We summarize our findings in Fig.\ \ref{diaOOP} where one can see that the
solitons exist in the whole entire region of propagation constant $\mu$ in the first Bragg gap, $\mu\in(4.2,5.46)$.
This smooth transition indicates that the OOP NN dipole solitons emerge out
of the Bloch band waves; see e.g. \cite{peli04} and \cite{shi07} for
a relevant discussion of the 1D and of the 2D problem respectively, in
the case of the cubic nonlinearity. Nonetheless, the OOP NN dipoles are
typically unstable due to a real eigenvalue pair.
Notice that this is in agreement with the prediction of the discrete
model (as can be seen from Table 1).



As the branch merges with the band edge, we observe an
interesting feature, namely that the configuration resembles that of a quadrupole with a $\pi$ phase difference between two neighboring solitons, which we call $+-+-$ quadrupoles below. This can be an indication that these structures bifurcate out of the Bloch band from the same bifurcation point. We elaborate this further in our discussion of the quadrupole structures in section 5.


\begin{figure}[tbh]
\begin{center}
\includegraphics[width=0.4\textwidth]{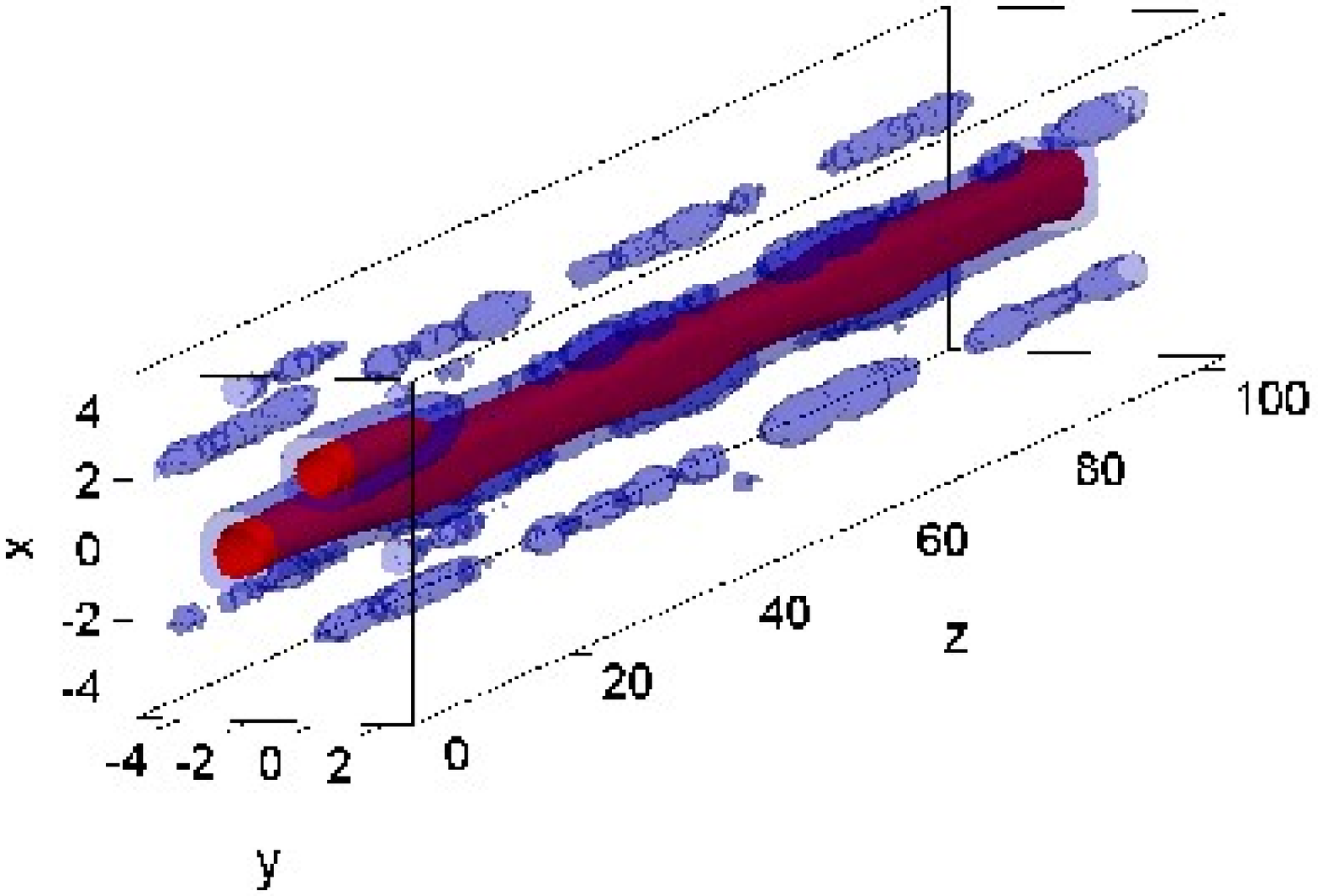}
\includegraphics[width=0.4\textwidth]{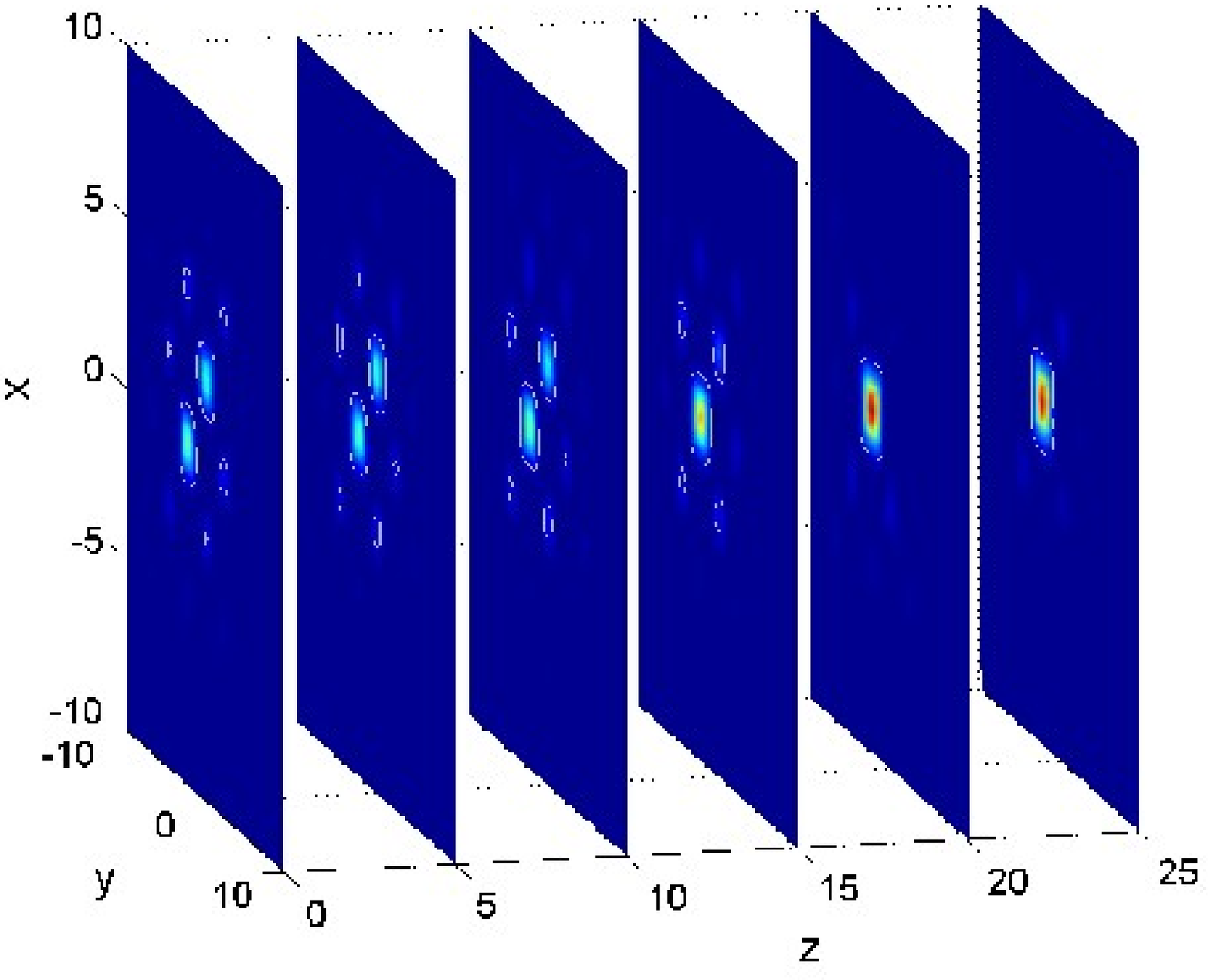}
\end{center}
\caption{(Color online) Similar to Fig.\ \ref{dyn_diaIP} but for the evolution of OOP NN
dipoles. Shown are the isosurfaces of height 0.05 (red) and 0.015 (blue) and
the contour plot slices at some select propagation distances. }
\label{dyn_diaOOP}
\end{figure}

In Fig.\ \ref{dyn_diaOOP} we present the instability dynamics
of an OOP NN dipole soliton perturbed by similar random noise perturbation
as in Fig.\ \ref{dyn_diaIP}. This type of dipoles is typically
more unstable than its IP counterpart, as is illustrated in the figure.
In particular, in this example of unstable evolution even at $z=10$, the
instability already manifests itself.
One similarity of the instability
of OOP NN dipoles with that of
the IP NN ones is that the dipoles tend to degenerate to a
single-site, fundamental gap soliton, which is stable in this setting.
We note in passing that similar evolution results for dipoles
and quadrupoles (but for short propagation distances) in the
focusing case are discussed
in \cite{yang04_4}.

\section{Next-nearest-neighbor Dipole Solitons}

We have also obtained dipole solutions that are not oriented along the
two nearest-neighboring lattice wells, but rather where
the two humps of the structure are located at two adjacent
next-nearest-neighboring lattice sites. These humps can once again
have the same phase or $\pi$ phase difference between them. We will
again use the corresponding IP and OOP designations for these
next-nearest-neighbor (NNN) waveforms. Notice here that NNN configurations
that we consider are among the ``closest'' next-nearest-neighbor
pairs, i.e., with the structures being aligned horizontally.
In principle, one can consider more remote pairs of next-nearest-neighbors
(e.g., along the diagonal), however the main qualitative stability properties
discussed below would not change in such a case.


\subsection{In-phase Next-nearest-neighbor Dipole Solitons}

\begin{figure}[tbh]
\begin{center}
\includegraphics[width=0.4\textwidth]{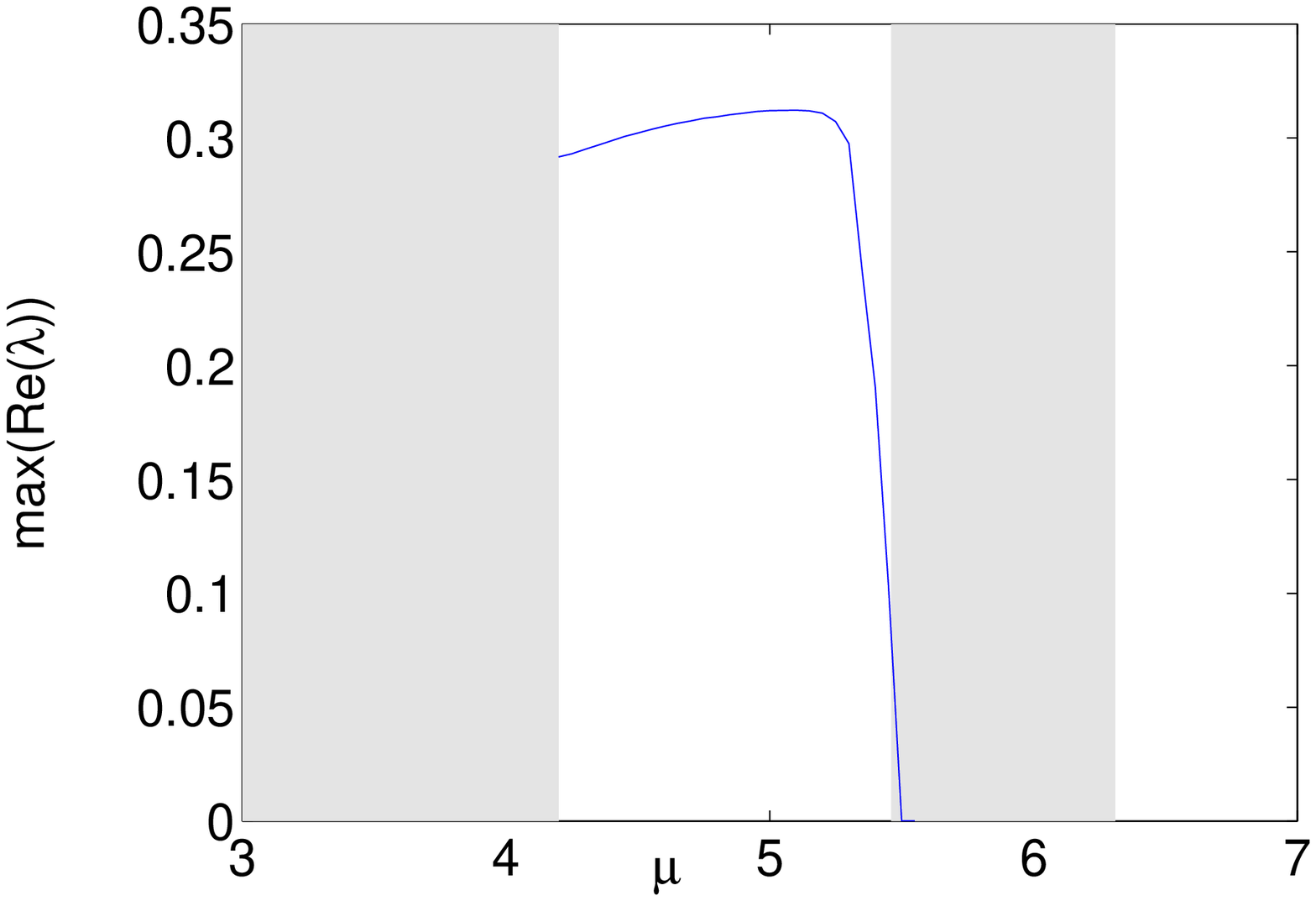}
\includegraphics[width=0.4\textwidth]{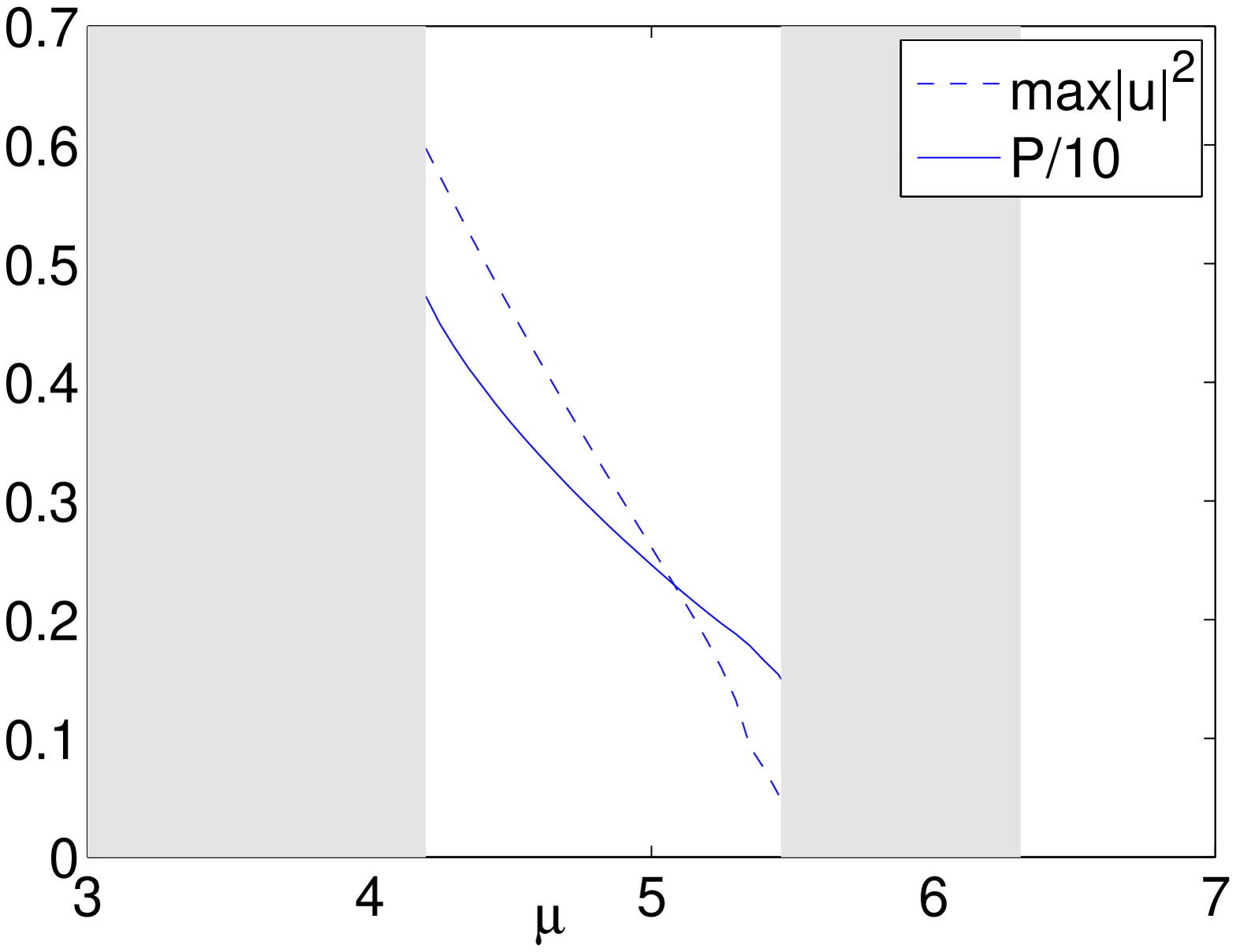}\\
\includegraphics[width=0.4\textwidth]{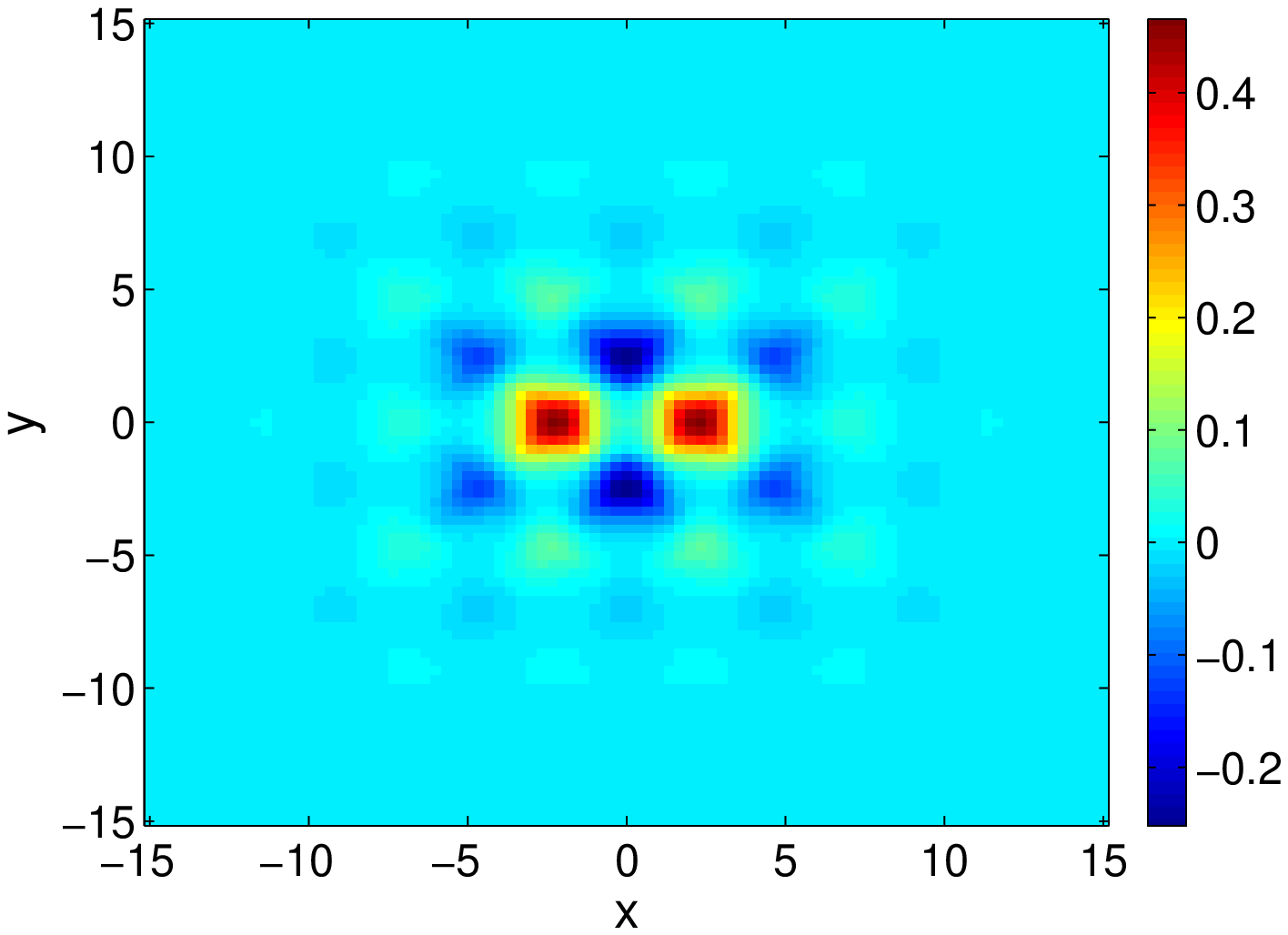}
\includegraphics[width=0.4\textwidth]{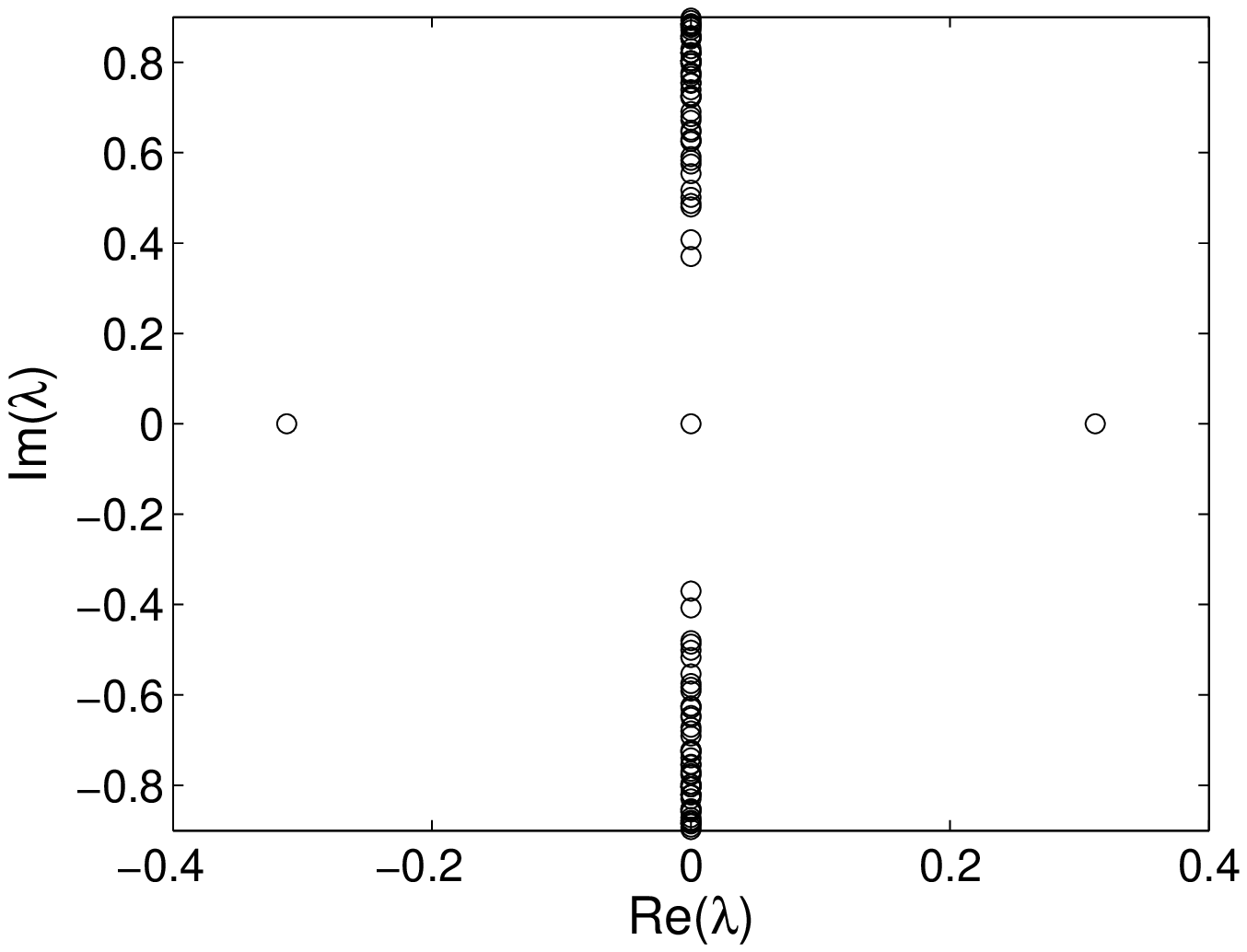}
\end{center}
\caption{(Color online) The top panels correspond to the same diagnostics
as in Fig.\ \ref{diaIP} but
for IP NNN dipole solitons. The bottom panels show the profile
$u$ and the corresponding spectral plane of the IP NNN dipole at $\mu=5.1$.}
\label{nondiaIP}
\end{figure}

\begin{figure}[tbh]
\begin{center}
\includegraphics[width=0.4\textwidth]{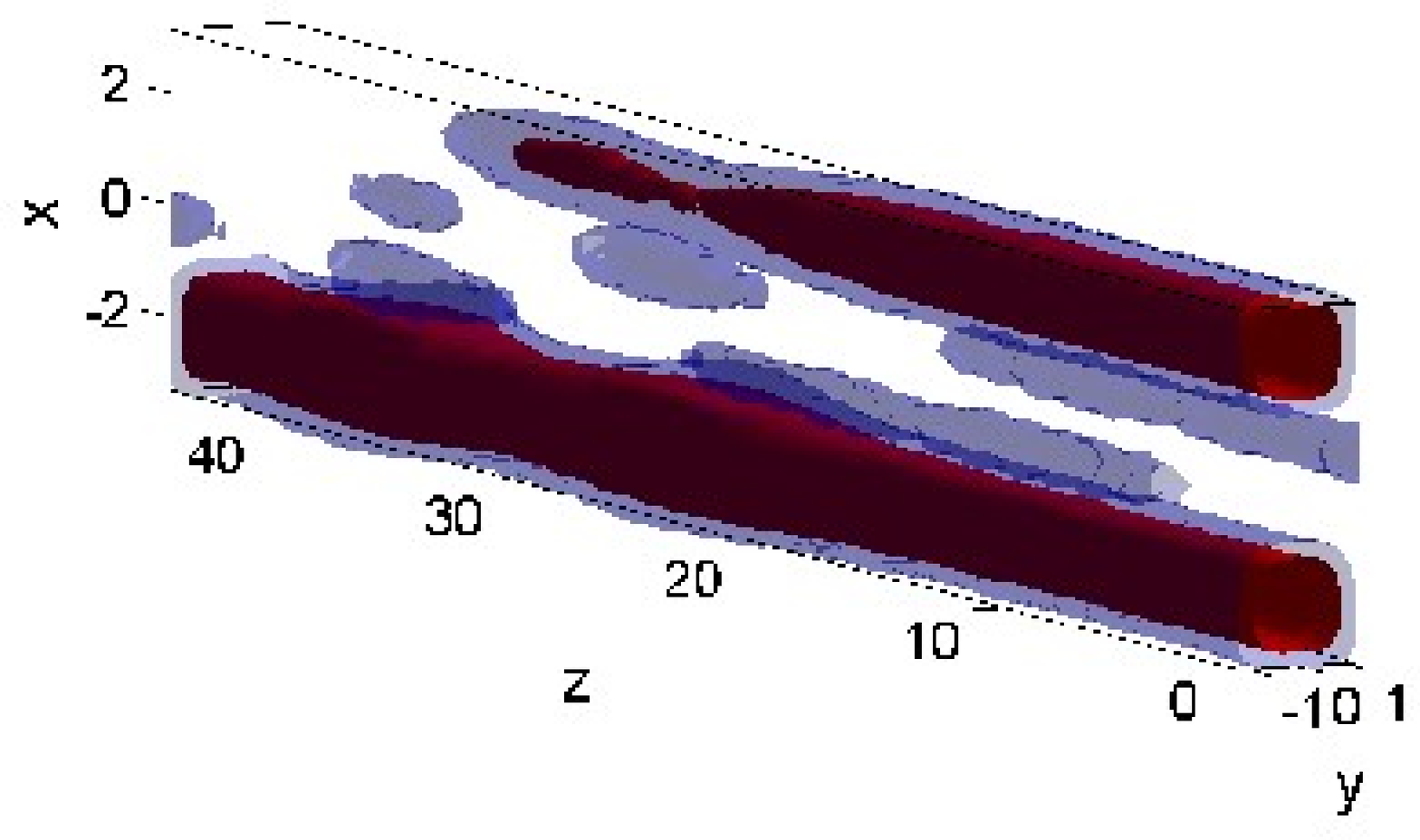}
\includegraphics[width=0.4\textwidth]{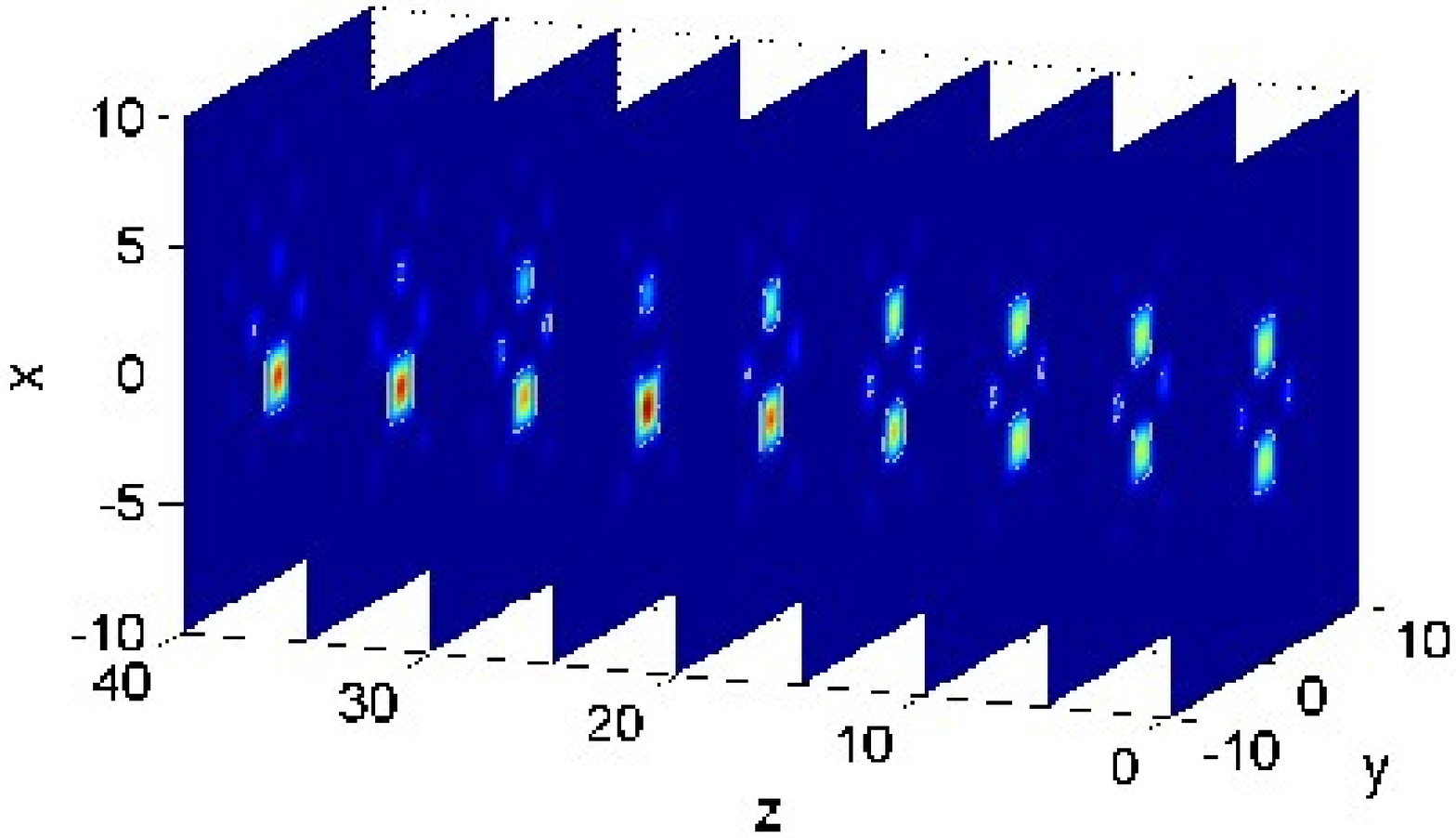}
\end{center}
\caption{(Color online) Similar to Fig.\ \ref{dyn_diaIP} but for the evolution of the
IP dipole shown in Fig.\ \ref{nondiaIP}. Presented are the isosurfaces of
height 0.1 (red) and 0.05 (blue) and the sliced contours at some selected
propagation distances. }
\label{dyn_non_diaIP}
\end{figure}

We have obtained this type of IP, NNN dipole solitons in a wide parameter
range. The stability, the power and peak intensity of these dipoles are
shown in Fig.\ \ref{nondiaIP}. These dipoles typically possess a real
eigenvalue (again in line with the discrete cubic model prediction of Table 1).

If one compares these IP NNN dipole solitons with the IP NN dipole solitons of
 Fig.\ \ref{diaIP}, there are clear differences, such as the typical
instability (with a real eigenvalue) of the former in contrast with the
typical stability of the latter (possessing
an imaginary eigenvalue of negative signature
that can potentially become unstable upon collision with another eigenvalue).
Another important difference is that the present IP dipoles emerge out of the
Bloch band edges, contrary to what is the case for
their IP NN counterparts where the solitons disappear through a saddle-node
bifurcation.
Interestingly, as the branch approaches
the upper band-edge, the profile of an IP NNN dipole becomes
similar to that of a $+-+-$ quadrupole (see section 5.1 below) and of an OOP NN dipole,
as the relevant limit is approached.

In Fig.\ \ref{dyn_non_diaIP}, we present the dynamical
evolution of the IP NNN dipole shown in Fig.\ \ref{nondiaIP} under
similar random noise perturbation as above. Here, we also see that the
instability appears earlier than the IP NN dipoles. The two humps
deform until they become one hump already at the propagation distance
$z \approx 20$, transforming the dipole solution into a single-hump
gap soliton.


\subsection{Out-of-phase Next-nearest-neighbor Dipole Solitons}

\begin{figure}[tbp!]
\begin{center}
\includegraphics[width=0.4\textwidth]{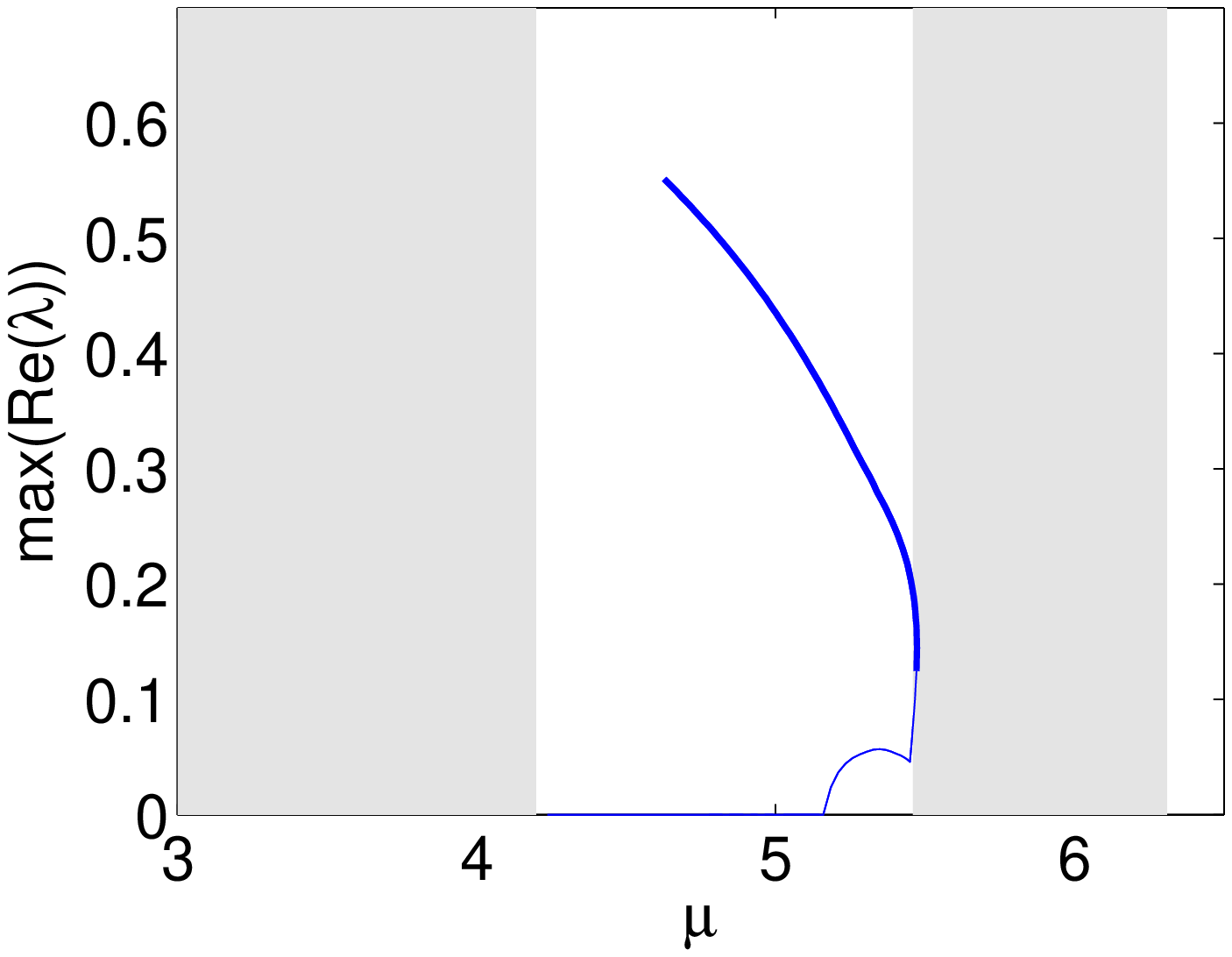}
\includegraphics[width=0.4\textwidth]{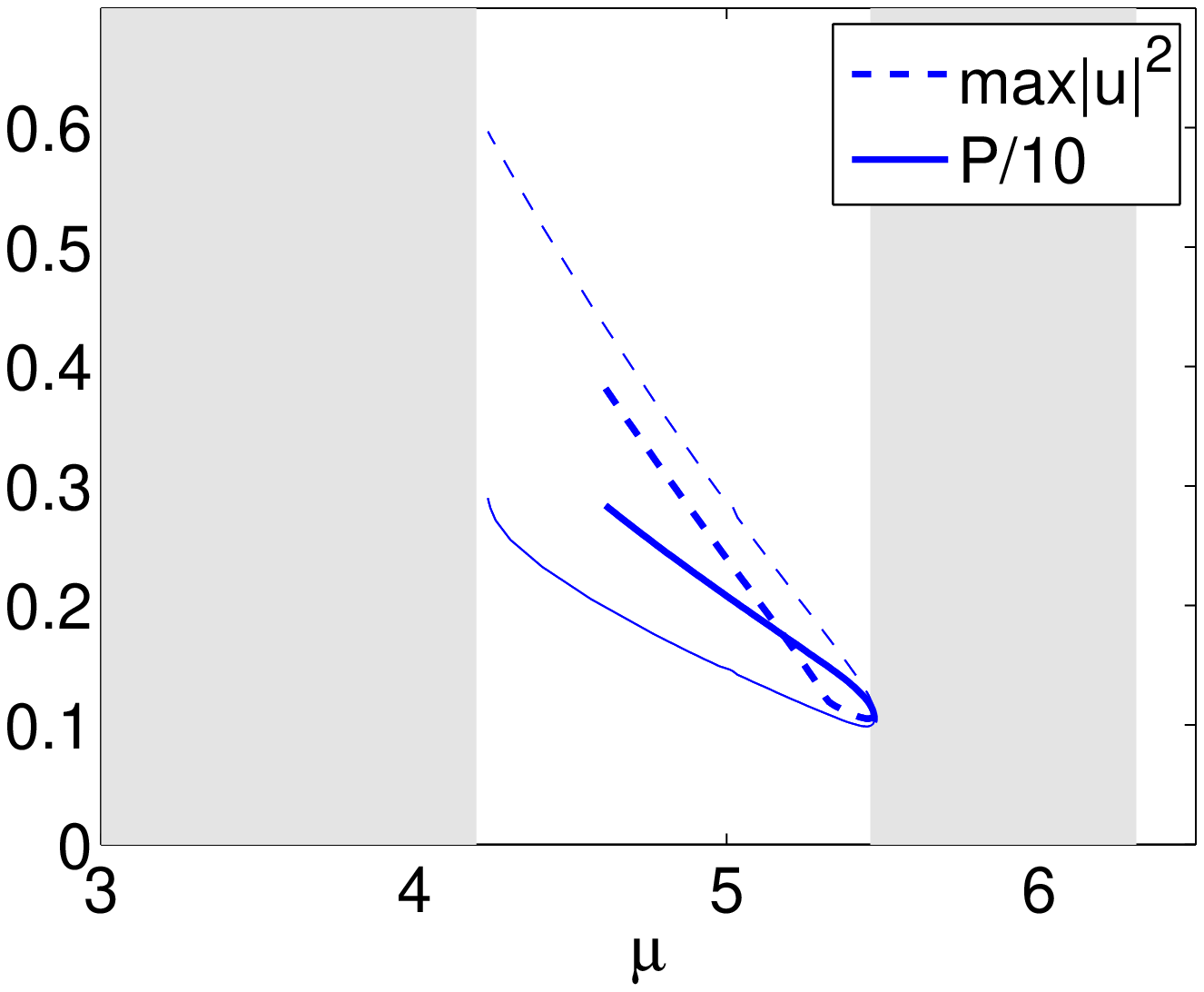}\\
\includegraphics[width=0.4\textwidth]{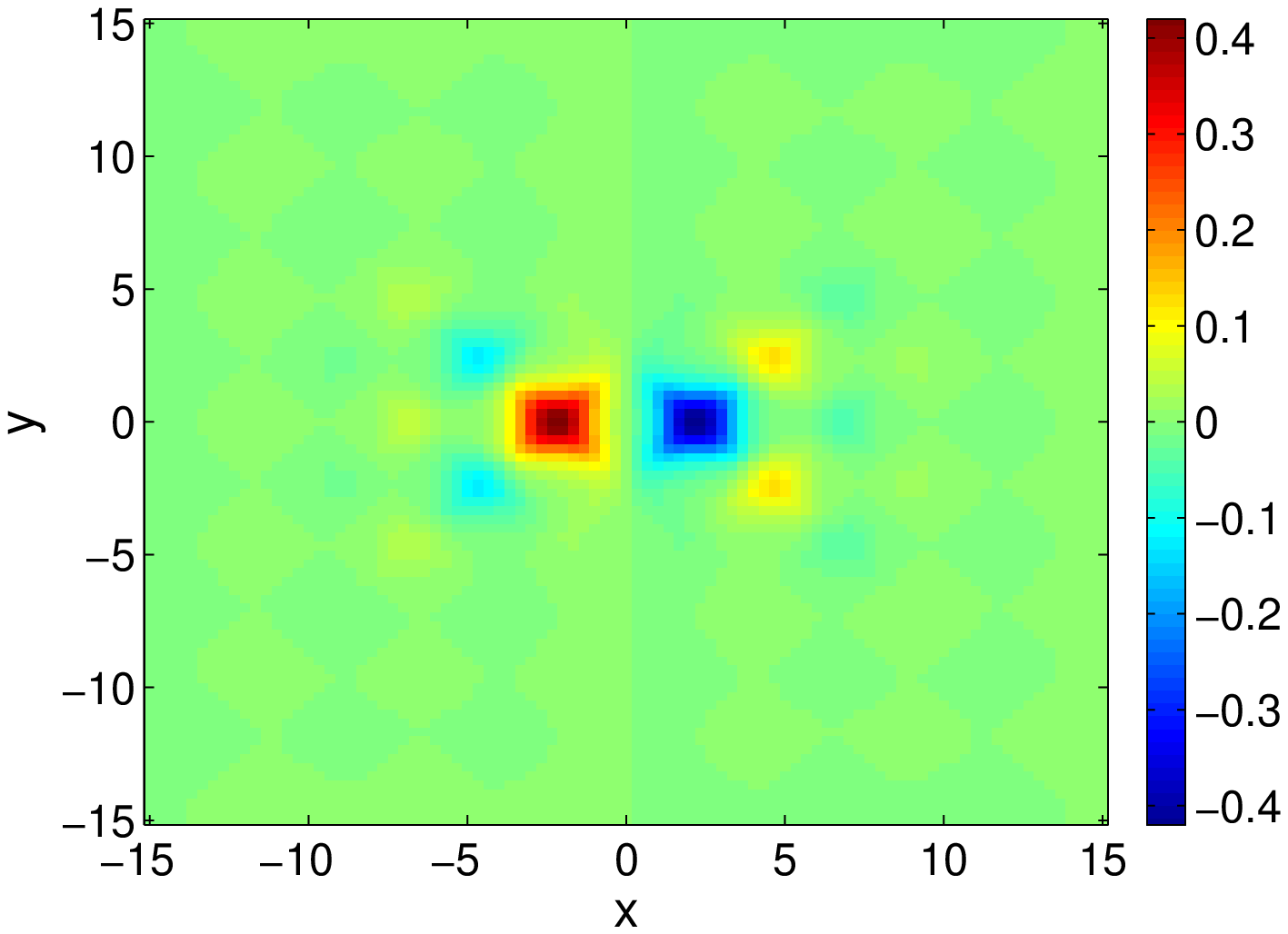}
\includegraphics[width=0.4\textwidth]{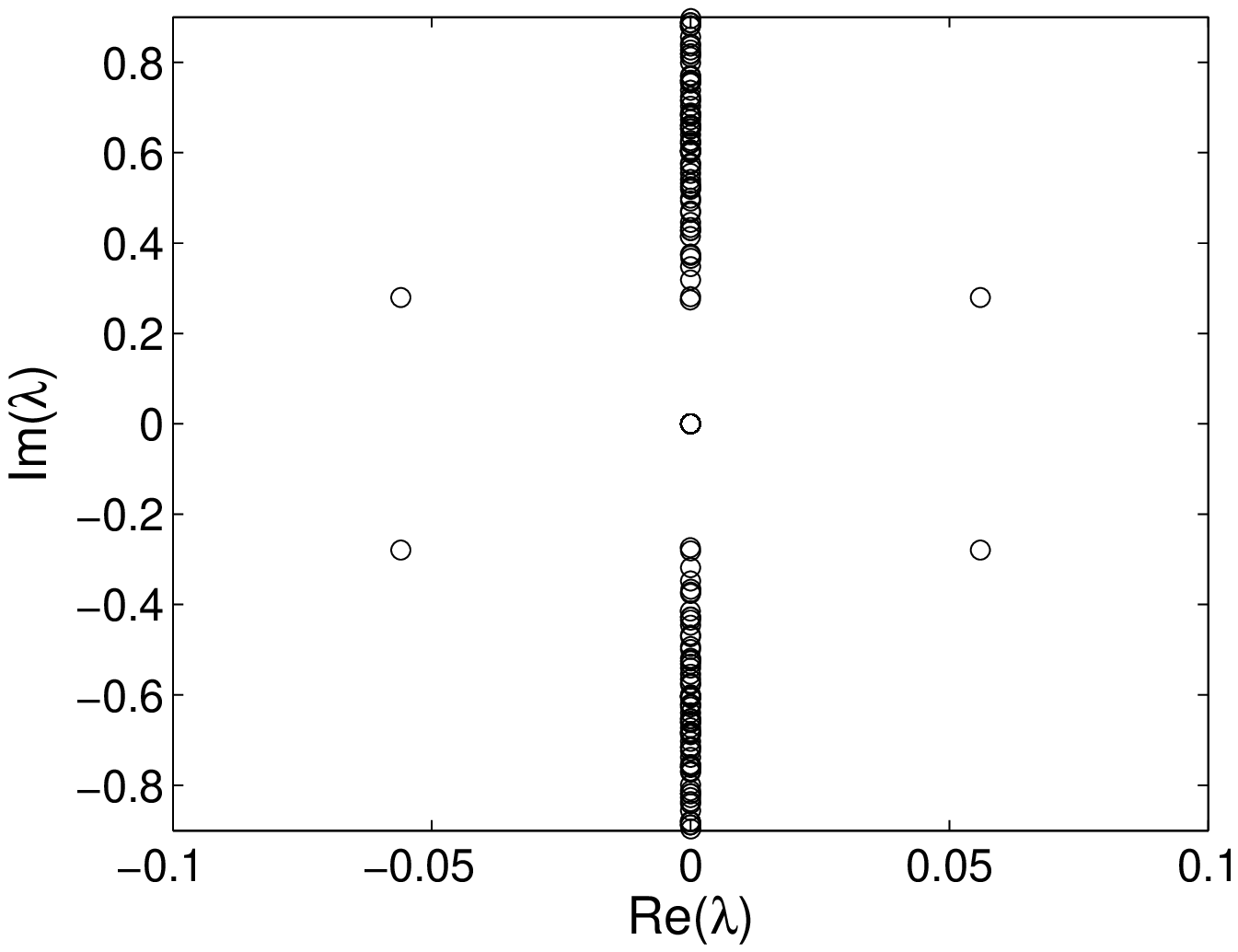}
\includegraphics[width=0.4\textwidth]{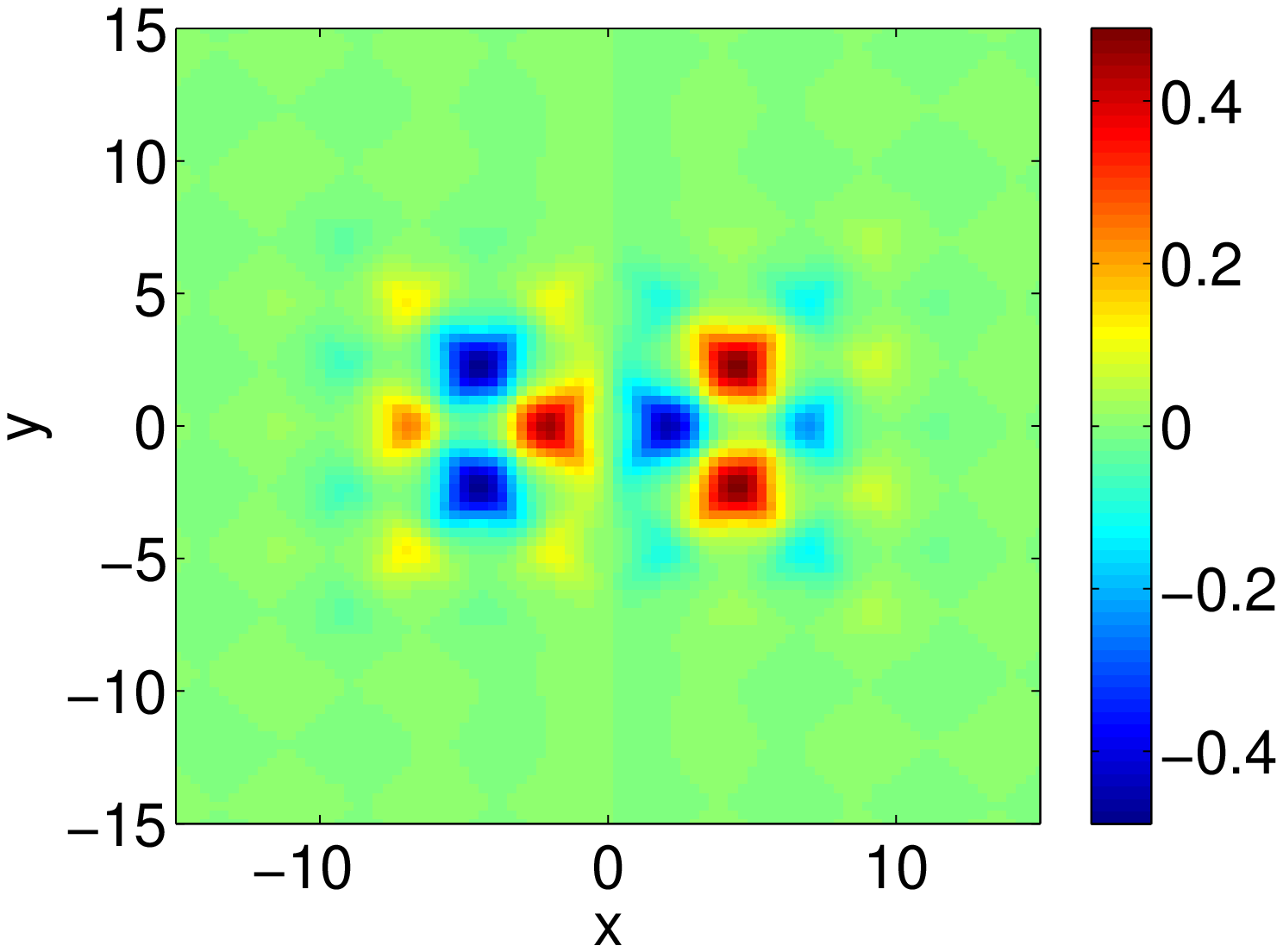}
\includegraphics[width=0.4\textwidth]{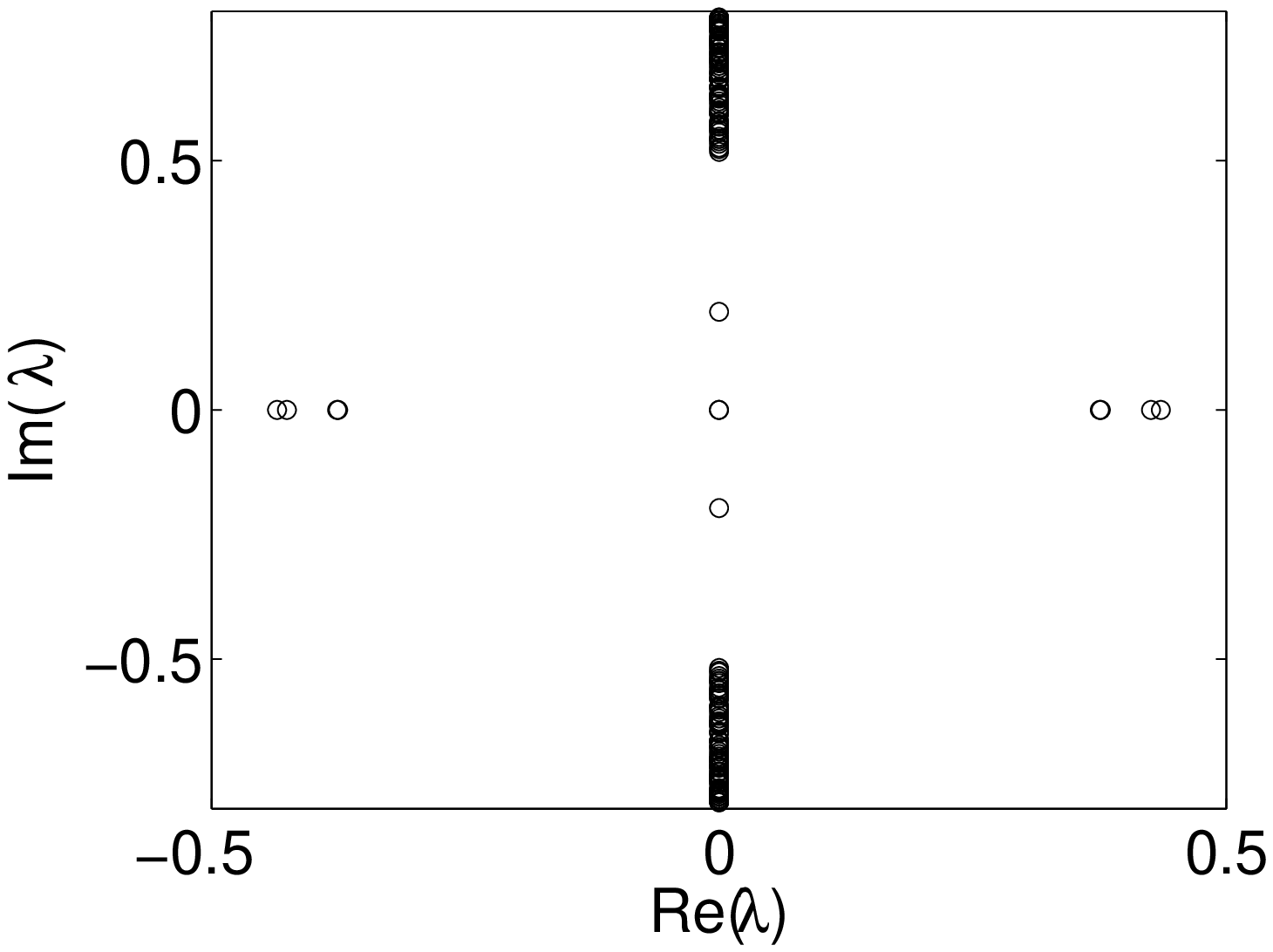}\\
\end{center}
\caption{(Color online) The top panels depict the largest real part of the critical
eigenvalue, as well as the power and the peak intensity of the OOP
NNN dipole solitons. The middle panels show the profile $u$ and the
corresponding spectra in the complex plane of the dipole at $\mu=5.3$,
and the bottom is the unstable saddle configuration, which collides with the OOP
NNN profile in a saddle-node
bifurcation, shown for $\mu=5$.}
\label{nondiaOOP}
\end{figure}

\begin{figure}[tbh]
\begin{center}
\includegraphics[width=0.4\textwidth]{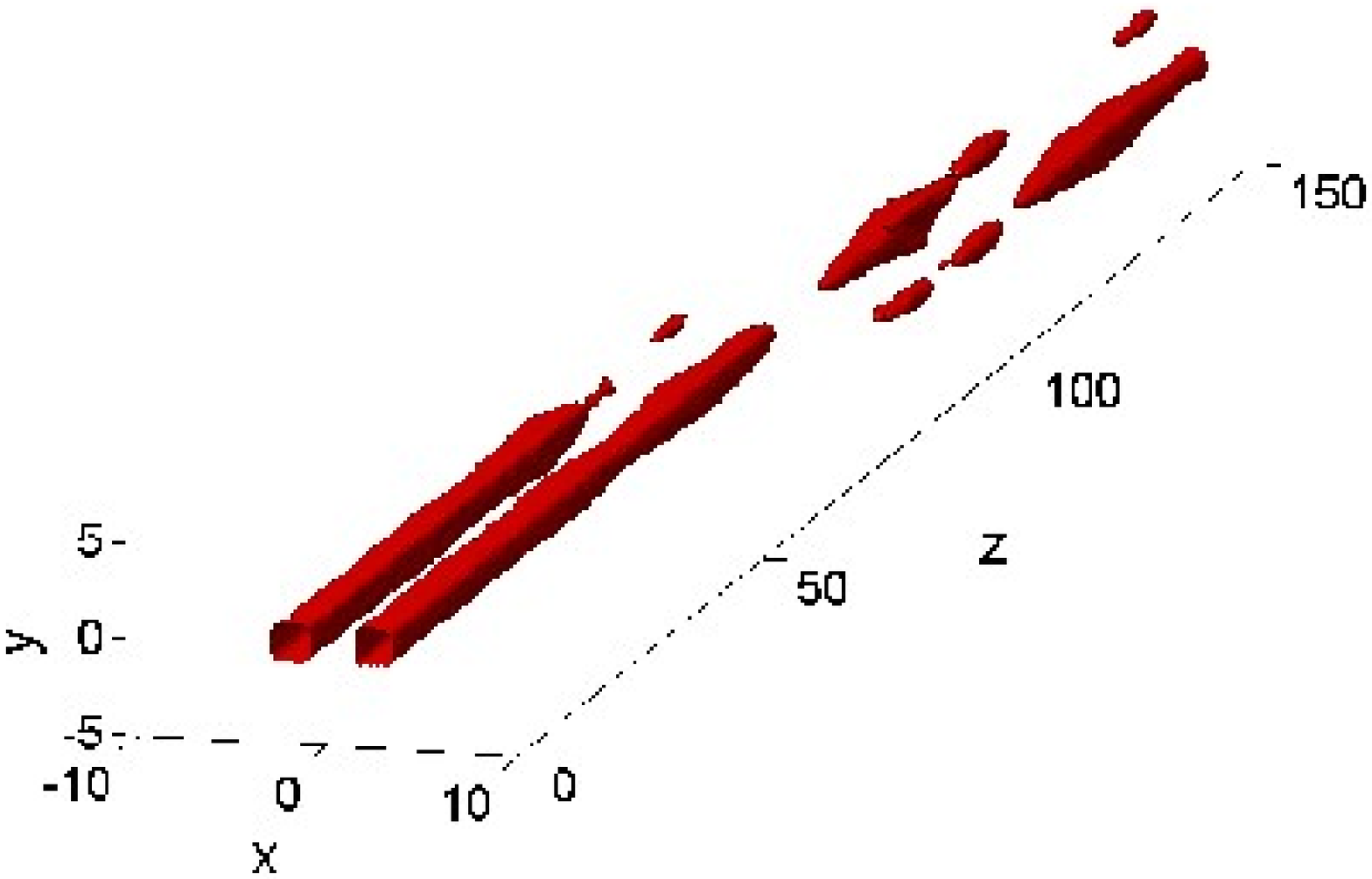}
\includegraphics[width=0.4\textwidth]{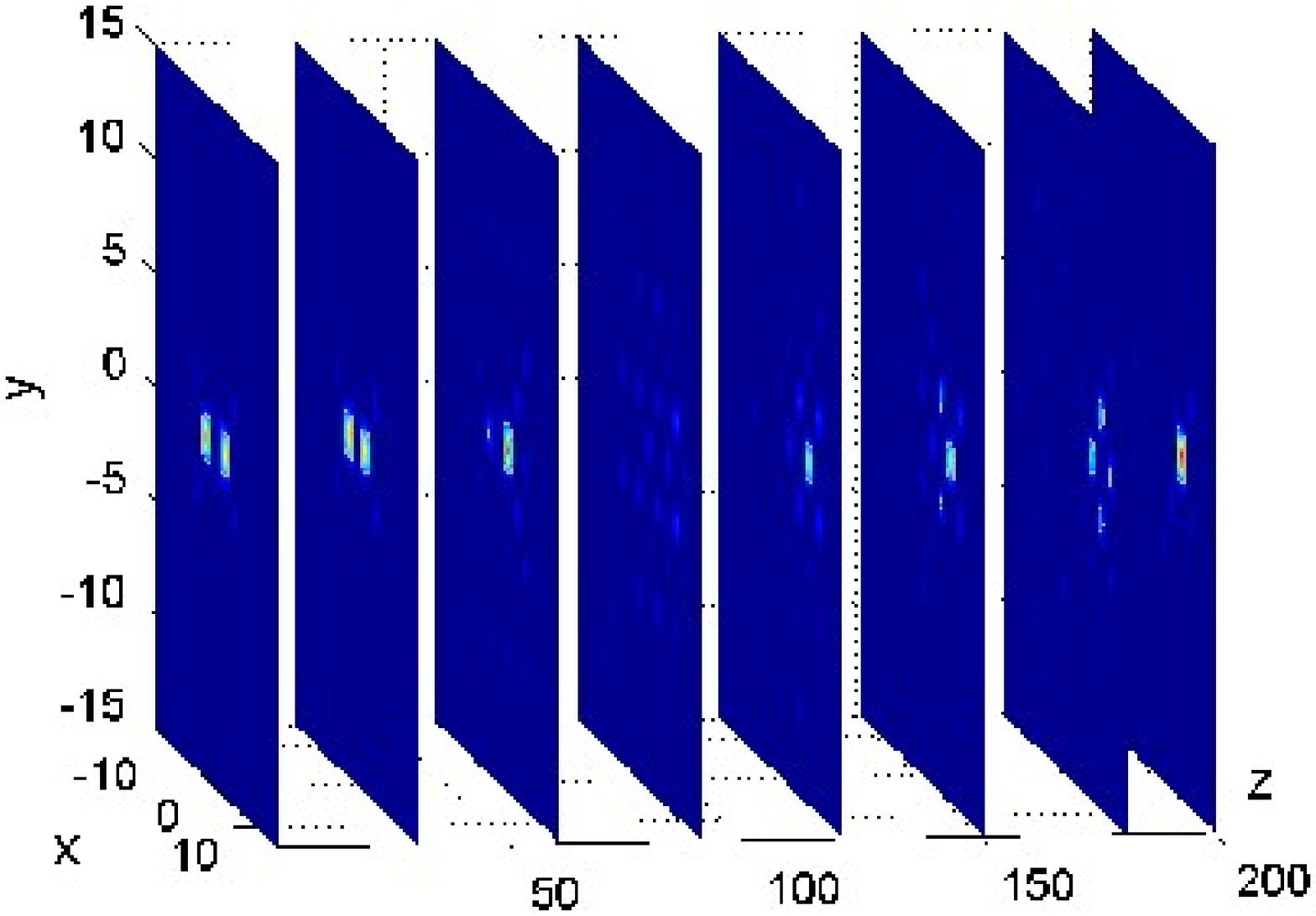}\\
\includegraphics[width=0.4\textwidth]{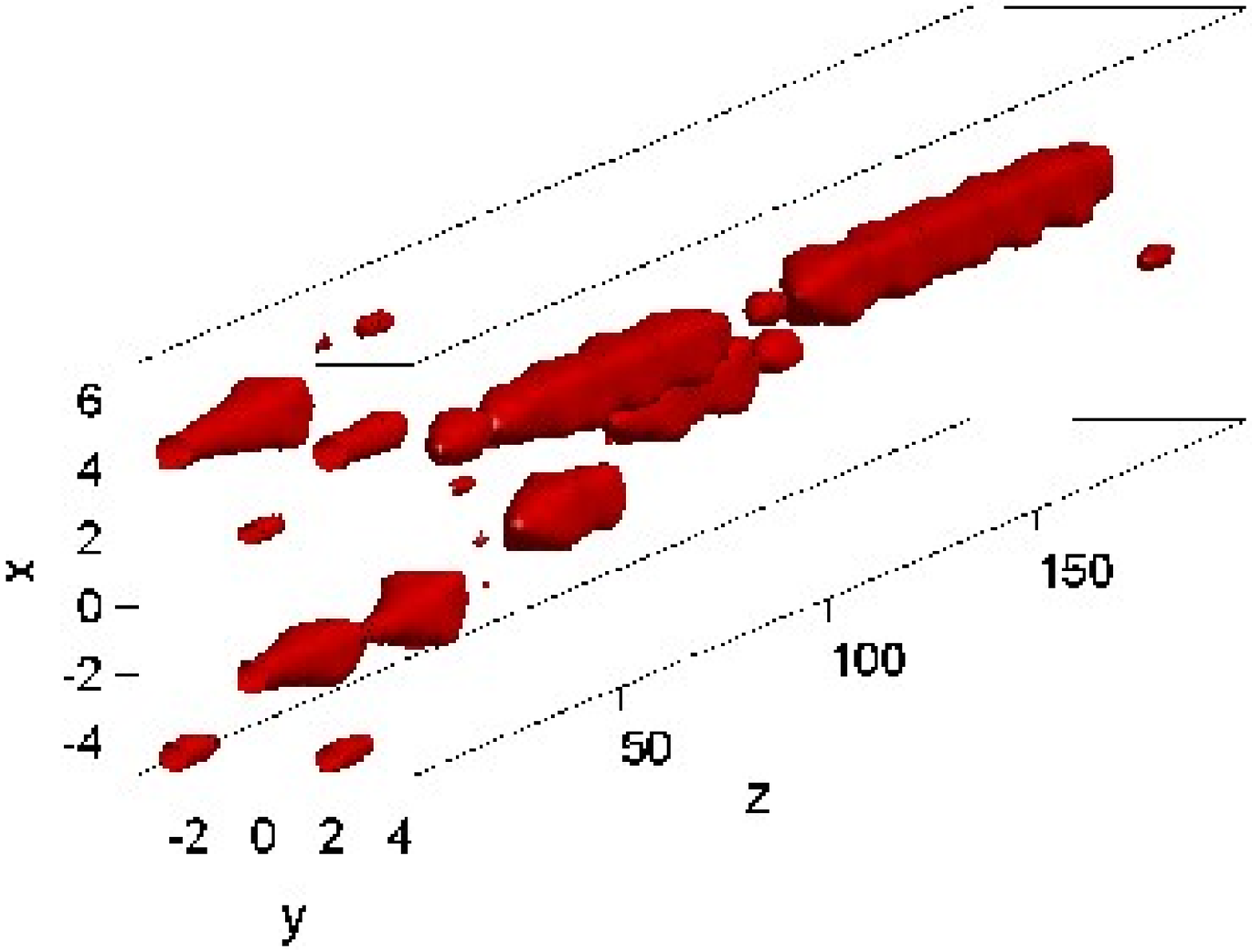}
\includegraphics[width=0.4\textwidth]{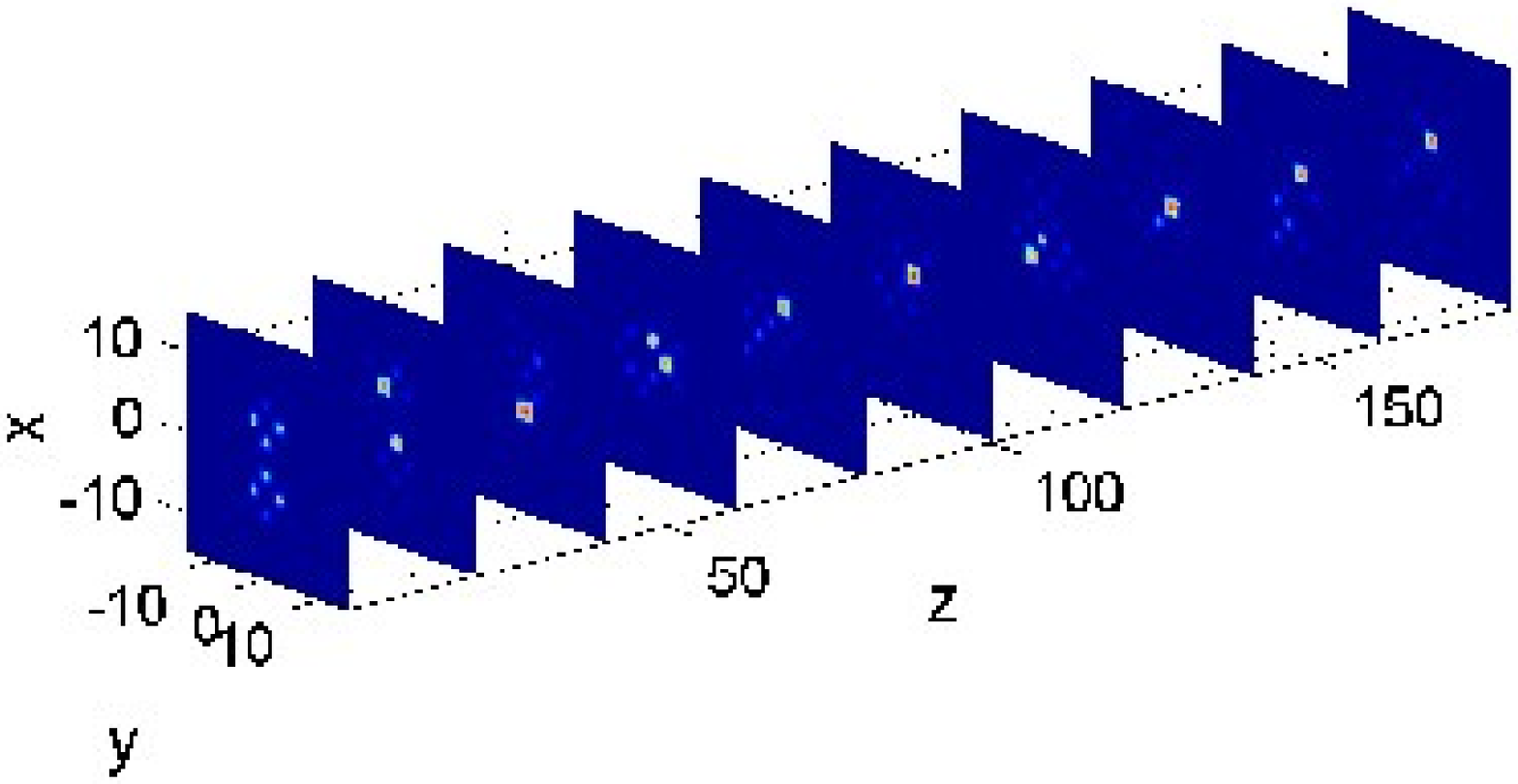}
\end{center}
\caption{(Color online) The evolutions of the OOP NNN dipole and its corresponding counterpart
for the same propagation constant $\mu$, i.e.\ $\mu=5.3$,
shown in Fig.\ \ref{nondiaIP}. Presented are the isosurface of height 0.1 and
the slices at particular propagation distances where one can see clearly
that the light tunnels away from its initial lattice wells.}
\label{dyn_non_diaOOP}
\end{figure}

We have also obtained OOP NNN dipole solitons. A typical profile of this
family of solutions for $\mu=5.3$ is shown in Fig.\ \ref{nondiaOOP}.
The power diagram of these solitons is presented in the top panel of
Fig.\ \ref{nondiaOOP}.
One important finding in this case
is that there is a relatively wide stability region,
i.e., typically these structures are stable (as indicated again by the
comparison with the results of Table 1). We have found that the stability
range for $E_0=8.62$ is $4.15\lesssim\mu\lesssim5.17$.
This class of solutions typically suffers an oscillatory instability
due to the presence of a single eigenvalue with negative signature and
its collision with the continuous spectrum, as shown in the middle right
panel of Fig.\ \ref{nondiaOOP} (cf. once again with the discrete
model prediction of Table 1).

For this solution also, we observe that similarly
to the IP NN dipoles,
it disappears at a non-zero peak intensity, in particular
for $\max(|u|^2) \approx 0.11$. The disappearance is because
of collision of this dipole with another configuration shown in
the bottom panels of Fig.\ \ref{nondiaOOP} in a saddle-node bifurcation.
It is relevant to note that the point of the bifurcation is very close to
the edge of the Bloch band, i.e.,\ for $\mu \approx 5.46$.

Subsequently, we simulate the dynamics of the instability and present it
in Fig.\ \ref{dyn_non_diaOOP}. One can see that for the OOP NNN, tunneling
occurs during  the propagation of the soliton along the $z$-direction,
finally leading to localization at a well different from the original
support of the dipolar structure. Regarding the counterpart solution of the other branch,
the evolution along the $z$-direction is a bit different as there is no tunneling away from
the original position. Yet, this configuration also leads to localization
at a single site in its final configuration.


\section{Nearest-neighbor Quadrupole Solitons}

We now turn to the examination of quadrupolar structures with
configurations having four lobes at four adjacent lattice wells.
Such structures turn out to exist in a large parameter region as well.
In particular,
we will focus our discussion on the case of such nearest-neighbor
quadrupoles.

Before we proceed with the discussion of our findings, we will explain the
notations that we use for the quadrupole configurations. Since there are
four humps with the phase difference between two neighboring wells being
either 0 or $\pi$, we have four possible configurations (up to rotational
and phase symmetries), i.e.\ $+-+-$, $+--+$, $++++$, and $+---$, where '+' and '-' represent the sign of the excitation of the main hump of $u$, i.e.\
whether it is positive (phase $0$) or negative (phase $\pi$), respectively,
at the four excited sites A, B, C and D of Fig. \ref{lattice}.

\subsection{$+-+-$ Nearest-neighbor Quadrupole Solitons}

\begin{figure}[tbh]
\begin{center}
\includegraphics[width=0.4\textwidth]{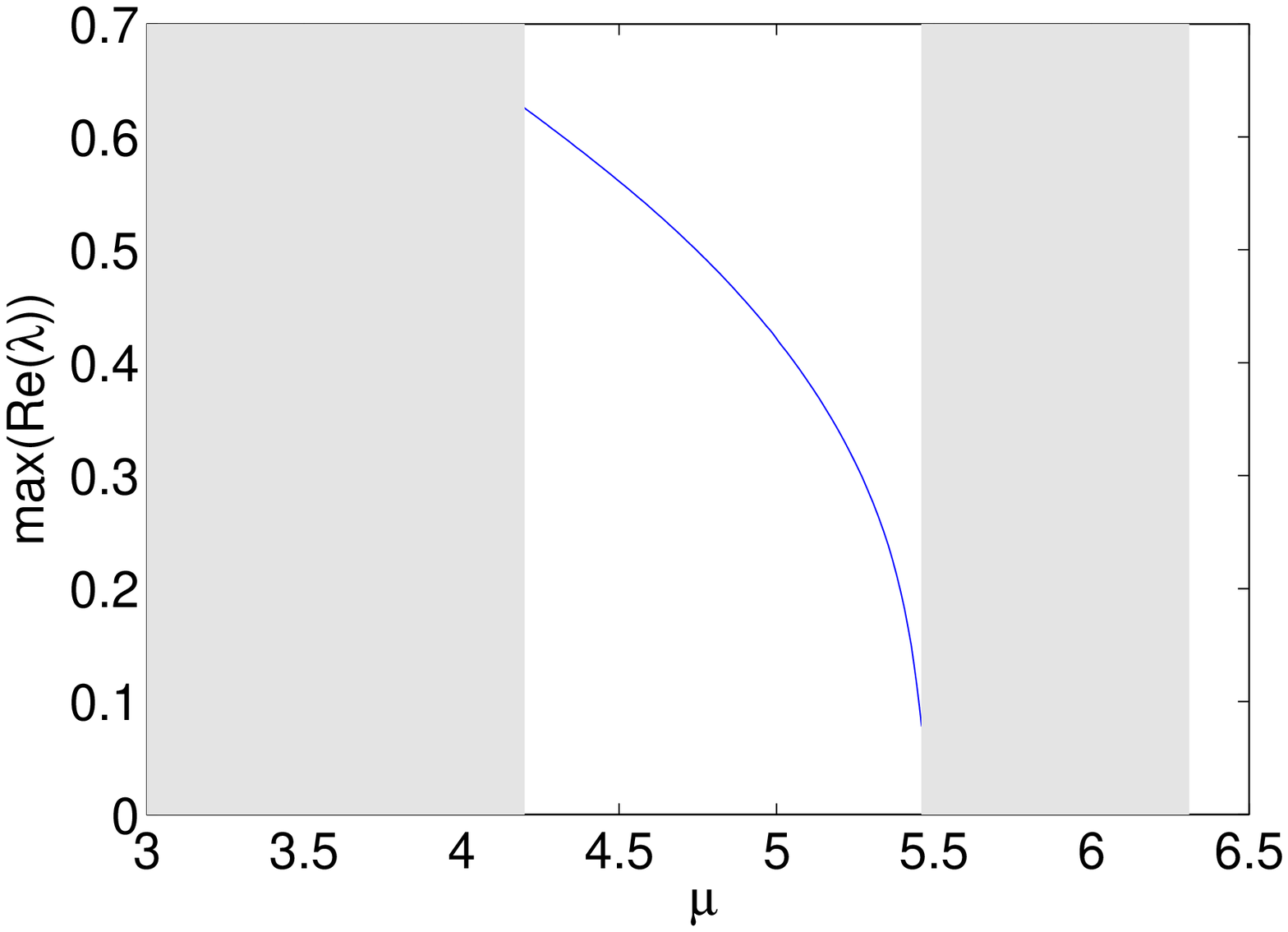}
\includegraphics[width=0.4\textwidth]{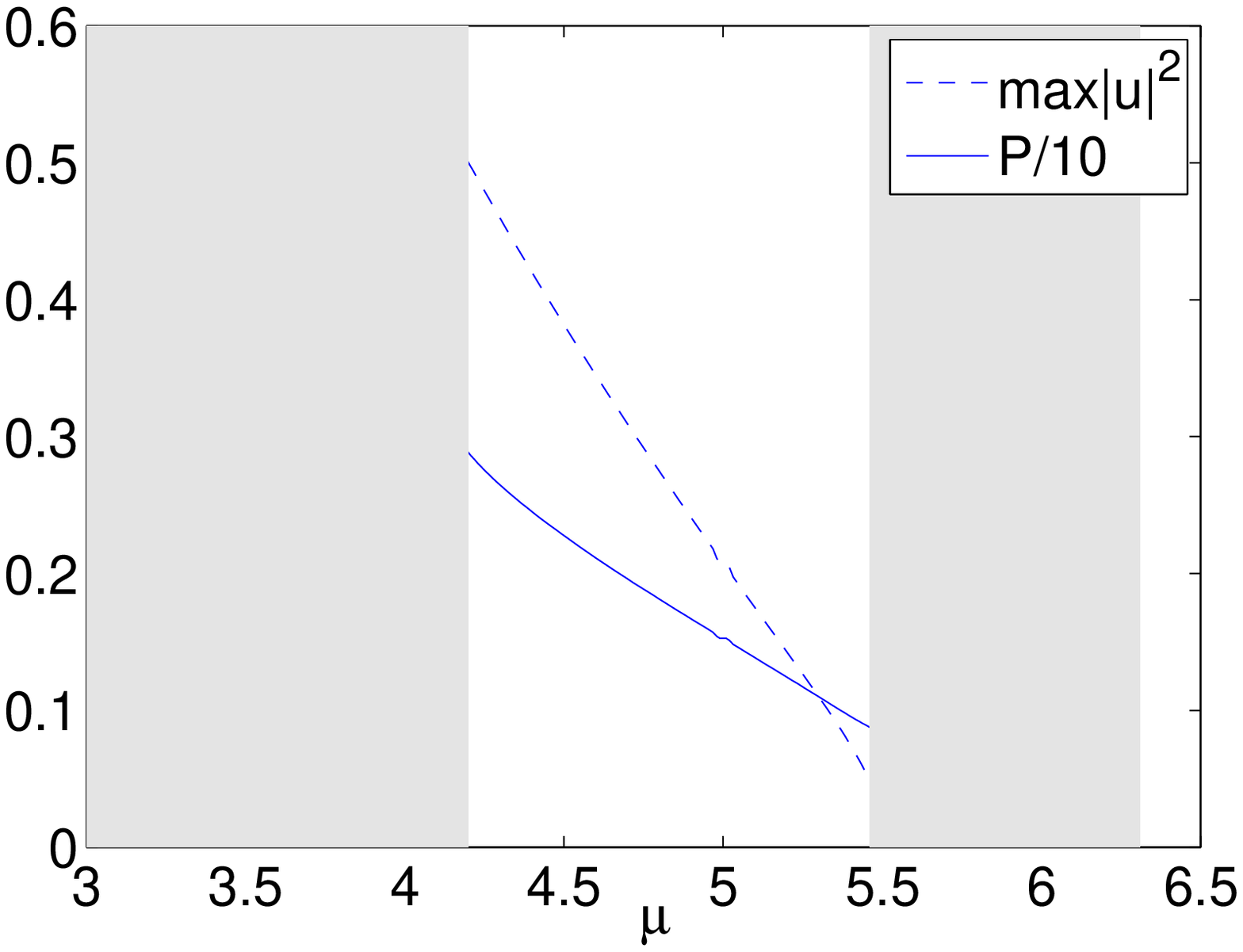}\\
\includegraphics[width=0.4\textwidth]{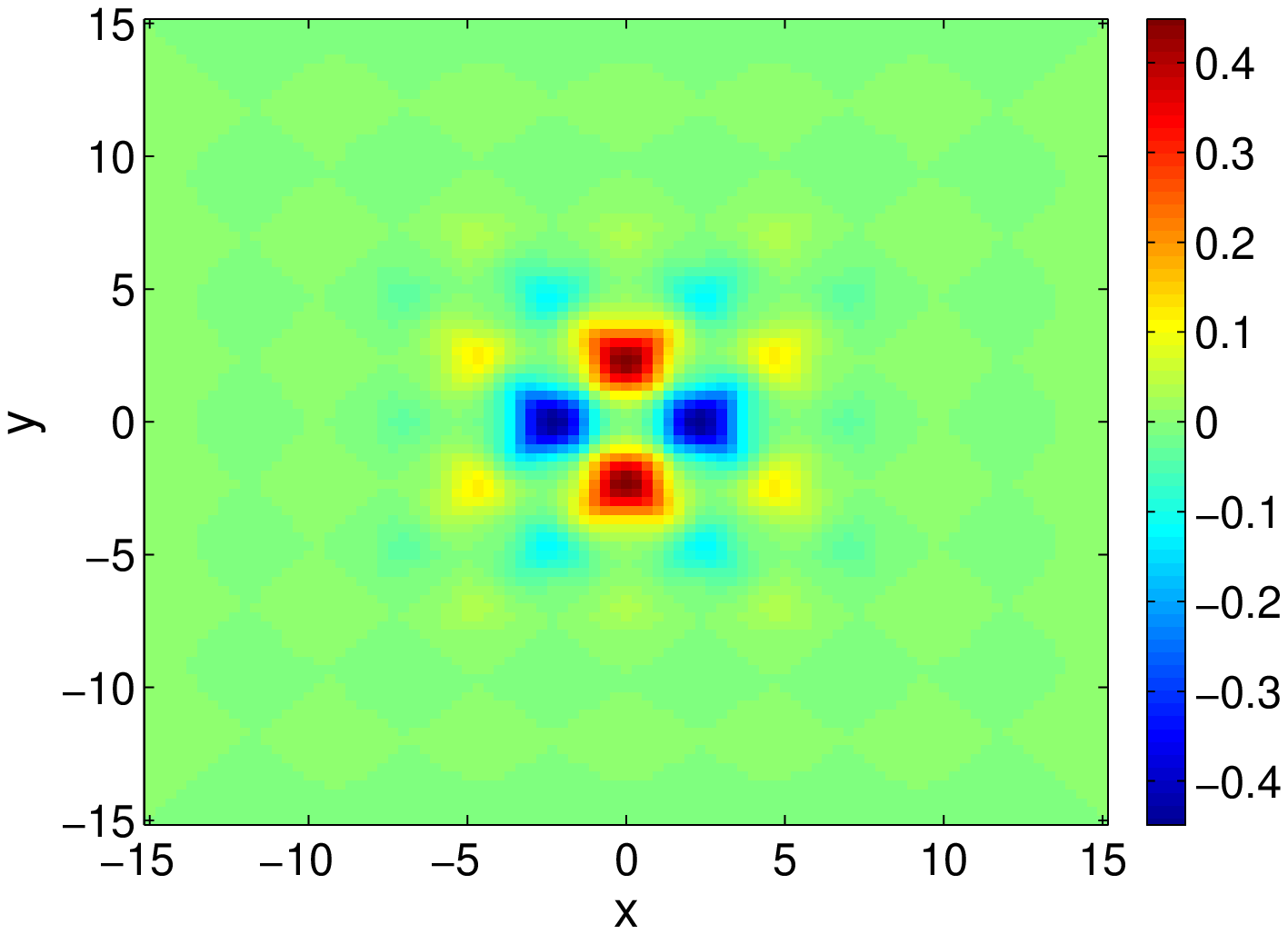}
\includegraphics[width=0.4\textwidth]{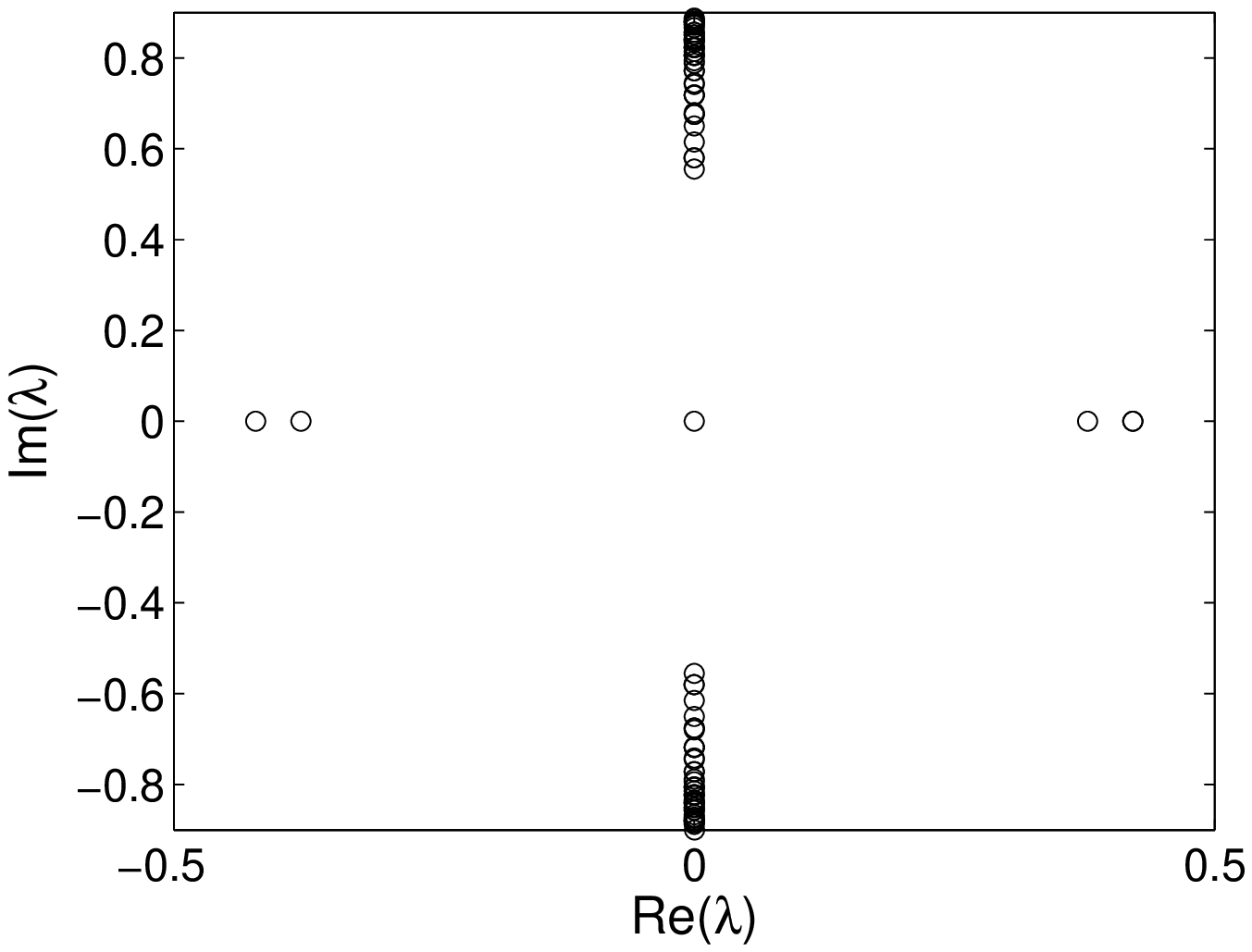}
\end{center}
\caption{(Color online) The maximum of the real part of the critical eigenvalue
(top left panel) and the power and the peak intensity of the $+-+-$
quadrupoles (top right panel). The bottom panels depict the profile $u$ of a
$+-+-$ NN quadrupole at $\mu=5$ and the corresponding linearization
spectral plane of eigenvalues.}
\label{quadUDUD}
\end{figure}

First, we consider the $+-+-$ type. In Fig.\ \ref{quadUDUD}, we present a
field profile of this structure for $\mu=5$.
We found that this configuration exists in the entire forbidden-band
$\mu\in(4.2,5.46)$, i.e.,\ OOP NN quadrupoles also bifurcate from the edge of
the Floquet band. Nonetheless, we observe that a typical instability that these
structures have is 3 real eigenvalue pairs (two of which are coincident in
the bottom right panel of Fig.\ \ref{quadUDUD}), in line with the
predictions of Table 1.

\begin{figure}[tbh]
\begin{center}
\includegraphics[width=0.4\textwidth]{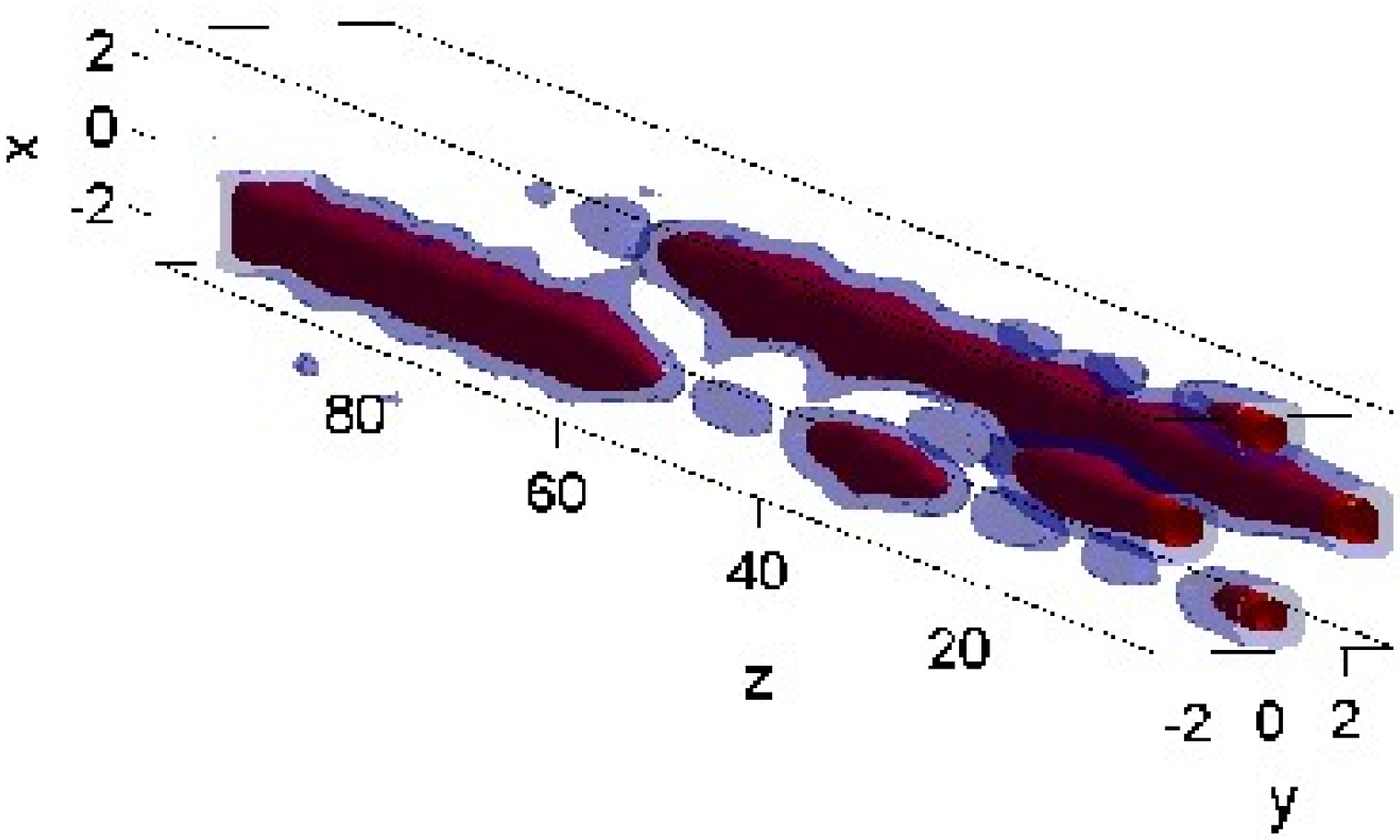}
\includegraphics[width=0.4\textwidth]{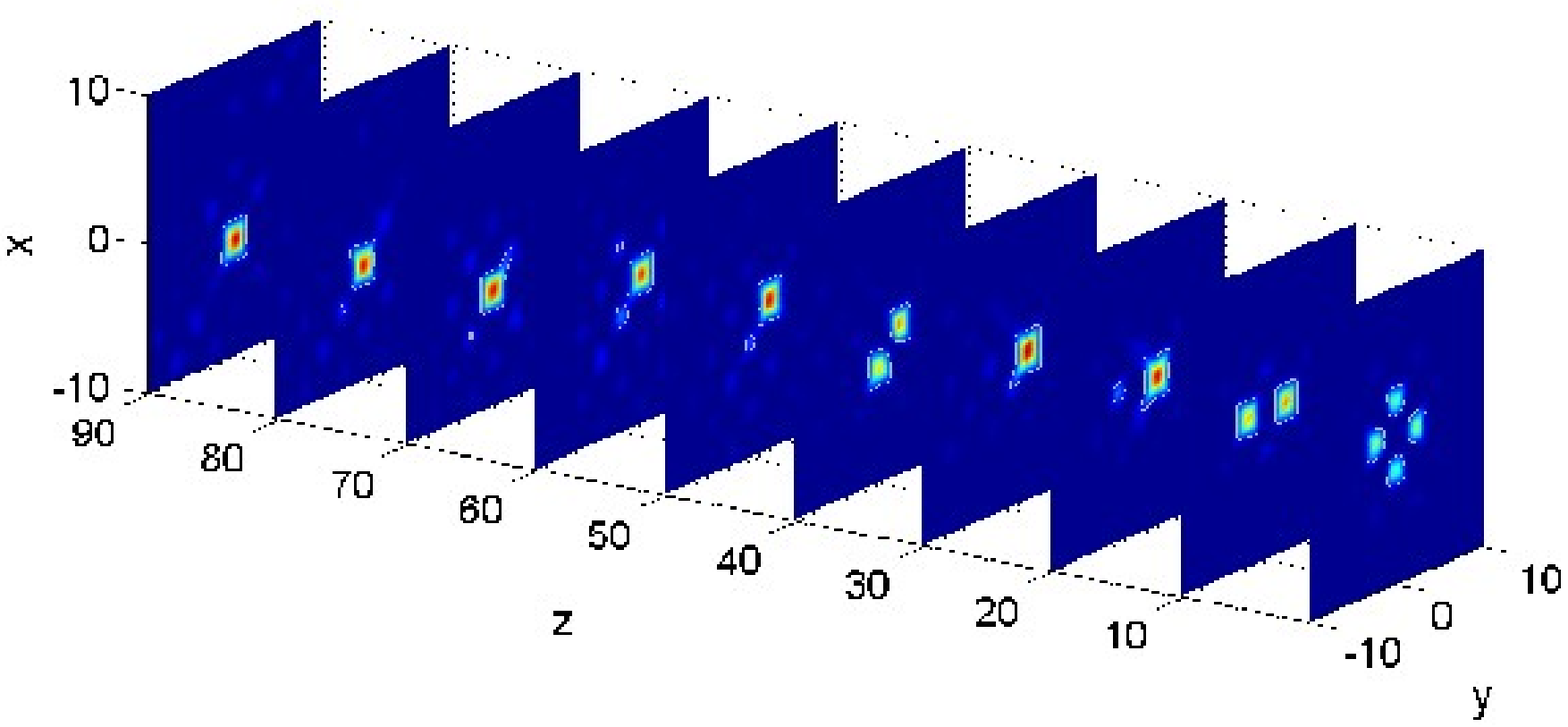}
\end{center}
\caption{(Color online) The evolution of the quadrupole presented in Fig.\ \ref{quadUDUD}.
Shown are the isosurfaces of height 0.15 (red) and 0.07 (blue) and its
slices at some selected propagation distances.}
\label{dyn_quadUDUD}
\end{figure}

In Fig.\ \ref{dyn_quadUDUD}, we present the evolution of the unstable $+-+-$
quadrupole shown in Fig.\ \ref{quadUDUD}. One can see that the configuration
is strongly unstable resulting in a breakup of the structure
already for $z \approx 10$ to a IP NNN dipole state.
Subsequent evolution also visits the other state associated
with the quadrupole branch, namely the OOP NN and eventually results
into a single hump gap soliton.

\subsection{$++++$ Nearest-neighbor Quadrupole Solitons}

\begin{figure}[tbh]
\begin{center}
\includegraphics[width=0.4\textwidth]{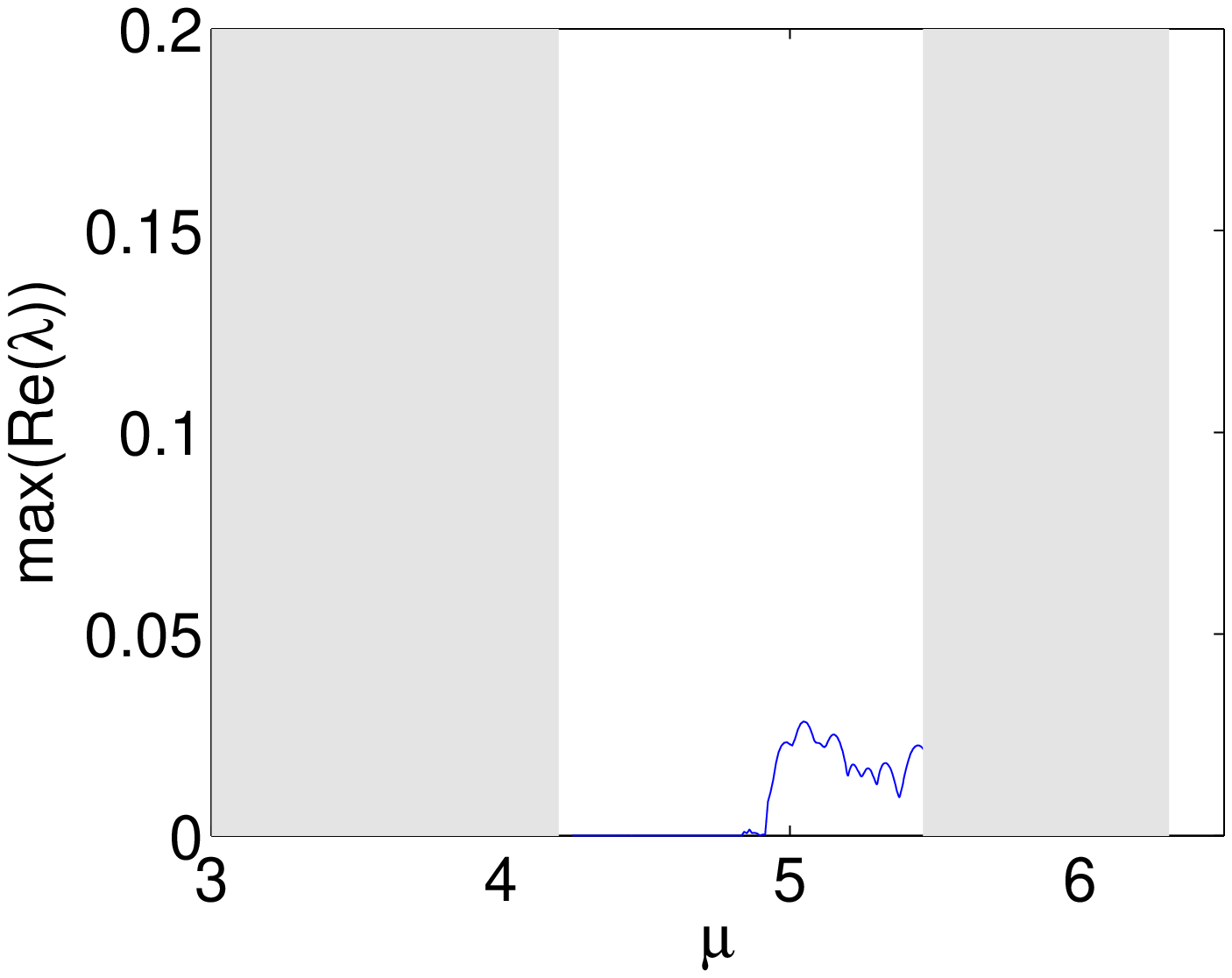}
\includegraphics[width=0.4\textwidth]{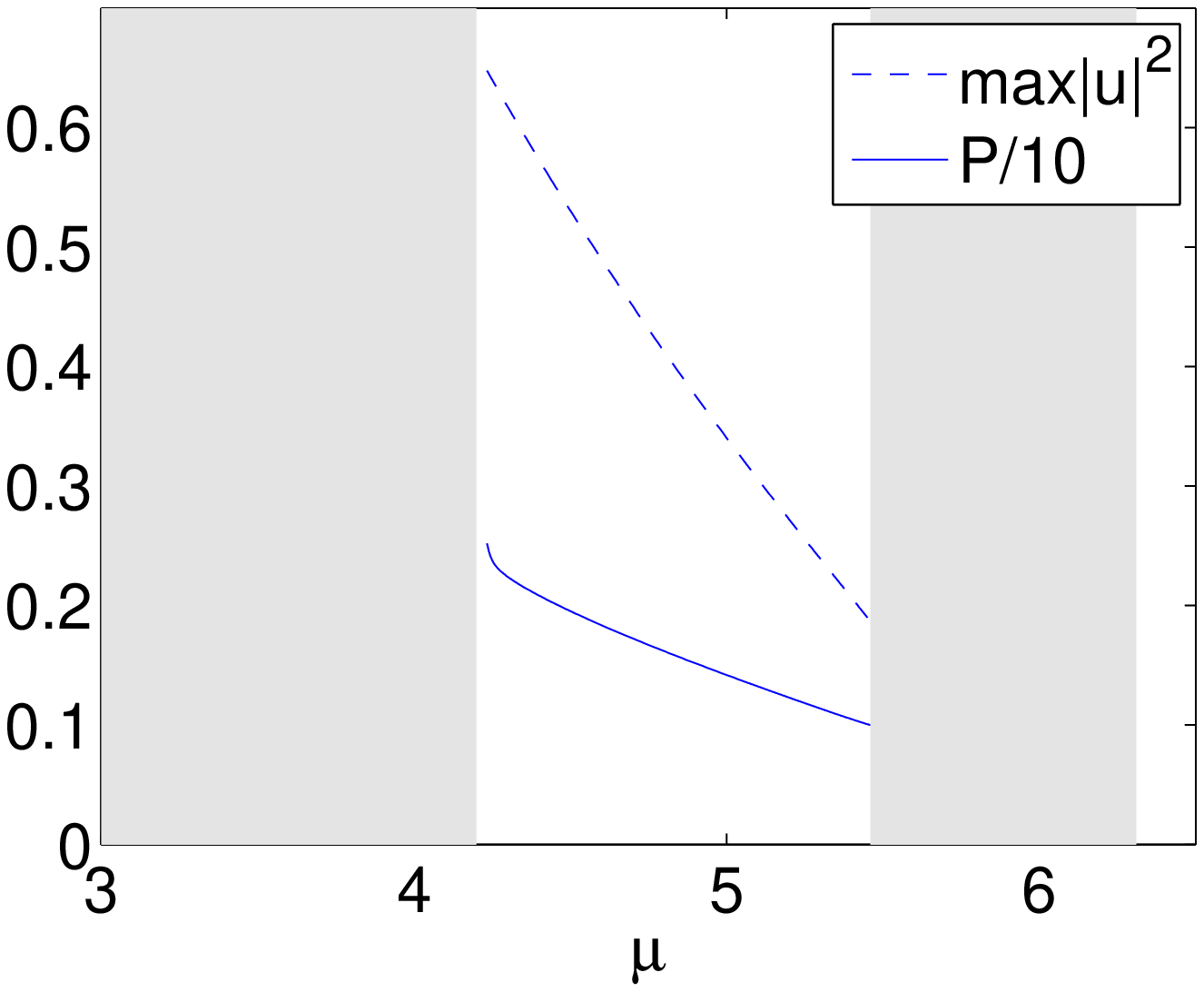}\\
\includegraphics[width=0.4\textwidth]{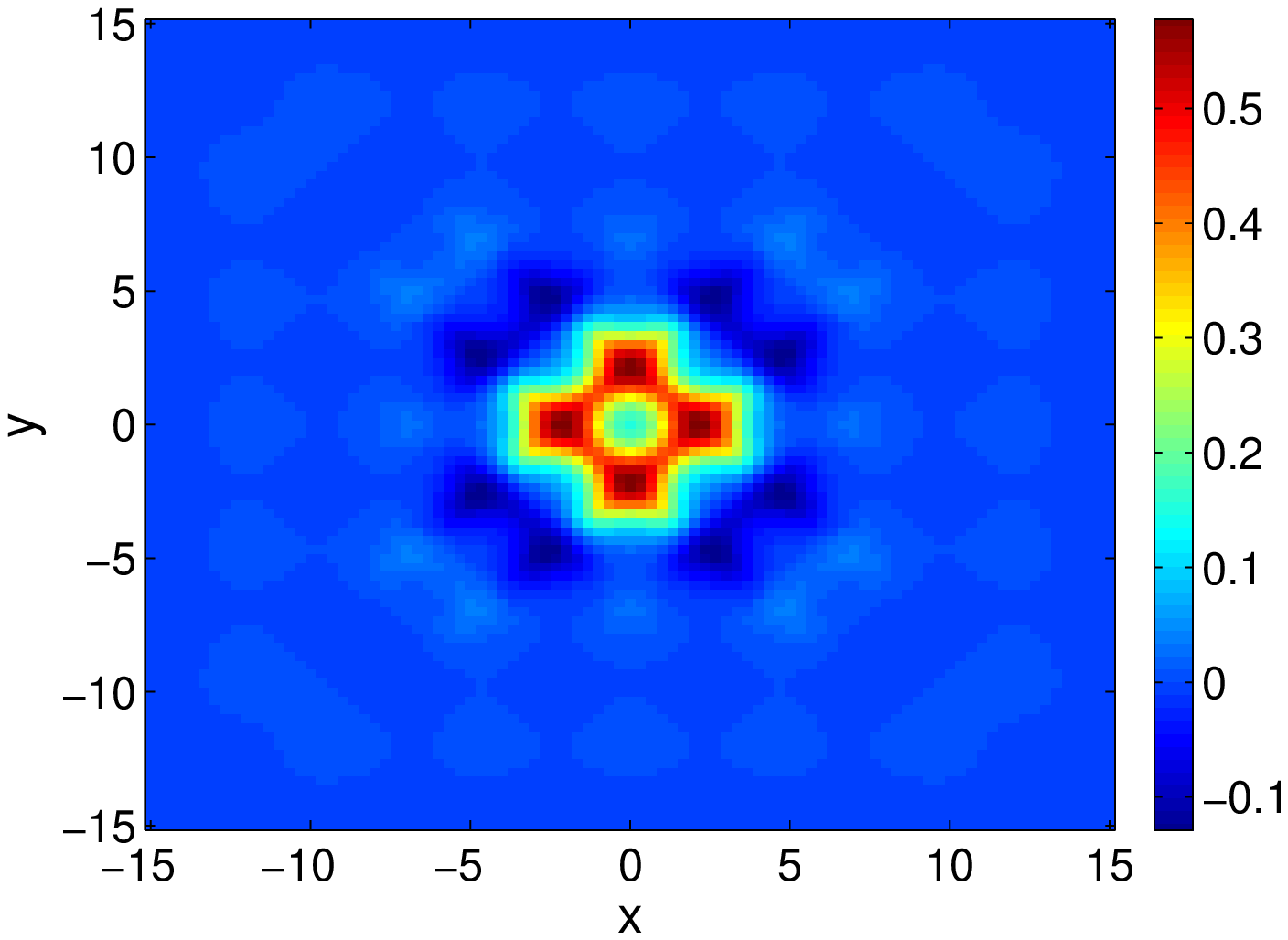}
\includegraphics[width=0.4\textwidth]{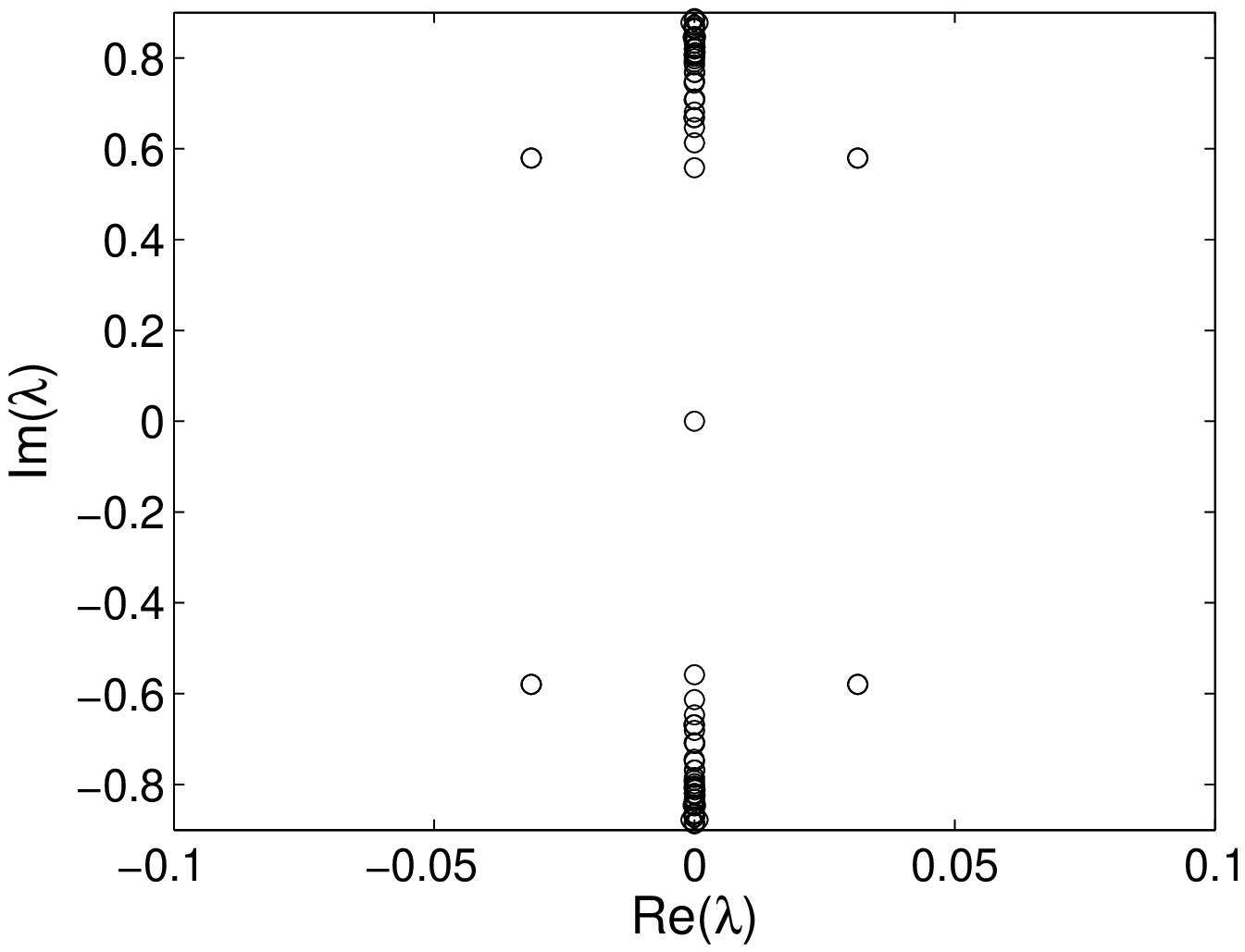}
\end{center}
\caption{(Color online) The same as Fig.\ \ref{quadUDUD} but for the $++++$ quadrupoles.
The top panels show the maximal instability growth rate of
perturbations (left panel) and the branch optical power and maximal intensity
(right panel), while the bottom panels show the solution profile (left)
and linear stability (right panel) for $\mu=5$.}
\label{quadUUUU}
\end{figure}

$++++$ quadrupoles have four in-phase humps at adjacent lattice sites.
A typical example of this sort is shown in Fig.\ \ref{quadUUUU} at $\mu=5$.
We have analyzed the stability of this configuration, finding that it
is stable in a wide parametric range, namely for
$4.09\lesssim\mu\lesssim4.93$, in line with what is predicted by
the cubic nonlinearity discrete model in Table 1.
Similarly to the IP NN and the OOP NNN branch this branch disappears
at a non-zero peak intensity (in this case, for $\max(|u|^2)\approx0.16$).

The dynamics of the configuration when it is unstable has also been
simulated; as a particular example, we integrate the field profile shown in
Fig.\ \ref{quadUUUU} in $z$. Even though it is unstable, we found that
propagation even for 100 units of (dimensionless) length, the profile still
resembles its initial condition. This seems to indicate that such structures
should be rather straightforward to observe experimentally. For sufficiently
long propagation, the dipole will eventually be destroyed through an
oscillatory instability.

\begin{figure}[tbh]
\begin{center}
\includegraphics[width=0.4\textwidth]{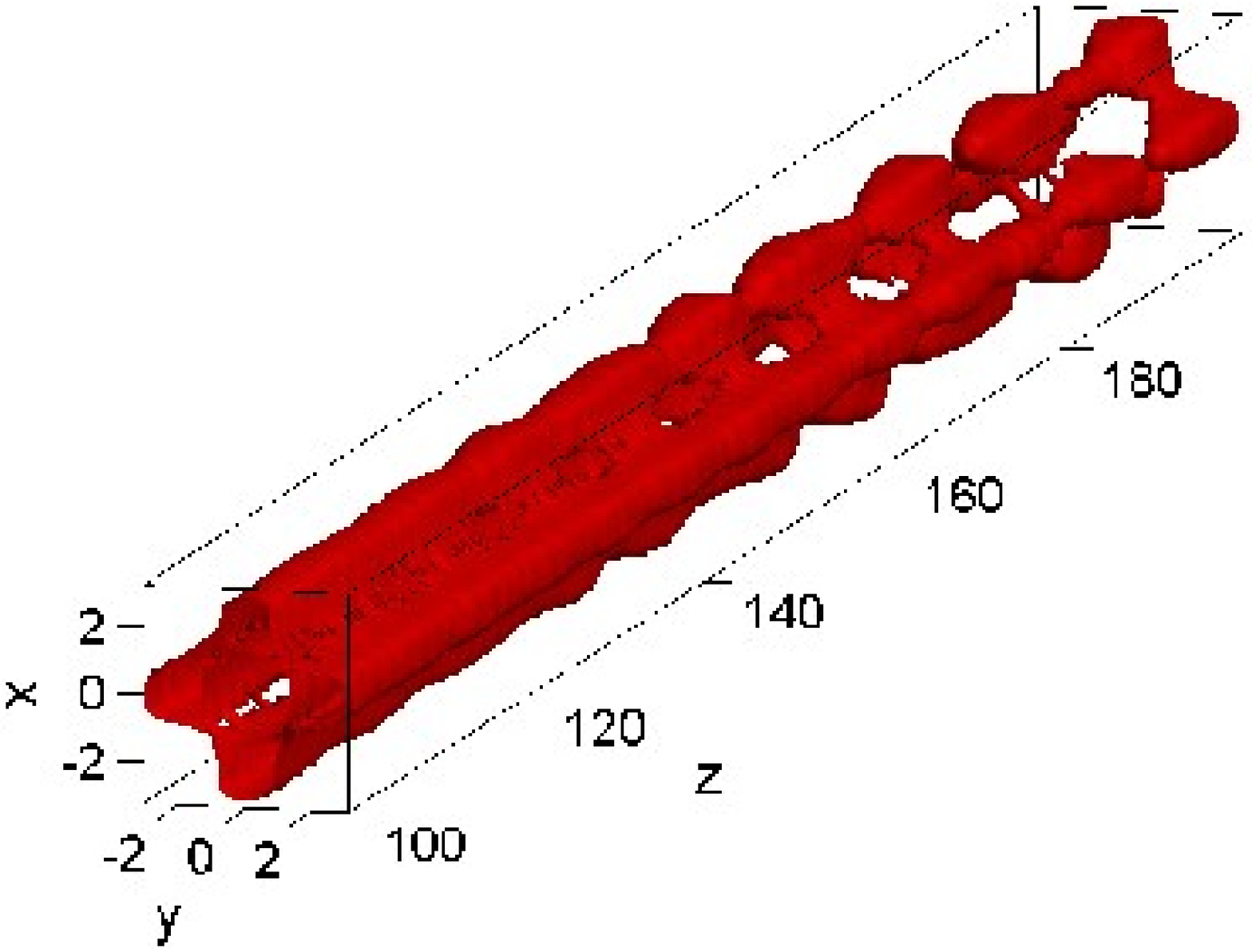}
\includegraphics[width=0.4\textwidth]{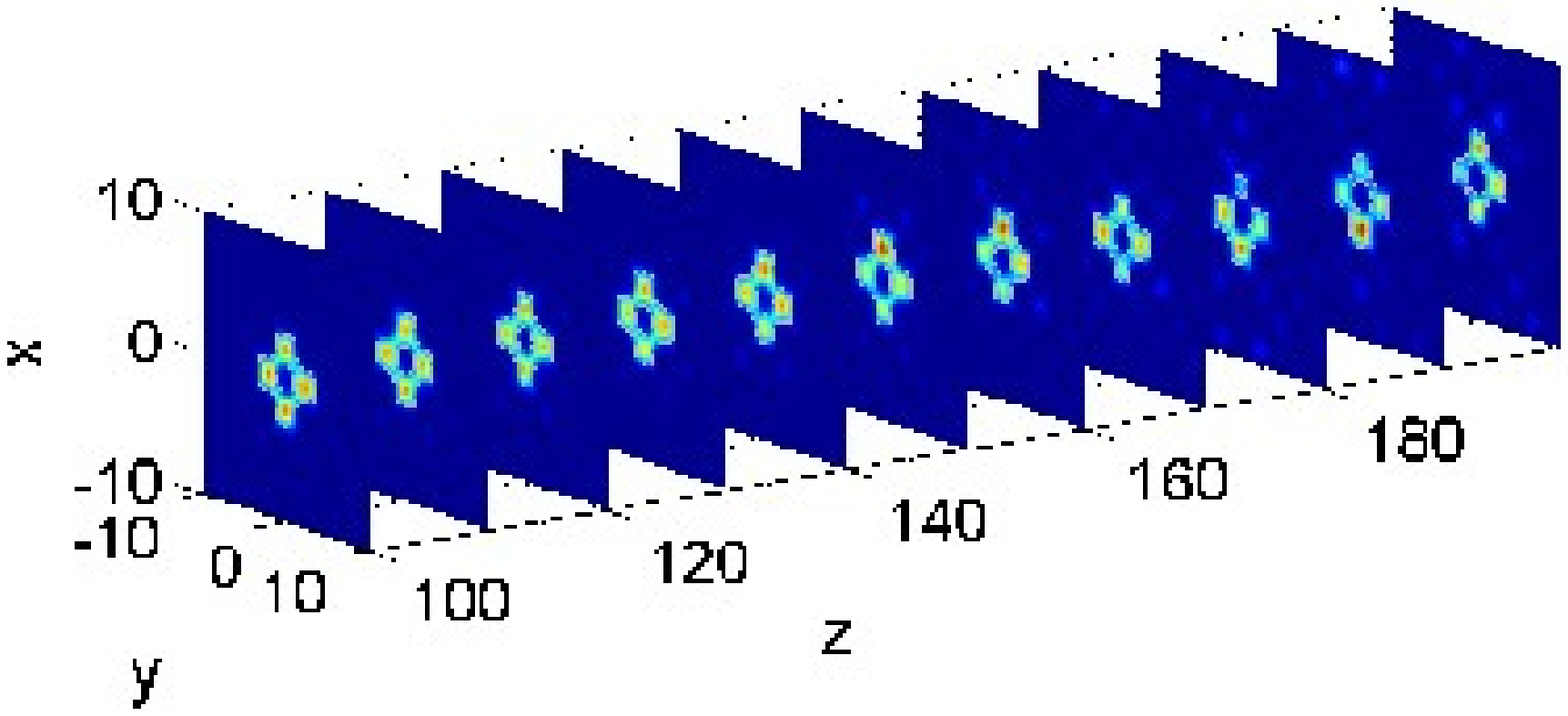}
\end{center}
\caption{(Color online) The evolution of the quadrupole presented in Fig.\ \ref{quadUUUU}.
Shown are the isosurfaces of height 0.2 (red) and 0.15 (blue) and its slices
at particular propagation distances.}
\label{dyn_quadUUUU}
\end{figure}

\subsection{$+--+$ and $+---$ Nearest-neighbor Quadrupole Solitons}

\begin{figure}[tbh]
\begin{center}
\includegraphics[width=0.4\textwidth]{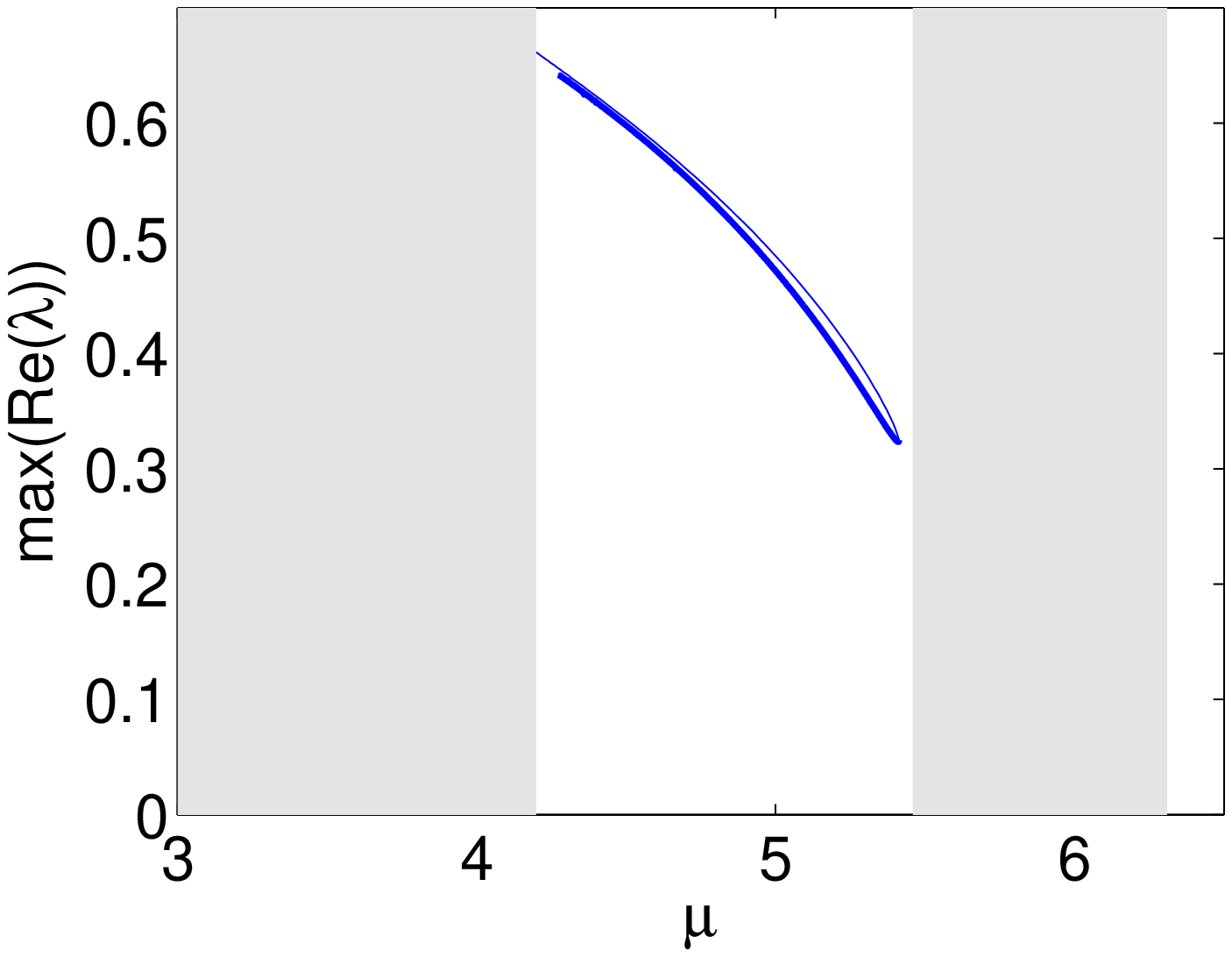}
\includegraphics[width=0.4\textwidth]{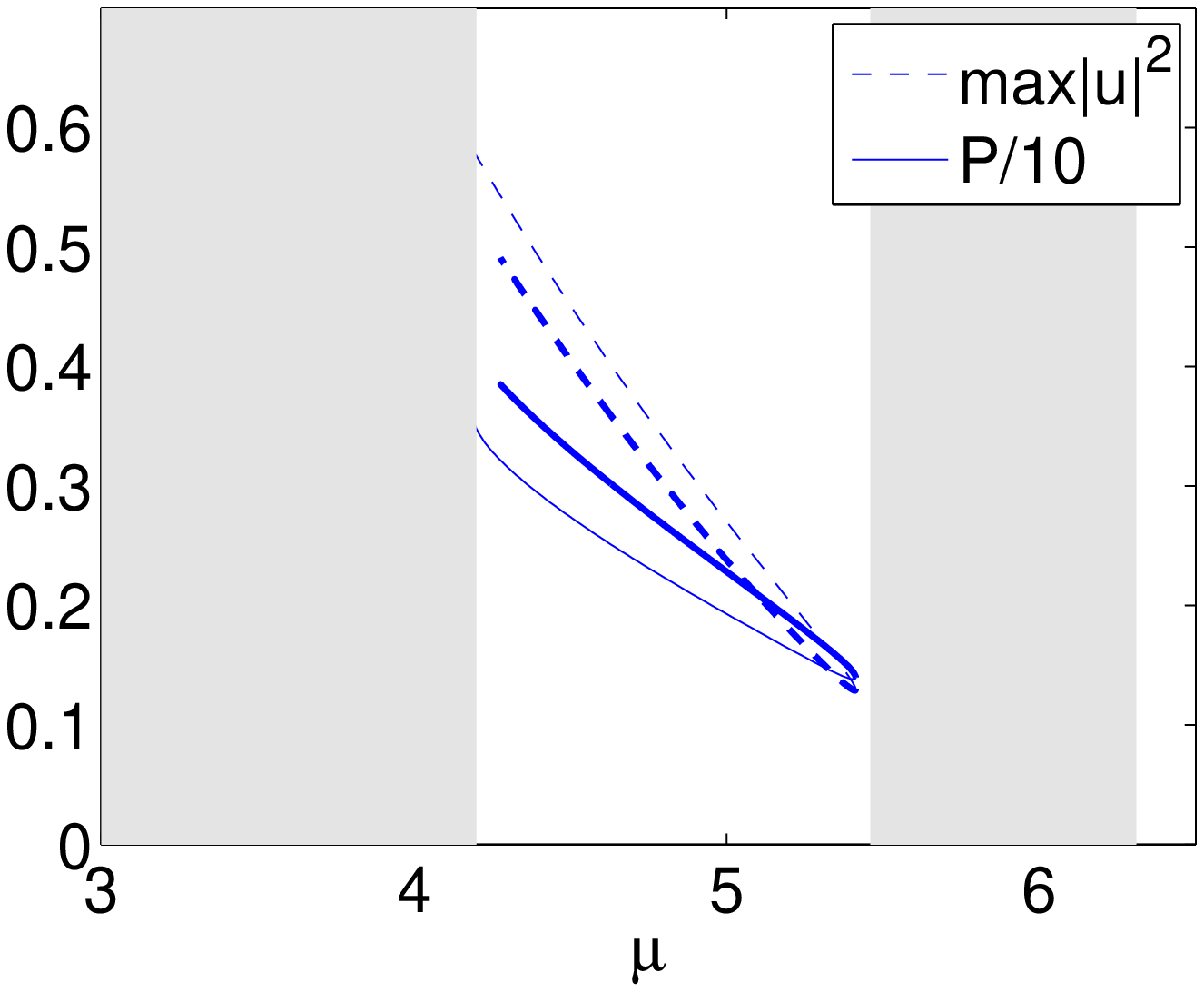}\\
\includegraphics[width=0.4\textwidth]{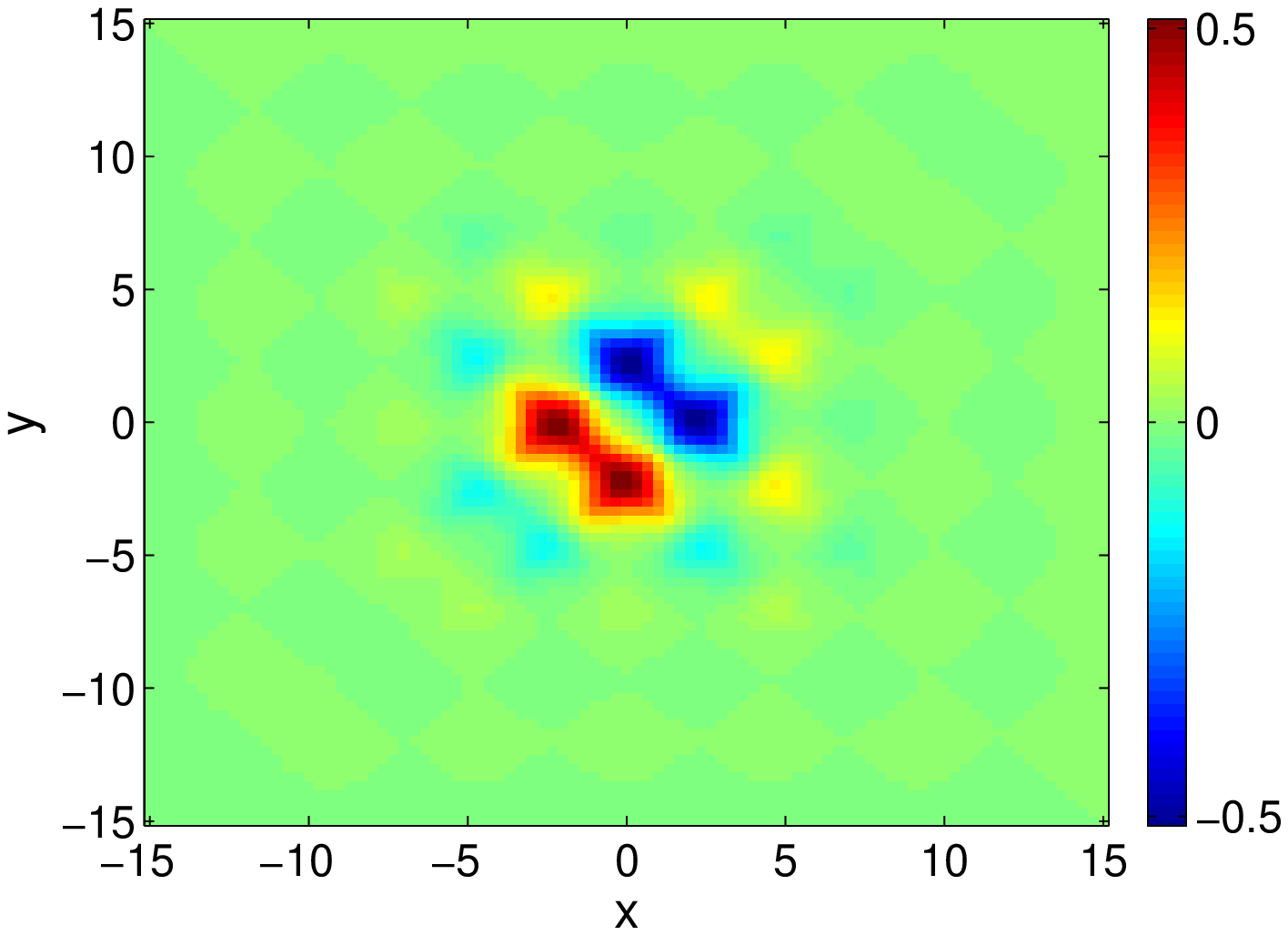}
\includegraphics[width=0.4\textwidth]{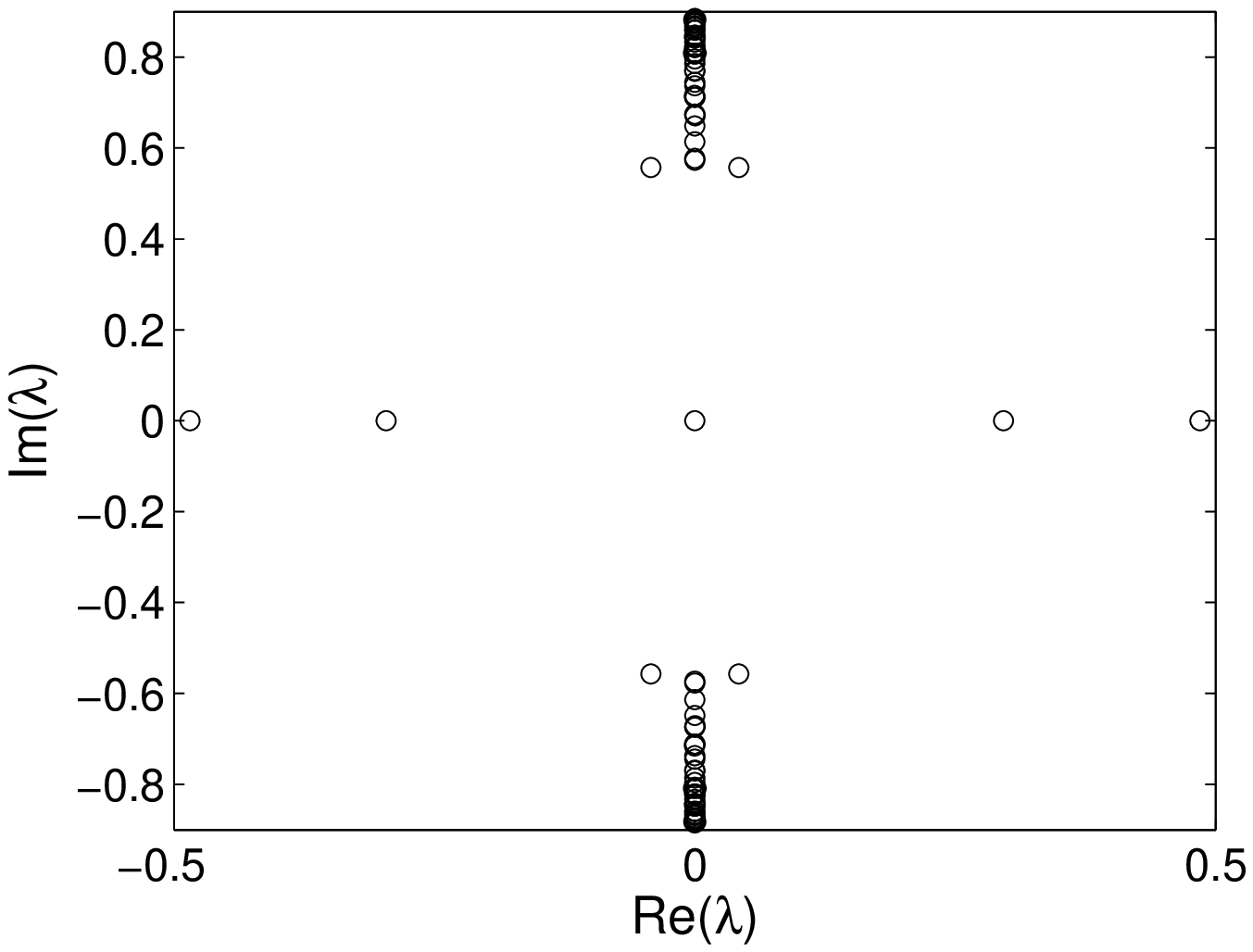}
\includegraphics[width=0.4\textwidth]{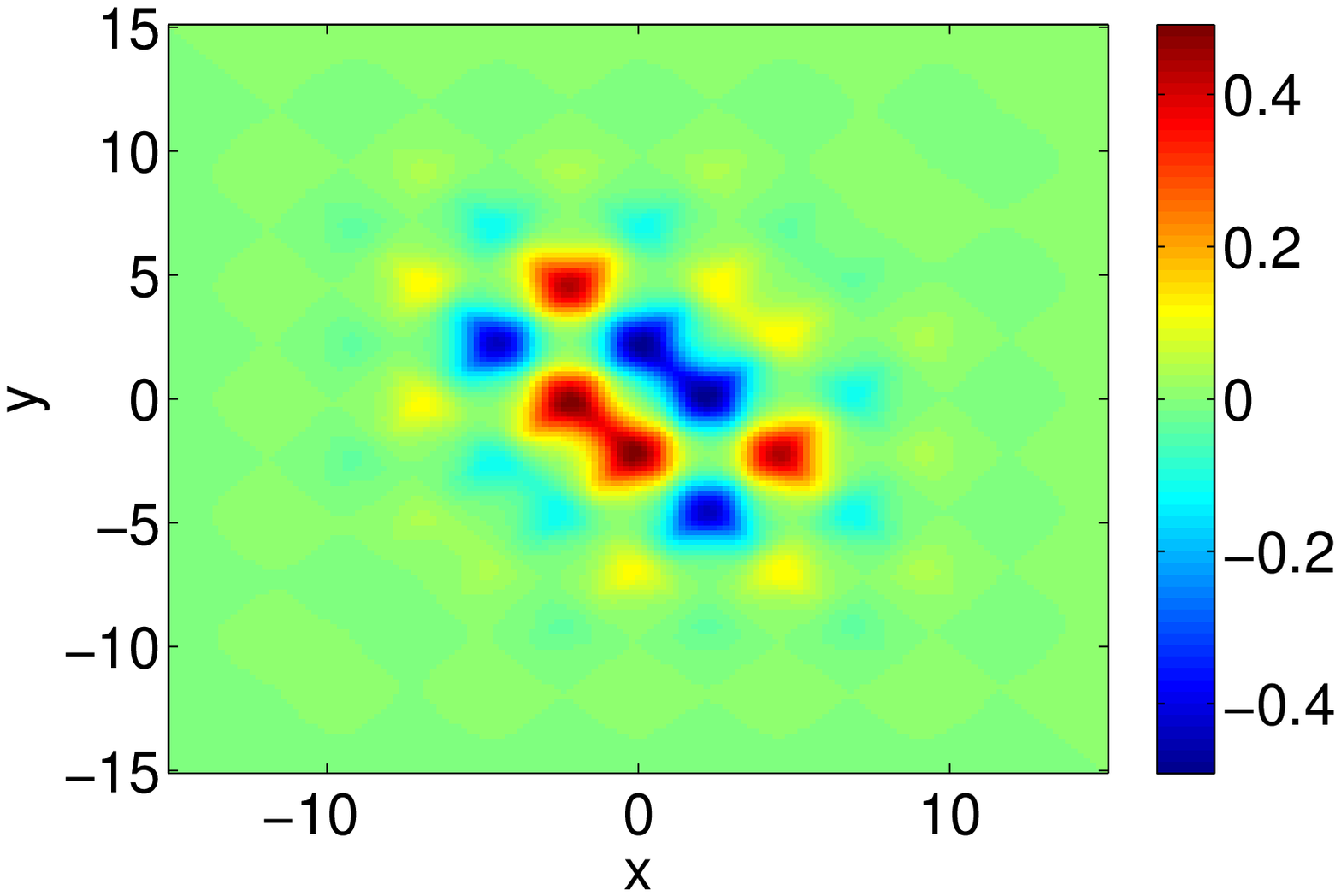}
\includegraphics[width=0.4\textwidth]{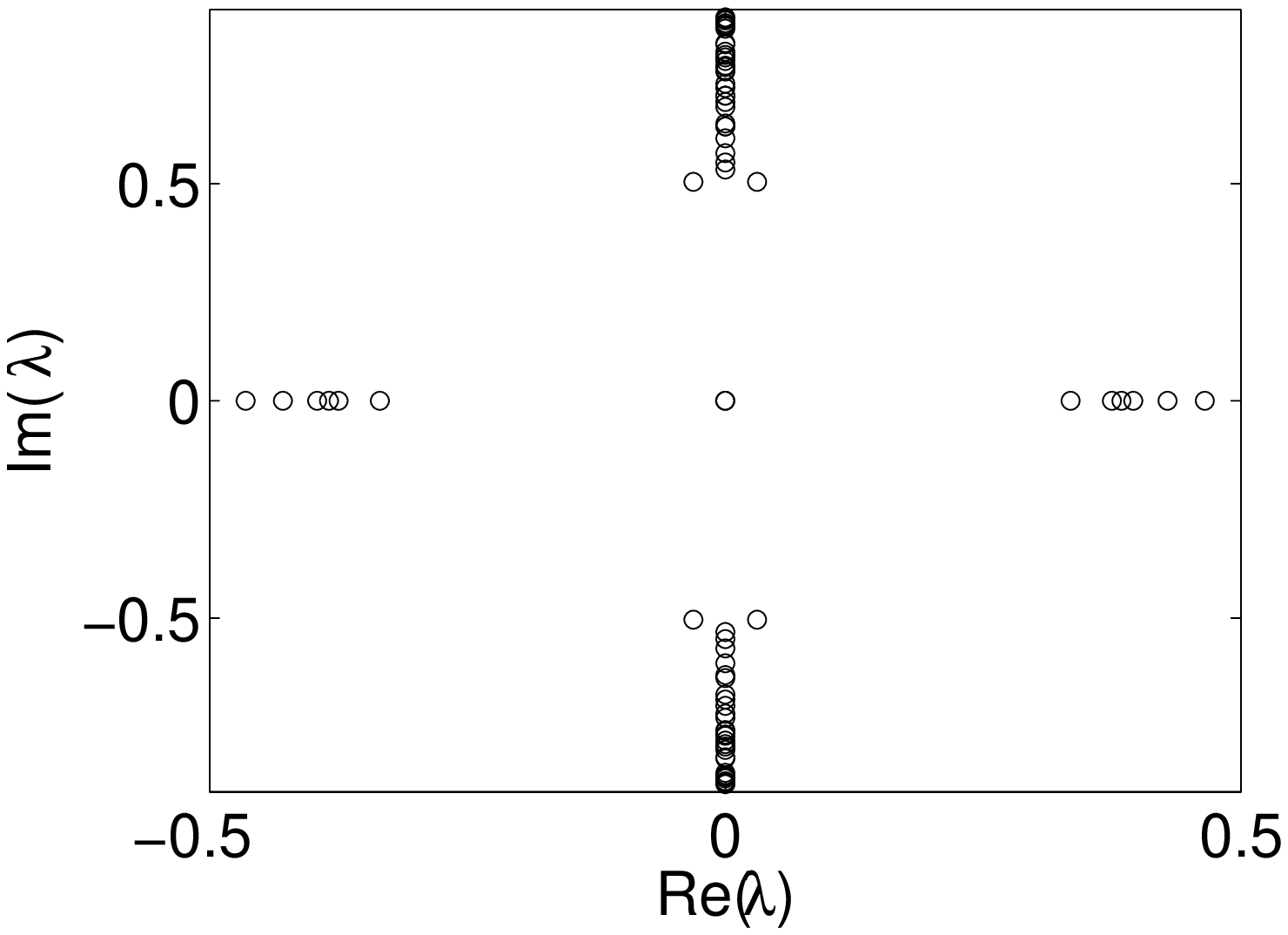}\\
\end{center}
\caption{(Color online) The top four images are the same as Fig.\ \ref{quadUDUD} but for the $+--+$ quadrupoles; once again the
thin lines of the top row correspond to the solution profile and spectral plane
of the middle row, while the bold lines of the top row to the
solution profile and spectral plane indicated in the bottom row.}
\label{quadUDDU}
\end{figure}

One can also examine other types of quadrupole configurations such
as $+--+$ or $+++-$. However, one then typically finds, as may be
inferred by their non-symmetric profile, that these configurations
are either always unstable (as is e.g., the case with $+--+$) or
are not straightforward to obtain numerically in a steady state
form (as is e.g., the case with $+---$). As a representative example
of these asymmetric profiles, we show in the middle panels of
Fig. \ref{quadUDDU} the case
of the $+--+$ profile, which is always unstable due to two real
eigenvalue pairs (and potentially also oscillatorily unstable).


Note that this configuration is equivalent to the superposition of two IP NN (or two
 OOP NN, etc.) dipoles.
Another saddle node bifurcation happens in this case as the solution approaches
 the band edge, leading to the collision of this branch
 with a more extended saddle configuration shown in the
bottom panels of Fig.\ \ref{quadUDDU}.



\begin{figure}[tbh]
\begin{center}
\includegraphics[width=0.4\textwidth]{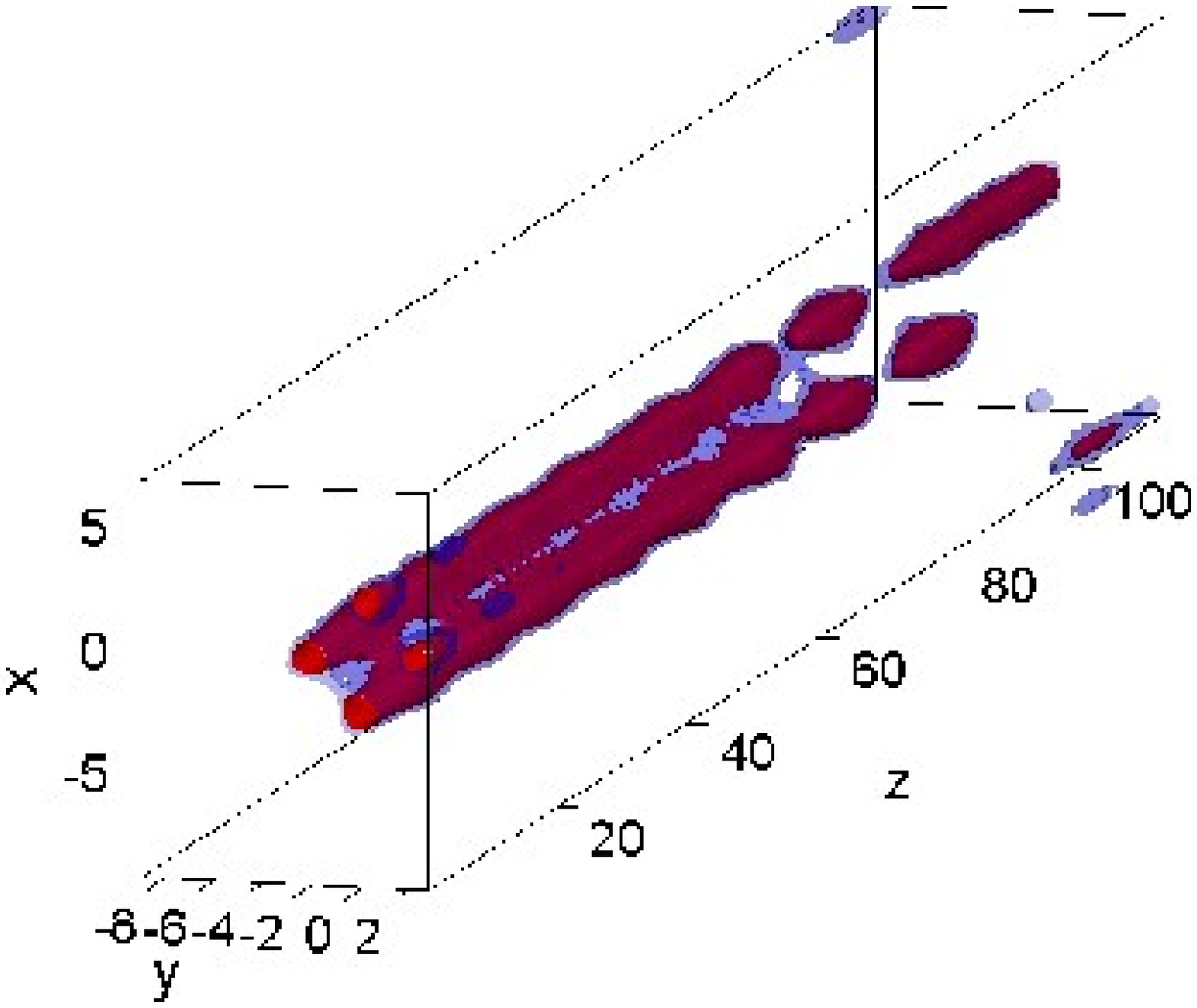}
\includegraphics[width=0.4\textwidth]{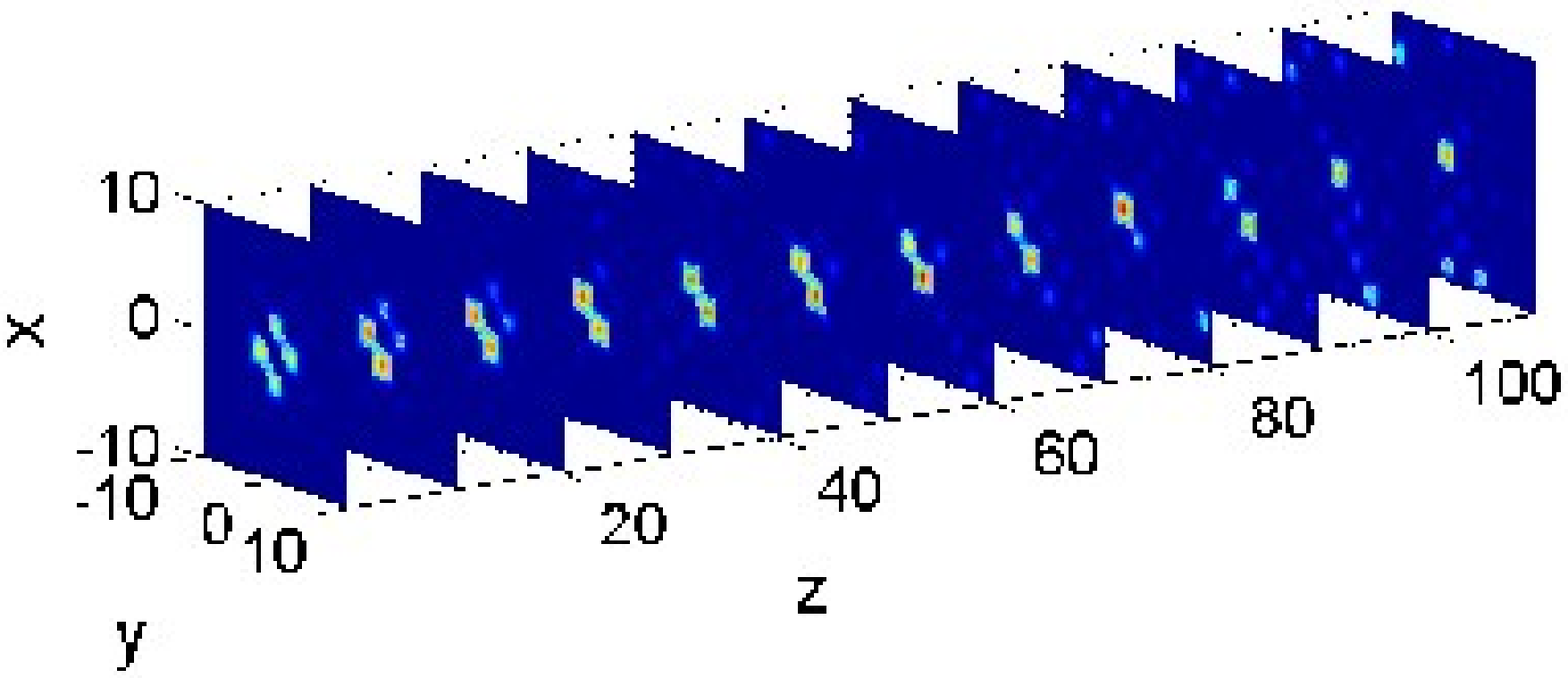}\\
\includegraphics[width=0.4\textwidth]{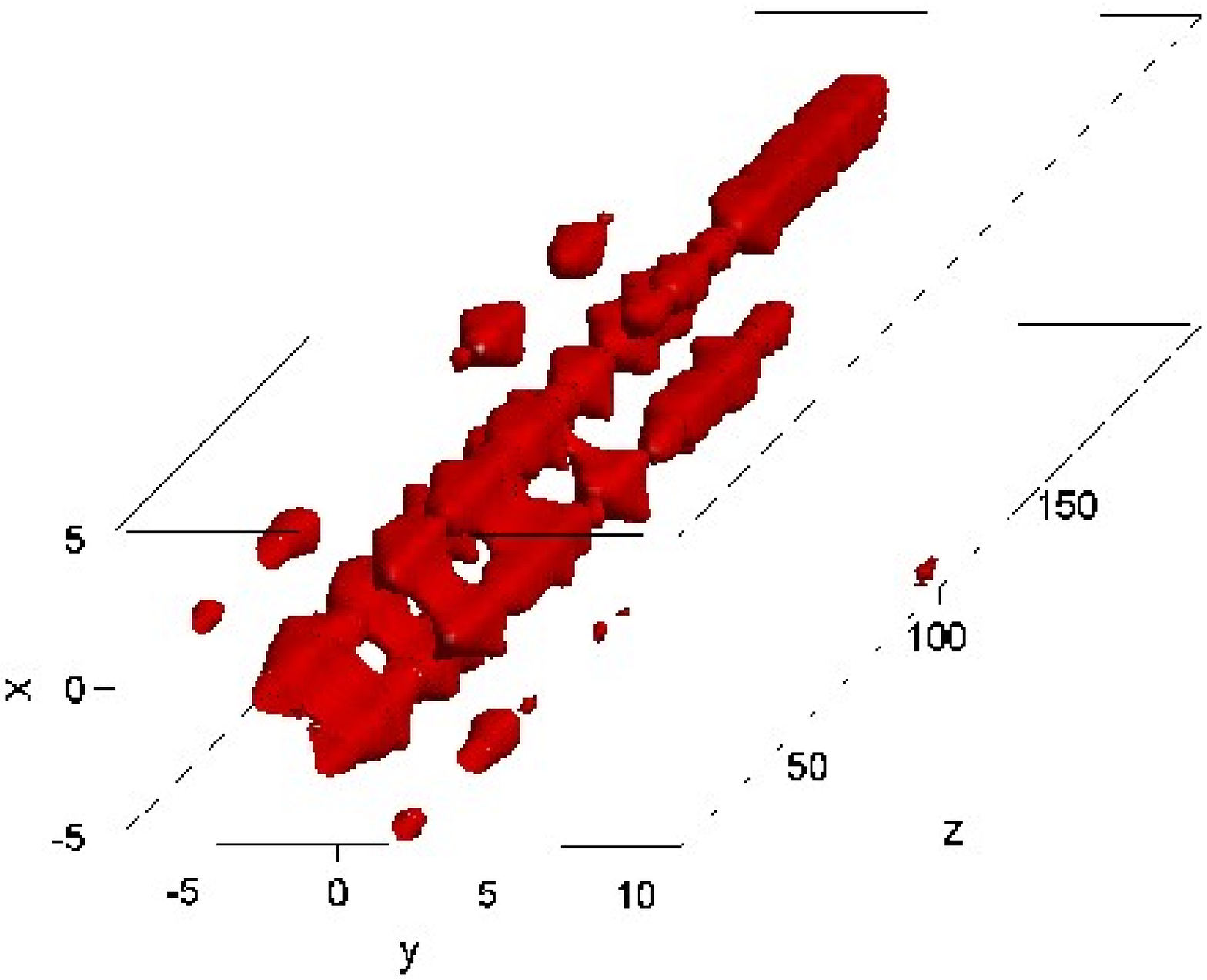}
\includegraphics[width=0.4\textwidth]{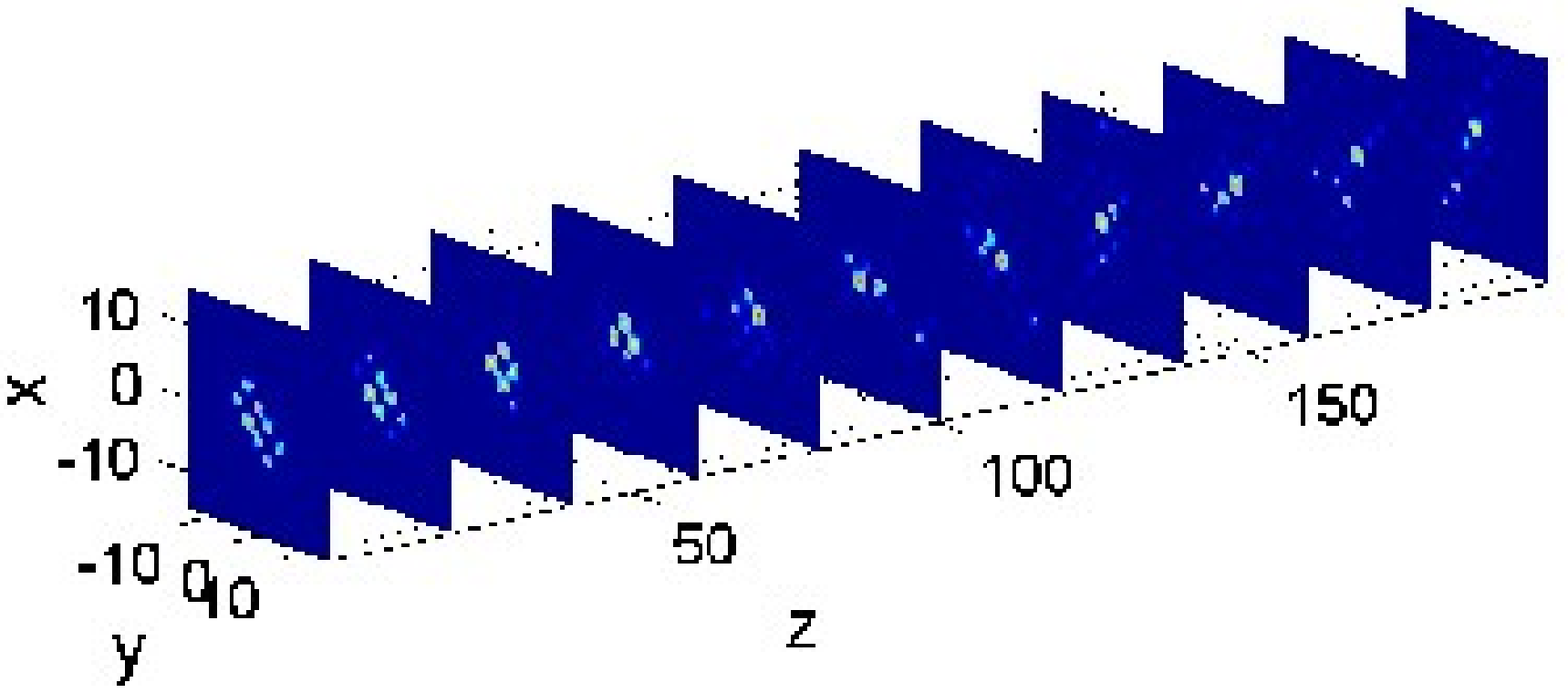}
\end{center}
\caption{(Color online) The evolution of the quadrupoles presented in Fig.\ \ref{quadUDDU}. Top panels
show the isosurfaces of height 0.2 (red) and 0.15 (blue) and the contour at select propagation distances
of the configuration in the middle panels of Fig.\ \ref{quadUDDU}. Bottom panels are the corresponding
figures for the counterpart soliton. The isosurface shown is of height 0.1. Both configurations eventually give rise to single-site localization.}
\label{dyn_quadUDDU}
\end{figure}

In this case also, observing the
dynamics of the instability in Fig.\ \ref{dyn_quadUDDU}, we have seen
the relatively fast degeneration of the mode into a single site solitary
wave profile.
The corresponding evolution dynamics for quadrupoles
in the focusing case (but for short
propagation distances) can be found in \cite{yang04_4}.

\section{Experimental Results}

In our experiments, we use a setup similar to that used in \cite{chenmac}.
A partially spatially incoherent beam (of wavelength 488 nm) is produced
by use of a rotating diffuser. A negatively biased photorefractive
crystal (SBN:60 6x10x5mm$^3$) provides a self-defocusing nonlinearity.  An
amplitude mask is also used to spatially modulate the otherwise uniform beam
after the diffuser, in order to produce a 2D periodic lattice. The mask is
then imaged onto the input face of the crystal, thus creating a pixel-like
input intensity pattern with a spatial period of about 25 $\mu$m. The
ensuing lattice beam (represented by $I_{ol}$ in our theoretical
model) is diagonally oriented and ordinarily polarized, thus the
induced waveguide arrays remain invariant during propagation. An
extraordinarily polarized beam splitting from the same laser output is used
as the probe beam (this is the complex field $u$ that we monitor in
our analysis above). The probe beam (intensity about 6 times weaker than that
of the lattice beam) is sent into a Mach-Zehnder interferometer to create a
dipole-like or quadrupole-like input pattern whose phase is controlled with
the piezo-transducer (PZT) mirrors. The input/output intensity patterns and
the k-space power spectra of the lattice and soliton-forming beams are
monitored with CCD cameras.

\begin{figure}[tbh]
\begin{center}
\includegraphics[width=0.6\textwidth]{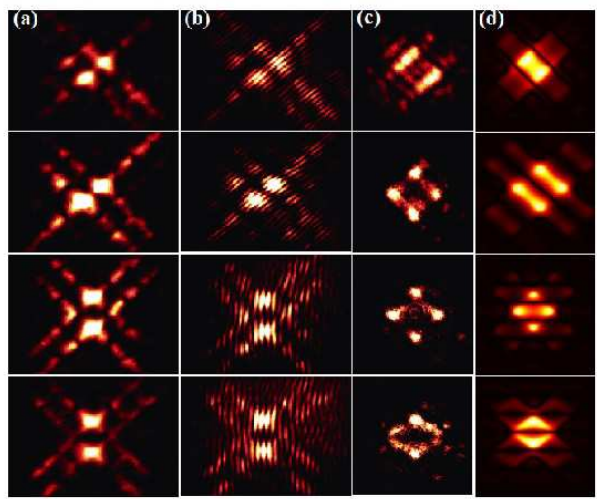}
\end{center}
\caption{(Color online) Observations of dipole like gap soliton. From top to
bottom: IP NN, OOP NN, IP NNN, and OOP NNN. Panel (a) shows the output
intensity patterns, (b) shows the interferagrams of these patterns with a
tilted plane wave, (c) shows the spatial (Fourier) spectra and, (d) shows
those spectra from theoretical calculations of the corresponding solutions.}
\label{zhig1}
\end{figure}

By employing a defocusing nonlinearity, the two-dimensional pixel-like
intensity pattern induces a backbone waveguide lattice whose intensity
minima correspond to the waveguide sites \cite{moti1,chenmac2}.
Launching two narrow Gaussian
beams into two nearest-neighbor or next-nearest-neighbor waveguide sites
either with in-phase or with out-of-phase relations, the probe beams evolve
into dipole-like gap solitons through the 10 mm SBN crystal under a proper
strength of nonlinearity. Typical results are shown in Fig. \ref{zhig1},
where the  output intensity patterns, upon propagation through the
crystal (first column), show two main bright spots. This indicates that
the energy of the probe beams is mostly localized in the waveguides that were
initially excited. The interferograms between these output patterns and a
tilted plane wave show that these two main spots remain in-phase or
out-of-phase, maintaining their initial phase relation, although the
secondary intensity peaks adjacent to the main ones are always out-of-phase
with the primary spots. The spatial spectra (Fourier transform) of the
patterns (third column) are in good agreement with those obtained from the
Fourier transform of the corresponding theoretical solutions (last column).
It is important to note here that although some of these configurations
such as the OOP NN and the IP NNN dipoles (second and third rows
in Fig. \ref{zhig1}) have been found to {\it always} be unstable,
the limited propagation distances inside the crystal (10mm which corresponds to
$z \approx 2.73$ in our dimensionless units) are too short for the
instability to develop to an observable degree in the experiment, while it
is clearly shown in the theoretical analysis of soliton solutions in
the previous sections.

\begin{figure}[tbh]
\begin{center}
\includegraphics[width=0.6\textwidth]{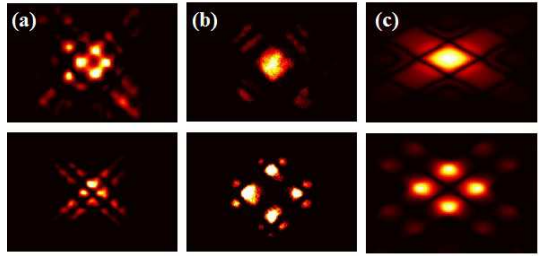}
\end{center}
\caption{(Color online) Observations of in-phase (top) and out-of-phase
(bottom) quadrupole-like gap solitons. Panel  (a) shows the output intensity
patterns, (b) shows the spatial spectra, and (c) shows those spectra from
theoretical calculations of the corresponding profiles.}
\label{zhig2}
\end{figure}

We have also performed experiments to excite quadrupole-like gap solitons,
with four narrow Gaussian beams launched into four adjacent waveguide
sites (i.e., four adjacent intensity minima) for both in-phase and
out-of-phase conditions. Under a proper level of defocusing nonlinearity,
self-trapping is observed with four principal intensity peaks localized in the
waveguides initially excited [Fig. \ref{zhig2}(a)] for both in-phase (top) and
out-of-phase (bottom) conditions. Furthermore, the measured spatial
spectra [of Fig. \ref{zhig2}(b)] for these quadrupole-like gap solitons are
also in good agreement with the corresponding ones obtained from the
theoretical solutions [as shown in Fig. \ref{zhig2}(c)].

\section{Conclusions}

In this communication, we examined in detail theoretically and
numerically the existence, stability
and dynamics of multipole lattice solitons excited with a
saturable defocusing photorefractive nonlinearity.
In this connection, we have obtained a wide array of relevant
structures, including dipoles and quadrupoles, examining
both the different possible configurations (in-phase, as well
as out phase profiles), as well as cases where the excited
sites are at nearest or at next-nearest-neighboring lattices sites.

We have found configurations such as the
in-phase, nearest-neighbor dipole,
the out-of-phase, next-nearest-neighbor dipole or the in-phase
nearest-neighbor quadrupole which are typically stable (although
they may incur oscillatory instabilities). We have also identified
those solitons including e.g., the in-phase, next-nearest-neighbor
dipole, the out-of-phase, nearest-neighbor dipole or the out-of-phase
nearest-neighbor quadrupole which are typically unstable due
to exponential instabilities and real eigenvalues.
The stability results have been found to be in direct agreement
with the results of the analysis of the prototypical
discrete model of this type, namely the discrete nonlinear Schr{\"o}dinger
equation.
Furthermore, we have also identified an interesting set
of bifurcations that are associated with the parametric continuation
and termination of some of the above branches.
The dynamical instabilities encountered in the present work have been
monitored through direct integration of the relevant dynamical
equation. In most cases, the result of evolution has been the
degeneration of the structure to a robust single site solitary
wave, although more complicated oscillatory evolutions
are also possible.

 Our experimental results agree with the theoretical predictions. Different
dipole and quadrupole configurations were realized in the experiment in a
photorefractive crystal with limited propagation length, even the
most unstable ones. This is simply because the growth rate of the dynamical
instabilities is such that the instabilities will manifest themselves
typically for
propagation distances on the order of 100-1000mm, but they are not appreciable
within the short distance comparable to the length of our crystal.


Since the framework of defocusing equations has been studied
far less extensively than their focusing counterparts, it would
be particularly interesting to extend the present considerations
to other structures. Perhaps the most interesting example would
be the study of discrete vortices, not only of single
but also of multiple charge which would be an interesting endeavor
both from a theoretical, as well as from an experimental point of
view. Such structures could be directly connected to ones
studied theoretically also in other fields, including e.g.
weakly interacting Bose-Einstein condensates; relevant examples
of vortices of charges 1 and 2 are given by the tightly bound structures of
Figs. 8 and 10 respectively of \cite{sakaguchi}.
Additionally, the present considerations of defocusing
nonlinearity may be extended
to other settings, including the case of radial lattices of
\cite{rings}. In the latter case, one would expect to observe modulationally
stable, radial beam waveforms with or without vorticity
\cite{kart,us}. Such studies are currently in progress and
will be reported in future publications.

\vspace{5mm}

{\it Acknowledgements}. This work was supported by the 973 Program, NSFC and
PCSIRT in China, and by US NSF and AFOSR (ZC). We thank J. Yang for
discussions.
PGK gratefully acknowledges support from NSF-DMS-0505663,
NSF-DMS-0619492 and NSF-CAREER.

\end{document}